\documentclass[
 aip,
 amsmath,amssymb,
preprint,
]{revtex4-2}

\usepackage{graphicx}
\usepackage{dcolumn}
\usepackage{bm}

\usepackage[utf8]{inputenc}
\usepackage[T1]{fontenc}
\usepackage{mathptmx}
\usepackage{etoolbox}

\usepackage{hyperref}
\hypersetup{hidelinks}

\usepackage[version=4]{mhchem}
\usepackage{siunitx}
\usepackage{longtable,tabularx}
 \usepackage{subfigure}
 \usepackage{subfigmat}
 \usepackage{array}

\makeatletter
\def\@email#1#2{%
 \endgroup
 \patchcmd{\titleblock@produce}
  {\frontmatter@RRAPformat}
  {\frontmatter@RRAPformat{\produce@RRAP{*#1\href{mailto:#2}{#2}}}\frontmatter@RRAPformat}
  {}{}
}%
\makeatother

\begin{document}

\title[Unsteady Compression Ramp Flows]{On Flow Unsteadiness in Strongly Separated High-Speed Ramp Flows using Kinetic and Data Driven Methods}
\author{Irmak T.~Karpuzcu}
\email{itk3@illinois.edu}
 \affiliation{University of Illinois, Urbana-Champaign, IL 61801,USA}
\author{Mert Senkardesler}%
\affiliation{University of Illinois, Urbana-Champaign, IL 61801,USA}%
\author{Deborah A. Levin}
\affiliation{University of Illinois, Urbana-Champaign, IL 61801,USA}%

\begin{abstract}

The gas dynamics of high speed flows over several compression-expansion ramp geometries with large separation regions were investigated using a particle kinetic approach and data driven methods. These large separation regions allow study of the role of scaled angle predicted by triple deck theory on recirculation strength, separation size, and appearance of secondary recirculation regions, all of which lead to flow unsteadiness. Examining four different free stream conditions for the same geometry using the direct simulation Monte Carlo (DSMC) approach, we show that the stability of single recirculation in the separation region is more sensitive to wall temperature and free stream velocity than physical ramp angle. Stability in this context refers to appearance of secondary separation regions. In addition, configurations with similar types of recirculation regions from literature considered, we find that general trends regarding unsteadiness, such as appearance of secondary recirculation regions, mostly hold; however, the flows are too complex to be characterized by only scaled angle. Thus, stochastic flowfields generated by DSMC were investigated using data driven methods, namely spectral proper orthogonal decomposition and dynamic mode decomposition. We found that both methods yield similar frequency components and mode shapes for spatio-temporal coherent structures of unsteady DSMC flows. With the help of these structures, the main mechanisms of unsteadiness were identified as oscillations of the reattachment shock, Kelvin-Helmholtz like instability of the shear layer, and c-shaped mode of the separation bubble. This is the first time SPOD has been applied to a DSMC generated flowfield and compared with DMD.

\vspace{1em}
\noindent{\small\textit{This article may be downloaded for personal use only. Any other use requires prior permission of the author and AIP Publishing. This article appeared in I.~T. Karpuzcu, M. Senkardesler, and D.~A. Levin, Phys. Fluids \textbf{37}, 096136 (2025) and may be found at \href{https://doi.org/10.1063/5.0281770}{https://doi.org/10.1063/5.0281770}.}}
\end{abstract}

\maketitle

\section{Introduction}

The compression ramp geometry is readily encountered in many supersonic and hypersonic vehicles at the engine inlet, fin junction and control surfaces causing complex interactions that create the well known Edney-IV structures.\cite{edney1968}  The interactions of shocks, shear layers, and recirculation regions can create well-known flow unsteadiness as well as trigger laminar to turbulent transition when the incoming boundary layer is laminar.\cite{DollingreviewSWBLI,ClemensReviewSWBLI,GAITONDE201580} In particular, when the separation bubble is large, earlier work has shown that the interaction of leading edge shock, separation shock and the shear layer can cause three dimensional perturbations to grow for a supersonic flow over a nominally two dimensional (i.e. infinite spanwise length) compression ramp geometry~\citep{karpuzcu2024linear}.
The focus of the current contribution is on two-dimensional flows over such configurations where the separation bubble is sufficiently large that its size is limited by the size of the flat plate and the ramp. Further destabilization of these large separation bubbles causes  complex flow phenomenon such as the  appearance of secondary recirculation regions.   In some sense the flow over a  different geometry, known as a ``tick,'' is similar to these compression ramp cases because it achieves leading edge separation through the sudden flow expansion  rather than through boundary layer interactions.\cite{chapman1958investigation,Khraibut_Gai_Brown_Neely_2017,GaiTickPOF2018,Tumuklutick2019,GaiTick2019} In contrast, the compression ramp  achieves separation through flow compression.

A key and widely used approach to studying the flow dynamics over compression ramps is that of
 triple deck theory.~\cite{neiland1969theory,stewartson1970laminar,rizzetta1978triple,cassel_ruban_walker_1995,EXPOSITO2023}  The appearance of more than a single recirculation in the separation region of a compression ramp was investigated by Gai and Khraibut~\cite{gai_khraibut_2019} and Exposito et al.~\cite{Exposito_Gai_Neely_2021} who determined  the scaled angle limits where secondary recirculation regions will occur.  This was similar to the work of Smith and Khorrami~\cite{Smith_Khorrami_1991}, Korolev et al.~\cite{KOROLEV_GAJJAR_RUBAN_2002}, Neiland et al.~\cite{neiland2008temperature}, Shvedchenko~\cite{Shvedchenko_2009} and Egorov et al.~\cite{egorov2011three}, who quantified the scaled angle at which secondary vortices form in the separation for the steady flowfields.  However, less attention was given to the unsteady nature of these flows and the specific mechanisms that might cause the unsteadiness. Furthermore, the configuration considered in this work is not solely a compression ramp, but rather a {\it compression-expansion ramp} that was selected because it allows flow expansion as soon as reattachment occurs for the flow conditions of interest.   Thus the characteristics of the flow unsteadiness can be quite complex. Moreover, it is unknown whether these secondary recirculation regions appearing in two dimensional simulations will eventually cause three dimensionality in the flow if a three dimensional simulation were carried out~\cite{gai_khraibut_2019}. Nonetheless, we focus only on two-dimensional simulations of the compression ramp configurations for this study. This choice is primarily driven by our goal of conducting a parametric study to examine how various flow parameters affect the unsteadiness characteristics of compression ramp flows, while maintaining computational feasibility. Although we recognize that certain configurations may exhibit three-dimensional effects\cite{karpuzcu2024linear,KarpuzcuLevinDC2025}, performing such a parametric study in 3D would be computationally expensive. Therefore, one of the motivations of this work is also to identify which configuration is most compelling and merits further investigation in a full three-dimensional setting.

To study the unsteady nature of laminar flows around a compression-expansion ramp we seek a high-fidelity approach that can capture high gradient regions without assuming continuum and is time accurate by nature, such as the direct simulation Monte Carlo (DSMC)~\cite{Bird} method to resolve the complex shock interactions.  Traditionally, DSMC is used to compute rarefied steady-state flowfields some with high enthalpy involving thermochemistry since its collision based approach is well suited for incorporating the best modeled and measured chemical kinetic cross sections.   Use of DSMC for  time-accurate unsteady flows is more recent~\cite{tumukluPoF,tumukluPoF2,TumukluWPOD2019,sawant_etal_2022,KarpuzcuCompRampSciTech2023,karpuzcu2024linear} because the numerical particle requirements for  instantaneous DSMC flowfields to be statistically convergent is much higher than  time averaged flowfields. However, with the help of the efficiently parallelized software and increase in the computational power it has been made possible. Several canonical hypersonic flows over geometries such as double cones~\cite{tumukluPoF,tumukluPoF2}, double wedges~\cite{TumukluWPOD2019,sawant_etal_2022} and compression ramps~\cite{KarpuzcuCompRampSciTech2023,karpuzcu2024linear} have been of interest and shown to be computed in a time accurate manner with DSMC. Even more, the ultimate unsteady flow phenomenon turbulence has been well captured with DSMC in situations where the Navier-Stokes equations fundamentally cannot capture the smallest scales which correspond to the smallest turbulent scales where the energy cascade terminates and viscous dissipation dominates.~\cite{GallisTurbolenza,GallisDSMCvsNS}.

The usual method to detect flow unsteadiness and determine relaxation rates is to put numerical probes at key locations in the DSMC computational grid to collect the time variation of macroparameters such as number density, pressure, and velocities.  However, to determine the global spatial coherence of the disturbance an approach is needed that provides an integrated picture of the variation of local perturbations. Coherent structures refer to spatially organized disturbances in the flowfield that exhibit orderly patterns over extended regions, governed by shared physical phenomena (e.g., high velocity gradients in shear layers or shock interactions). Unlike local point instabilities detectable via probes, these structures are captured by modal decomposition methods, which solve for unique spatial mode shapes and associated temporal frequencies that best reconstruct the unsteady flowfield. Spatial coherence ensures the disturbances are of physical origin, as they manifest consistently in regions of common flow physics rather than as isolated or noise-induced artifacts. This naturally leads to the use of data driven methods  to analyze and extract coherent flow structures from the unsteady flowfields. The most popular ones were formulated by Lumley~\citep{lumey2012stochastic}, that includes the proper orthogonal decomposition (POD) and its variations originally formulated to capture coherent flow structures within turbulent flows which are stochastic in nature. Schmid~\citep{SCHMID_2010} developed dynamic mode decomposition (DMD) to extract coherent structures as well as the transient growth/decay rates and unsteady frequencies attached to those structures.  The spectral proper orthogonal decomposition (SPOD) methodology proposed and verified by Schmidt and Colonius aimed at extracting flow features as well as the spectral content of the flowfield, thus allowing in a better understanding of the spatio-temporal covariance of unsteady flowfields~\citep{Schmidt_Towne_Rigas_Colonius_Brès_2018,Schmidt_Colonius_2020}. The relation between POD, SPOD, resolvent analysis and DMD were discussed and it was shown that for stationary flows, SPOD gives the same modes as DMD and resolvent analysis~\citep{Towne_Schmidt_Colonius_2018}. Of these SPOD is the less computationally demanding method, thus making it a suitable candidate for when the data size is limited.

The stochastic nature of DSMC flowfields and the treatment of instantaneous sampling have previously been of interest \cite{GallisDenoiseReport2006,WagnerMonteCarlo2005,karabulut2009stochastic,WagnerStochastic2011}, alongside studies on uncertainty quantification of the sampling process and de-noising of DSMC data \citep{StefanovDSMC2011,KumardenoiseDSMC2018}. These studies focus on post-processing the noise in sampled DSMC macroparameters, which arises either from having very few particles in local collision cells in some cases or from limited time sampling in others. They show that data-driven methods involving the Fourier transform, similar to POD formulations, can be effective in de-noising DSMC flowfields. However, the application of data-driven methods, such as the aforementioned POD-based approaches, to DSMC flowfields  for both de-noising macroparameters  and extracting coherent structures from unsteady flows is relatively recent. For example, Tumuklu et al. \citep{TumukluWPOD2019} in their work applied right orthogonal proper decomposition (WPOD) to de-noise and predict the state evolution of the DSMC flowfields for a hypersonic flow over a double wedge geometry. Since this was done mainly to extract flowfields of the steady state results from the snapshots and de-noise the flowfield, it did not involve spectral calculations that estimate the unsteady frequency bands. In contrast, Klothakis et al. \cite{KlothakisDMD} applied DMD to DSMC flow around a roughness element as well as Sawant et al. \cite{sawant_etal_2022} who utilized DMD to extract the modes pertaining to a linear stability analysis for 3D simulations of flow over a double wedge and showed that the unstable modes agreed with linear stability theory.

In this work we use the high fidelity particle-kinetic DSMC method to generate time accurate high speed flowfields over compression ramps. First, we will investigate the destabilization of the flow with the appearance of the secondary recirculation regions and their relation with the scaled angle quantity of the triple deck theory defined as~\cite{stewartson1970laminar};
\begin{equation}
    \alpha = \alpha^* \frac{Re_L ^\frac{1}{4}}{0.332^\frac{1}{2}C^\frac{1}{4}(M_\infty^2 - 1)^\frac{1}{4}}, \hspace{20pt} C = \frac{\mu_w T_\infty}{\mu_\infty T_w} \label{eq:scaleang}
\end{equation}
where $\alpha^*$ is the angle of the ramp in radians, the Reynolds number is defined as,
\begin{equation}
\label{eq:Reynolds}
Re_L = \rho _\infty U_\infty L/\mu_\infty
\end{equation}
and $\mu$ is the dynamic viscosity, $T$ is the temperature and the subscripts $w$ denotes values at the wall, and $\infty$ at the freestream.
 Second, the flowfield unsteadiness will be investigated using data driven methods to extract coherent flow structures and identify the main mechanisms causing flow unsteadiness. The organization of the remainder of the paper is as follows. In Section~\ref{sec:FreeStreamDSMC} we present the free stream conditions and parameters for the four cases  that will be considered in this work as well as change in the salient flow features at steady state predicted by DSMC.  We augment these cases with from the literature to understand the ability of scaled angle to predict unsteadiness characterized by the presence of multiple recirculation regions, separation length, recirculation strengths in Sec.~\ref{sec:ScaledAngle}  Although scaled angle predicts general trends we find that the flows are too complex and a deeper examination of the flows is required.  This leads us to the discussion of the application of data driven methods to DSMC in Sec.~\ref{sec:datadrivenmethods}.  Then, in Sec.~\ref{sec:FlowUnsteadiness} we use DMD and SPOD on two of the four original compression ramp cases to assess the relative importance of Kelvin-Helmholtz type instabilities within the shear layers and the presence of c-shaped modes within the separation bubble and their effect on flow unsteadiness. Our conclusions are presented  in Sec.~\ref{sec:conc}.

\section{Effect of Scaled Angle on Compression Ramp Cases Computed with DSMC} \label{sec:FreeStreamDSMC}

We briefly discuss the well known DSMC method and its implementation for modeling the unsteady behavior of flow over the compression ramp as well as present four base time averaged flows with initial parameters selected by triple deck theory.

\subsection{Direct Simulation Monte Carlo}

Direct simulation Monte Carlo (DSMC) is a particle-kinetic method that models the Boltzmann transport equation~\cite{Bird}. In this method, a finite number of simulation particles are used to represent a certain number real particles, characterized by a parameter called \textit{FNUM}, which is generally a large number of order $\textit{FNUM} \gg 1$. \textit{FNUM} is the ratio of real number of particles to the number of simulation particles. DSMC separates the translational motion and collisions of simulation particles. For each time step, particles are first moved with a discrete jump like motion, i.e. they are assumed to move along a distance of their velocity vector times the time step. After that the collision pairs are selected and collisions are performed and the final state of the particles are determined.

DSMC generated flowfields are stochastic in nature because of several reasons. First, all the events during a DSMC computation are decided by the physics based probability of the event and a the random number generation. An event can be the collision of two particles, collision of the particle with a surface or how the energy is exchanged between the particles after the collision is performed. In addition to this, injection of the particles to the flowfield through the boundaries are handled by a sampling process from a Maxwellian probability density function by the method of acceptance-rejection, again introducing stochasticity to the flowfield.
Since DSMC uses simulation particles to generate the flowfields its fundamental output is  the spatial locations and velocity vectors of the computational particles. To obtain macro-parameters, such as number density, velocity, and temperature, computational particles from each time step need to be sampled over sampling cells.   To decrease the variance  of macro-parameters, this sampling is repeated over several time steps for the steady state solutions.

In this study, we performed DSMC simulations of both supersonic and hypersonic flows over compression ramps, considering a range of free stream conditions that produced different scaled angles. The simulations were carried out using the Scalable Unstructured Gas-dynamic Adaptive mesh Refinement solver, SUGAR-3D~\cite{Sawant2018}. All configurations share the same free stream conditions, with a translational temperature of T$_\infty=50$~K and number density n$_\infty=14.32\times10^{21}$, using molecular nitrogen as the working fluid. The flat plate starts at $x=0.01$~m, the ramp starts at $x=0.19$~m and the domain length and height are $0.30$~m  and $0.06$~m respectively. The flat plate length $L=0.18$m is used as the characteristic length. Using this length scale and free stream mean free path the Knudsen number is calculated as $Kn=\lambda_\infty/L=3.30\times10^{-4}$. The flat plate and ramp walls are isothermal walls with a temperature of $T_w=275K$ for the baseline configuration. The total number of particles used in the domain is about $ 200\times10^{6}$ with at least 150 simulation particles for each sampling cell, and 20 particles per collision cell at each time step for all cases. These values are larger than typical DSMC simulations of non-reacting steady state gas flows which typically employ about only four simulation particles per collision cell.  The sampling cell size is 0.2~mm with a time step value of $ 5.0\times10^{-8}s$ for all configurations as well.
SUGAR employs the majorant frequency scheme~\cite{majorant} for modeling collisions. To capture the temperature dependence of viscosity with greater accuracy, we employed the Variable Hard Sphere (VHS) model~\cite{Bird} with molecular nitrogen at a reference temperature of 273K, with molecular diameter and viscosity coefficient of $4.17\times10^{-10}m$ and 0.255, respectively.

The methodology for the numerical convergence for DSMC is essentially different than its partial differential equation (PDE) counterparts as it is a kinetic method. In DSMC, sampling grids hold no influence over the numerical accuracy of the solution whereas collision grids, that are constructed via adaptive mesh refinement in SUGAR so they have much smaller cell sizes than sampling grids, do have influence. Two main parameters that are the indicators of numerical convergence in DSMC are the ratio of the local mean free path to the collision cell size (LKn) and ratio of the local mean collision time to the time step (TR) of the solution that are calculated at each collision grid cell. LKn should be on the order of 1 and there should be at least 4 simulation particles in each of these collision cells so that the requirement for having at least 4 simulation particles in $\lambda_{local}^3$ is satisfied. On the other hand, TR should be greater than 1 at each time step. In our simulations we choose DSMC numerical parameters that satisfy these requirements. Moreover, DSMC simulations with half (100M) and twice (400M) the number of simulation particles of the baseline case of the current study (200M) are conducted and it was shown that the results were identical. Therefore, it can be said that there are enough particles in the flowfield, and numerical parameters are sufficient to resolve the flow physics of the configurations considered in this work

DSMC flow simulations are usually computed and studied in the steady state limit.  Since each computational particle represents a finite number of true molecules, $\textit{FNUM}\gg 1$, there are two sources of fluctuations, {\em i.e}. numerical and thermal.  Note that the latter fluctuations are the physical response of the flow due to the collisions of particles.  However,  DSMC is inherently time-accurate, an important feature for studying transient flows.  Regardless, any DSMC simulation must be based on a time step smaller than the locally varying mean collision time throughout the flow with at least four simulation particles in a cell volume of $\lambda_{local}^3$. Steady state DSMC solutions can be rendered with sufficient statistics by averaging (or sampling) instantaneous macroparameter values over a sufficiently large sample after the flow has reached steady state.   However, it is still challenging to extract {\it instantaneous flow structures} from DSMC flowfields because of the stochastic variation of flow macroparameters.   For this reason the use of data driven methods, one of the most efficient procedures to extract such coherent structures, will be discussed in Secs.~\ref{sec:datadrivenmethods} and \ref{sec:FlowUnsteadiness}.

\subsection{Scaled Angle as a Criterion for Flow Destabilization}

Using triple deck scalings, it has been shown that flow structures can be related to a parameter known as the scaled angle,~\cite{stewartson1970laminar} as defined in Eqn.~\ref{eq:scaleang}.

 A primary application of the scaled angle is to help predict the onset of secondary recirculation within a laminar separation bubble~\cite{gai_khraibut_2019}.  To date, Reynolds numbers considered in literature~\cite{gai_khraibut_2019} have been  much higher than those considered in this study and led to flow configurations where secondary circulation was expected to occur  $\sim \alpha>5$.
This work examines how free stream velocity, wall temperature, and ramp angle influence the separation bubble across four test cases summarized in Table~\ref{tab:CasesFreeScaled}. According to Eqn.~\ref{eq:scaleang}, these configurations yield scaled angles that exceed those explored in prior studies. As shown in the table, a reduction in wall temperature leads to a slight decrease in the scaled angle. Increasing the free stream velocity and ramp angle, however,  results in more significant changes with the former decreasing and the latter increasing the scaled angle for the last two cases in  Table ~\ref{tab:CasesFreeScaled}. Note that the scaled angle values in Table~\ref{tab:CasesFreeScaled} are calculated using the free stream conditions. The preliminary results for these cases were reported in Karpuzcu et al.~\cite{KarpuzcuRampAviation2022,KarpuzcuCompRampSciTech2023,Karpuzcu2024ScaledAngle}.

These free stream parameters were intentionally chosen so that the scaled angles are large, $\alpha>5$.
  In an    earlier study, Karpuzcu et al.~\cite{karpuzcu2024linear}, showed that the baseline configuration of Table~\ref{tab:CasesFreeScaled}  creates a large separation bubble that is susceptible to 3D perturbations. Having large separation bubbles over such short compression ramps creates  interactions with the leading edge shock as well as the complex interaction of the expansion waves and reattachment shock at the expansion corner. Since the compression ramp cases in the literature largely omit both these conditions, i.e. they have a longer flat plate region that ensures that  the separation bubble is far from the leading edge,  and longer ramps to achieve  reattachment far from the expansion waves created by the expansion corner, this study focuses on understanding these effects in terms of the stability of  2D separation bubbles.

\begin{table}[htb!]
\caption{Free stream properties and resulting scaled angles for each configuration.  Tw175: reduced wall temperature (175 K); M6: increased free stream speed (Ma=6); CR45: increased ramp angle (45$^{\circ}$).}
\label{tab:CasesFreeScaled}
\begin{center}
\scalebox{0.9}{
\begin{tabular}{cccccccc}
\hline
\hline
 &Baseline&Tw175& M6 & CR45\\
\hline
Scaled angle (-)&8.60&8.45&7.07&9.22\\
Ramp angle ($^{\circ}$) &42.0 & 42& 42 & 45.0\\
Free stream velocity ($m/s$)&432.0&432&864.0&432\\
Reynolds number (-)&$11,200$&11,200&$22,400$&11,200\\
Mach number (-)&3.0&3.0&6.0&3.0\\
Wall temperature (K)&275.0&175.0&275.0&275.0\\
Stagnation temperature (K)&140.0&140.0&409.9&140.0\\
\hline
\hline
\end{tabular}}
\end{center}
\end{table}

\begin{figure}[htbp]
\center
\subfigure[Baseline]{\label{fig:base}\includegraphics[trim=100 40 100 750,clip,width=0.47\linewidth]{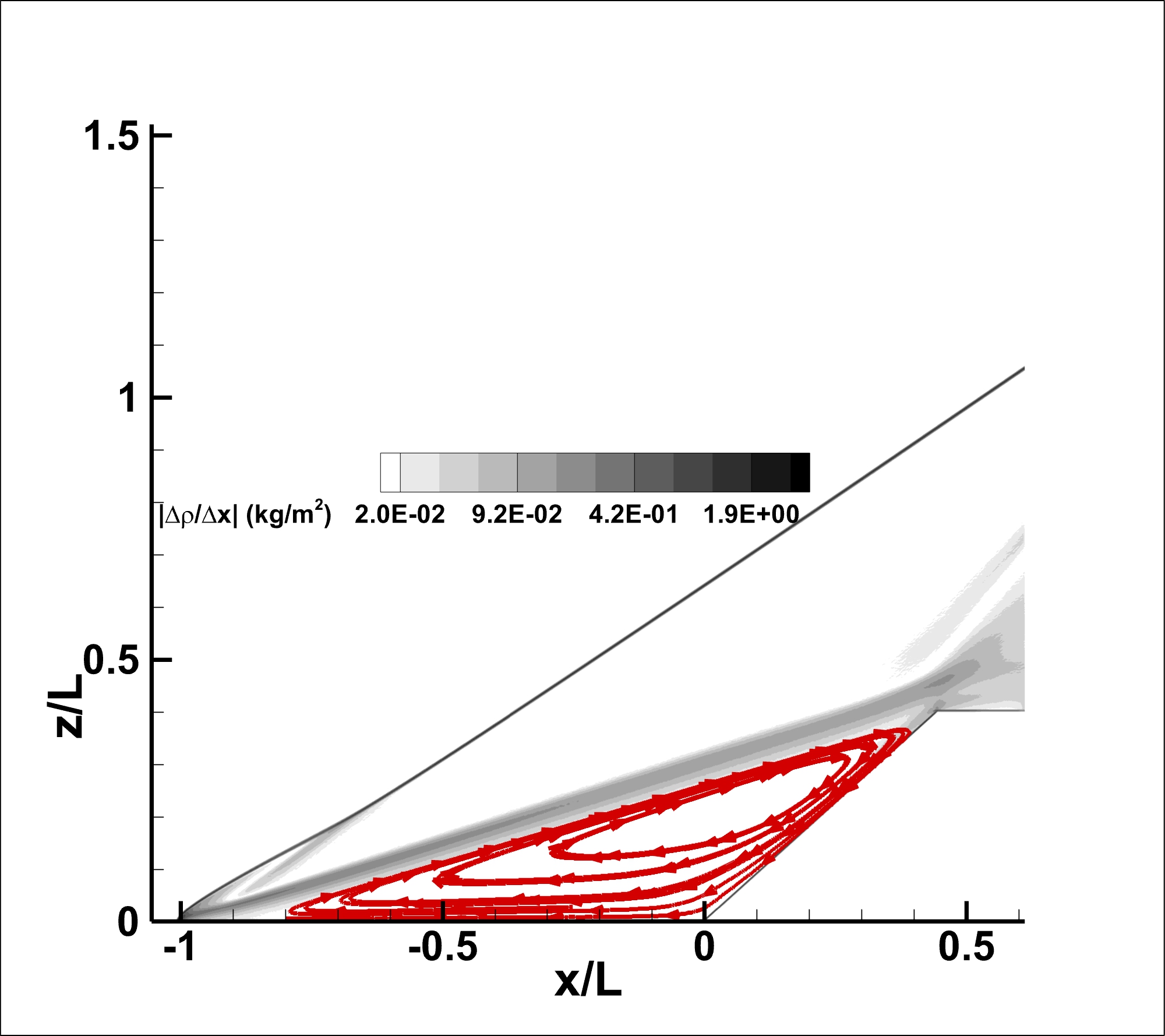}}
\subfigure[Tw175]{\label{fig:Tw175}\includegraphics[trim=10 100 5 60,clip,width=0.47\linewidth]{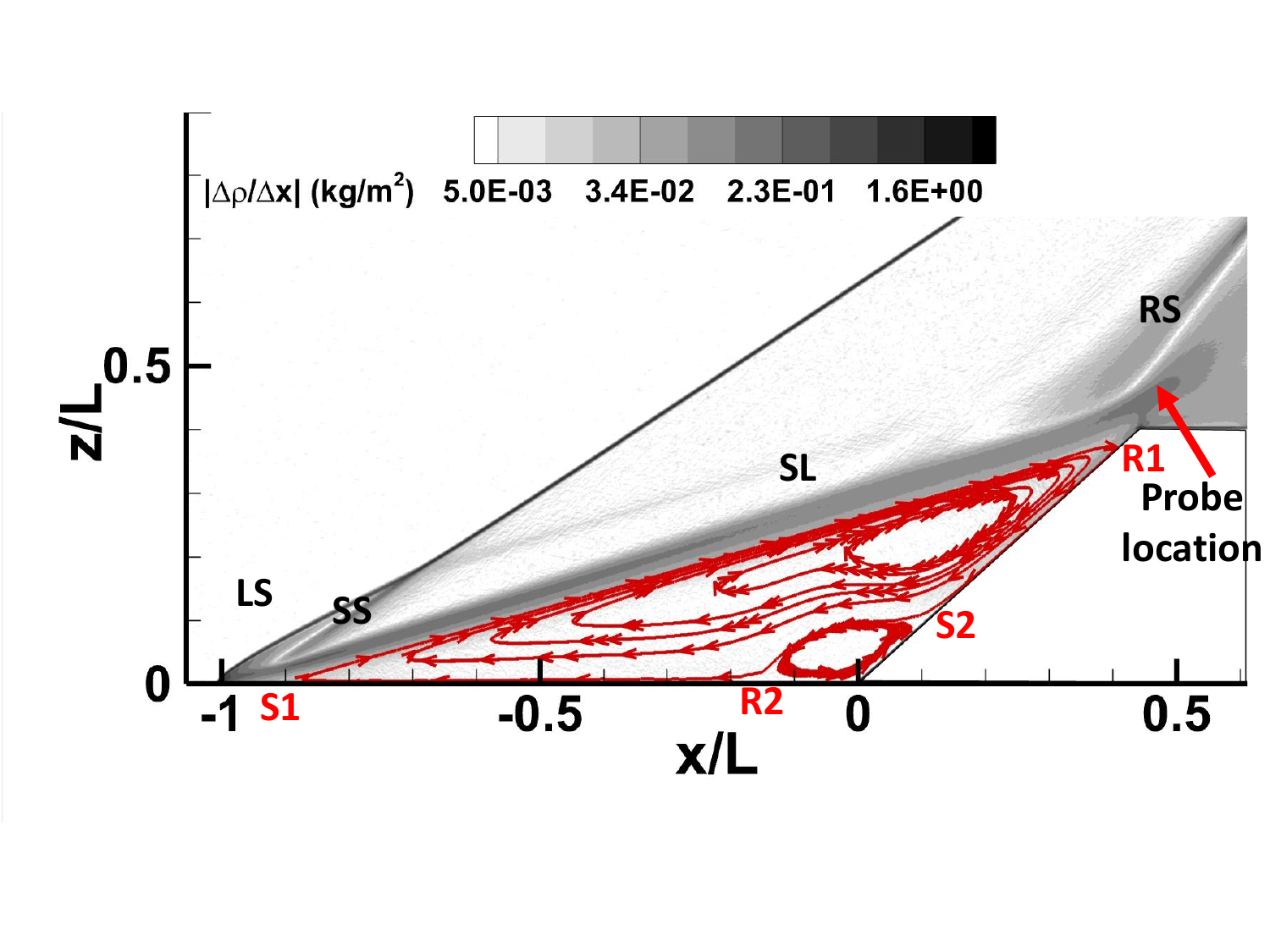}}
\subfigure[M6]{\label{fig:M6}\includegraphics[trim=10 100 5 80,clip,width=0.47\linewidth]{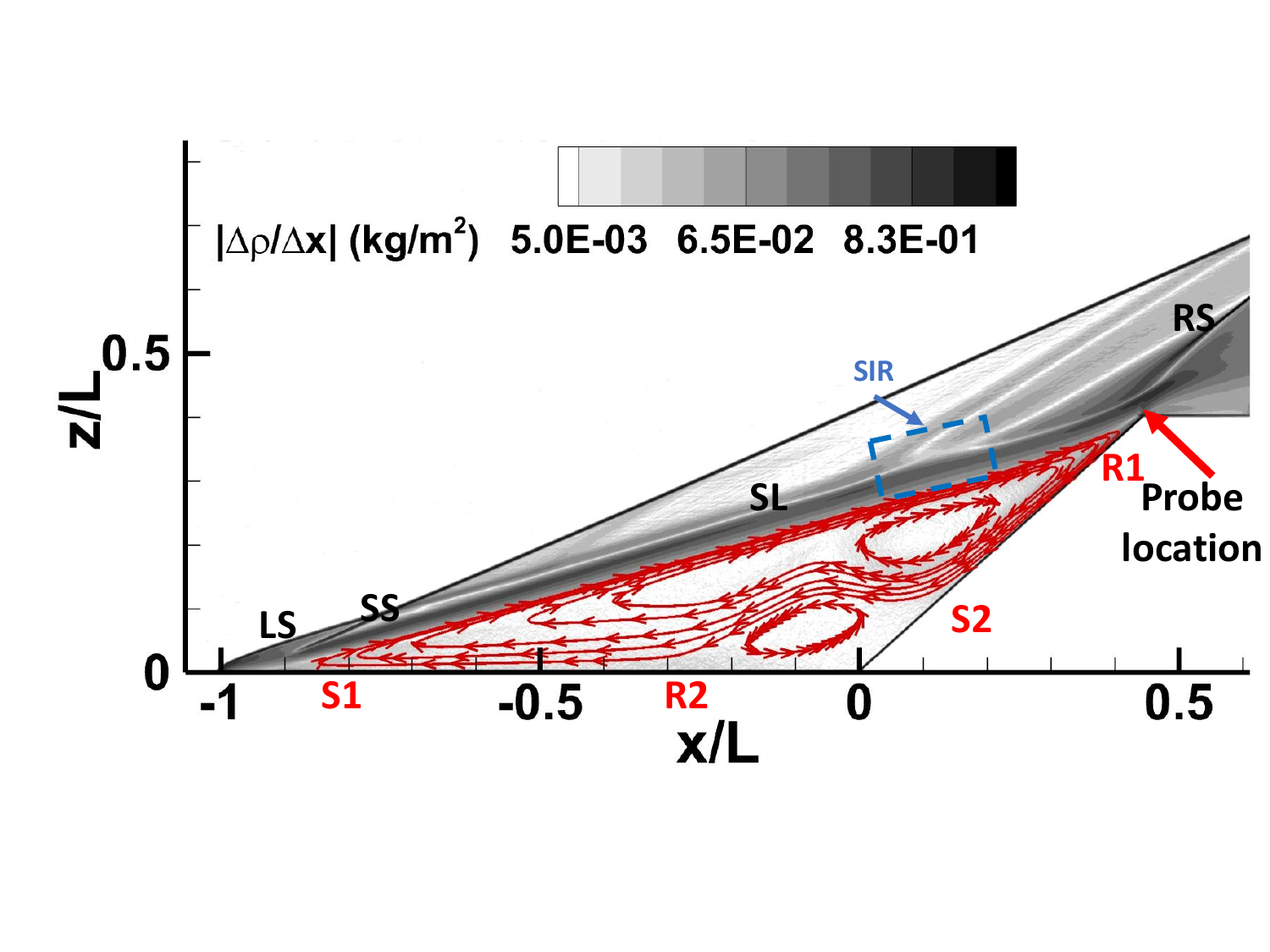}}
\subfigure[CR45]{\label{fig:Ramp45}\includegraphics[trim=10 140 5 40,clip,width=0.47\linewidth]{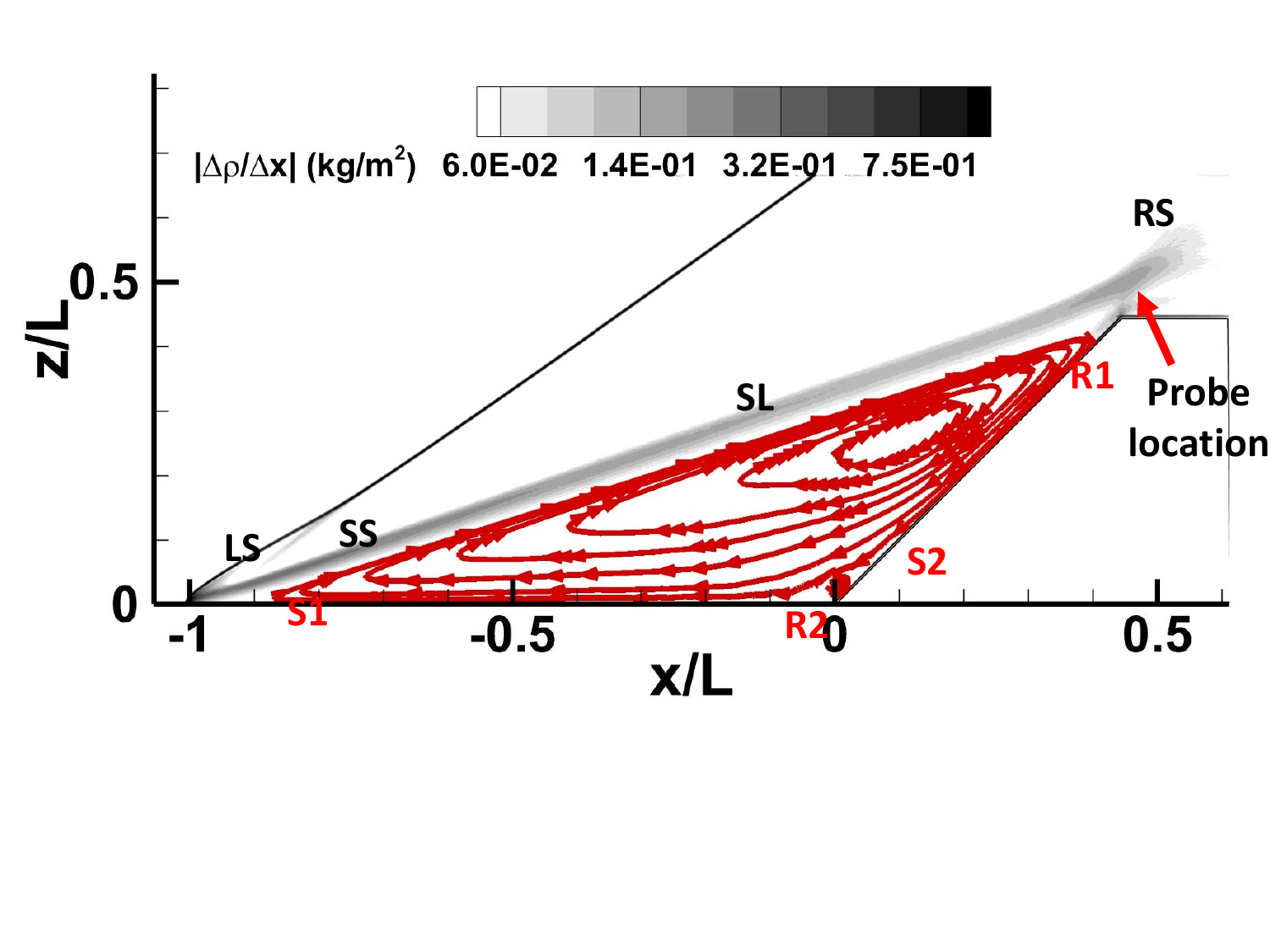}}
\caption{Numerical Schlieren images with streamlines for each ramp configuration given in Table~\ref{tab:CasesFreeScaled}.  Probe is located near the reattachment shock for all cases at x/L=0.41 and y/L= 0.37.}
\label{fig:Schlieren}
\end{figure}

Figure~\ref{fig:Schlieren} presents DSMC numerical Schlieren images (density gradient contours) that have been time averaged for at least 100,000 time steps after the flow establishes over the ramps for the four cases given in Table~\ref{tab:CasesFreeScaled}, and the preliminary results for these cases were discussed in Karpuzcu et al.~\cite{Karpuzcu2024ScaledAngle}.  It is evident that a defining feature of the compression ramp flows analyzed in this study is the presence of a large separation bubble in all configurations, with flow separation initiating near the tip of the flat plate.  In addition,  the common features for all four configurations are the canonical flow features of separated compression ramp flows at the leading edge shock (LS), followed downstream by the separation shock (SS) caused by the boundary layer separation due to the compression ramp and the separated shear layer (SL). For the cases with more than one recirculation, primary separation and reattachment points (S1, R1) as well as the secondary separation and reattachment points (S2, R2) are also indicated. In those cases, secondary separation occurs after flow reattaches at location R1 and starts moving upstream where it separates again at location S2 and then reattaches at location R2.  The compression ramp flow over the baseline configuration presented in Fig.~\ref{fig:base} shows that there is a
single and steady circulation present even though the scaled angle is much higher than five. This highlights the significance of the free stream Reynolds number or more precisely, the influence of rarefaction effects on the behavior of the separation bubble. These effects introduce additional relaxation mechanisms that help sustain a large separation region in the flow~\cite{INGER200742,Inger2008}. We observe that the baseline configuration represents a limiting case with a single steady circulation region.

When selected flow parameters are altered, secondary circulations emerge, indicating that further destabilization of the primary recirculation region leads to the appearance of secondary vortices.
 First, we consider the effect of the wall temperature as shown in Fig.~\ref{fig:Tw175} for configuration Tw175. The decrease in the wall temperature from $T_w/T_0=1.95$ to $T_w/T_0=1.25$ where $T_0$ is the stagnation temperature of the flow, results in a modest reduction in the scaled angle; however, the flowfield undergoes significant changes nonetheless. It can be seen from Fig.~\ref{fig:Tw175} that even though the scaled angle decreased slightly, secondary recirculation started to occur for this case. For configuration M6, the free stream velocity is increased, which causes both Mach number (from $M=3$ to $M=6$) and Reynolds number to increase and scaled angle to decrease compared to the baseline configuration. Even though the increase in the free stream speed directly affects the Reynolds and Mach numbers in the scaled angle calculation, it also alters the stagnation temperature and thus the total to wall temperature ratio. Thus results regarding M6 configuration is therefore is a combination of all of these effects. However, this again causes secondary circulations to appear in the bubble with a smaller scaled angle, see Fig.~\ref{fig:M6}. This suggests that at lower Reynolds numbers, the influence of Reynolds number or the free stream velocity on the flow becomes more dominant than that of the Mach number. The scaled angle (Eqn.~\ref{eq:scaleang}) depends on the Reynolds number only to the 0.25-th power, compared to the Mach number’s dependence of 0.5, predicting that a linear increase in both quantities leads to a lower scaled angle.  However, it appears that the flowfield responds as if the scaled angle is higher, i.e. a much more complex flowfield and strong unsteadiness becomes evident.  This suggests that in this flow regime the power of the Reynolds number in the scaled angle definition might also need revision. In addition, increase in the Mach number causes the leading edge shock to have a smaller shock angle which then interacts with the separation shock and results in a transmitted shock. This transmitted shock interacts with the separated shear layer downstream, as indicated by SIR in Fig.~\ref{fig:M6}, and creates Edney type interaction. This results in a bump in the shear layer and causes compression waves, which can be also observed in Fig.~\ref{fig:M6}. Lastly, we increase the ramp angle in configuration CR45 and again saw the secondary circulation appearing even though it is a very small region concentrated near the corner, see Fig.~\ref{fig:Ramp45}. A $3^\circ$ increase results in a substantial increase in the scaled angle however it was not as effective to create secondary recirculation regions compared to free stream speed and wall temperature. We have also considered other free stream conditions with reversed parameter direction, i.e. increase in the wall temperature or decrease in the ramp angles. However, all those cases resulted in more stable flowfields with smaller separation regions so we have not included them in the cases discussed in this work.  Interested reader may find these results for other ramp angles in Karpuzcu et al.~\cite{KarpuzcuRampAviation2022}.

\section{Ability of Scaled Angle to Predict Unsteadiness}\label{sec:ScaledAngle}

\begin{figure}[htbp]
\center
\subfigure[Separation length]{\label{fig:ScalevsSepLength}\includegraphics[trim=190 10 200 10,clip,width=0.47\linewidth]{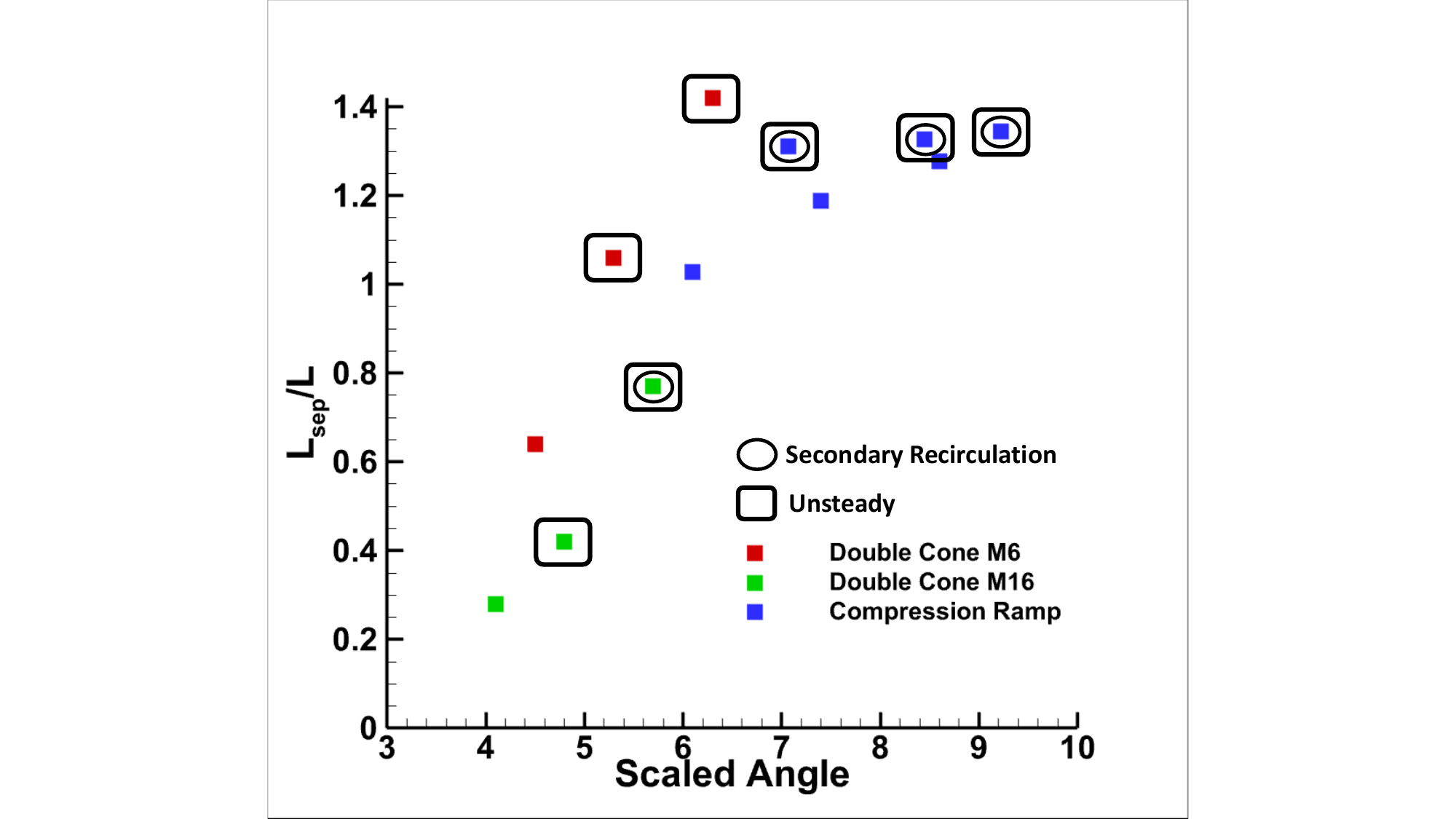}}
\subfigure[Recirculation strength]{\label{fig:ScalevsRecirc}\includegraphics[trim=190 10 200 10,clip,width=0.47\linewidth]{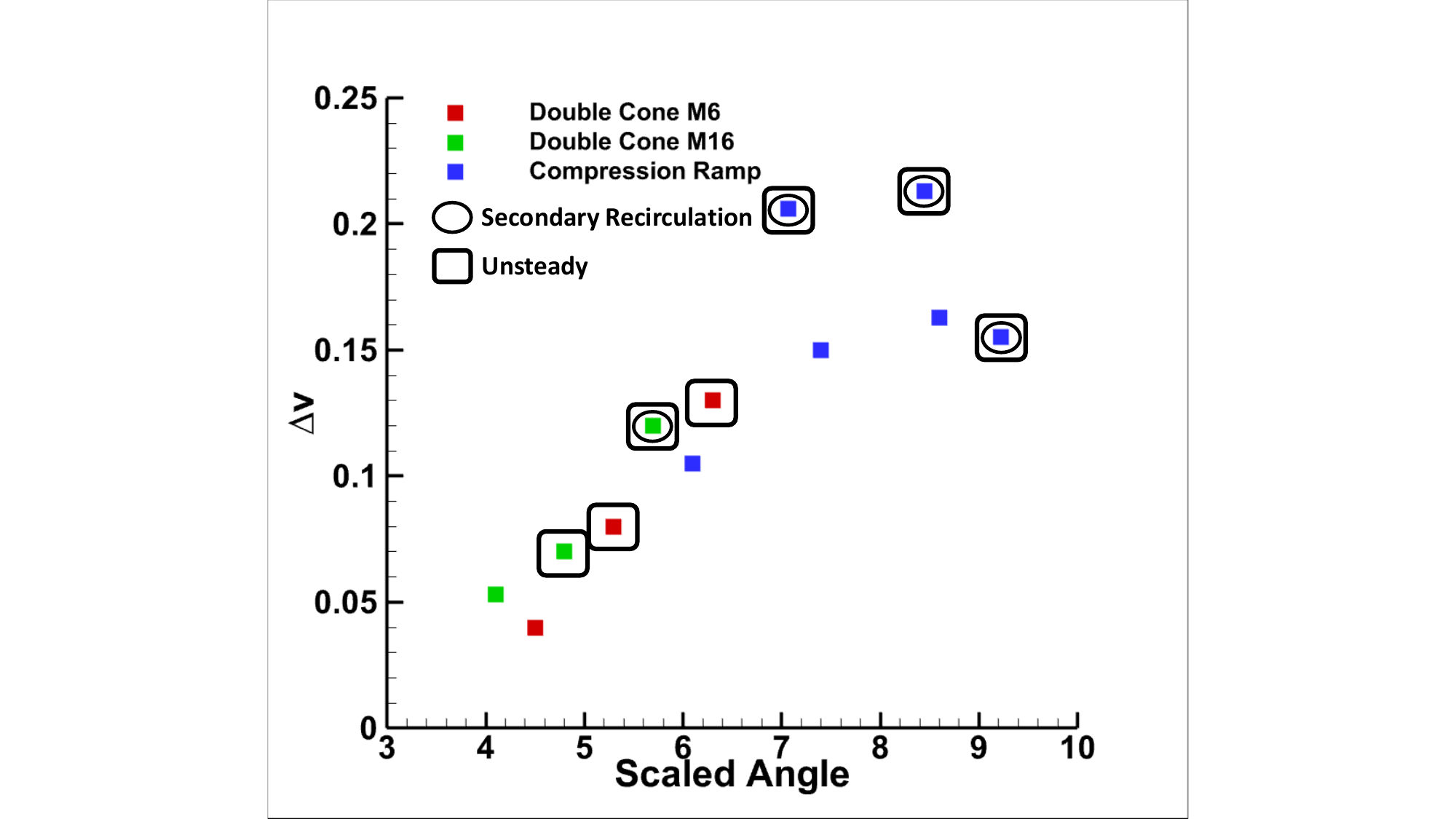}}
\caption{Variation of separation length and recirculation strength with respect to the scaled angle. Configurations with several recirculation regions are indicated by circled symbols. Rectangular symbols indicate the cases with unsteadiness. The data indicated by Double Cone M6 (red rectangles) is from Karpuzcu et al.~\cite{KarpuzcuDCM6}, Double Cone M16 (green rectangles) is from Tumuklu et al.~\cite{tumukluPoF,tumukluPoF2} and a portion of the data for the compression ramp cases is taken from Karpuzcu et al.~\cite{KarpuzcuCompRampSciTech2023}. Figure extended from Karpuzcu et al.~\cite{Karpuzcu2024ScaledAngle} with additional unsteadiness information.}
\label{fig:ScaledComparison}
\end{figure}

Despite the power of high-performance computing that enable high-fidelity simulations such as DSMC, we still seek to understand the general nature of the unsteady flow gas dynamics.  To test the ability of scaled angle to predict the degree of unsteadiness we
 combine our results  with the earlier studies by Tumuklu et al.~\cite{tumukluPoF,tumukluPoF2} and Karpuzcu et al.~\cite{KarpuzcuDCM6,KarpuzcuCompRampSciTech2023} which investigated flows over double cones under various free stream Mach and Reynolds numbers, as well as compression ramps with differing physical angles. The detailed description of the flowfields for the previously studied cases can be found in  Tumuklu et al.~\cite{tumukluPoF,tumukluPoF2} and Karpuzcu et al.~\cite{KarpuzcuDCM6,KarpuzcuCompRampSciTech2023}. In the case of the double cone configurations, scaled angles were determined by treating the physical angle between the two cones as an analogue to the ramp angle in compression ramp flows. Other parameters, such as Mach number and Reynolds number, were based on values taken downstream of the conical shock. We choose two characteristics parameters, separation length and recirculation strength, to compare the differences in flowfields in between the configurations, as shown in Fig.~\ref{fig:ScaledComparison}. Separation length, $L_{sep}$, is defined as the length of the straight line drawn between the separation point and the reattachment point and recirculation strength, $\Delta v$, is defined as the ratio of the magnitude of the maximum negative streamwise velocity within the separation bubble to the free stream velocity~\cite{Theofilis2000}. Firstly, as the scaled angle increases, separation length and recirculation strength also increases, as can be seen from Figs.~\ref{fig:ScalevsSepLength} and ~\ref{fig:ScalevsRecirc}. Comparing these two parameters, separation length seem to be asymptote to a value of 1.3, which indicates that it is restricted by the physical size of the ramp whereas recirculation strength seems to increase without an asymptotic behavior. An alternative explanation is that because the separation bubble cannot get physically larger, further destabilization of the separation bubble causes a higher recirculation strength. In this context, stability or destabilization refers to the presence or absence of secondary separation regions forming within the flow.

However, a key challenge lies in identifying the range of scaled angles that support a stable separation bubble with a single recirculation region without transitioning into multiple recirculation zones. As shown in Figs.~\ref{fig:Tw175},~\ref{fig:M6} and  ~\ref{fig:Ramp45} and the circled symbols shown in Figs.~\ref{fig:ScalevsSepLength} and ~\ref{fig:ScalevsRecirc} more than one  recirculation  was found to be present in several cases where configurations with single recirculation regions also present for similar scaled angles. For example, configurations with very high scaled angles such as $\alpha=8.60$, are shown to maintain a single separation region for the compression ramp cases whereas scaled angle values of $\alpha=5$ were found to result in several recirculation regions for the double cone cases, as reported in literature~\cite{gai_khraibut_2019,hao_occurancer_2021}. This behavior is attributed to the presence of temperature jump and velocity slip effects~\cite{INGER200742,Inger2008}, which are known to introduce relaxation into the flow. These mechanisms contribute to the stability of a single recirculation region in the two-dimensional flowfields observed in all cases shown in Fig.~\ref{fig:ScaledComparison}.

Another important outcome of the results shown in Fig.~\ref{fig:ScaledComparison} is that a scaled angle quantity that combines all the flowfield variables into one single quantity might not be sufficient to describe the stability of the flow over the compression corners since very similar scaled angles could result in single or several recirculation regions or steady or unsteady flows. Examining Fig.~\ref{fig:ScaledComparison} carefully, there are three points around the scaled angle value of $\alpha=6$, corresponding to cases from double cone cases of M6 and M16 as well as the compression ramp case at physical angle $\alpha^*=30^\circ$. Of all these three cases, one is stable and has a single recirculation region, another is unsteady with a single recirculation region and the last one has both several recirculation regions and unsteady. In addition, all resulted in very different separation lengths. Moreover, we used the conventional length scale of flat plate length in the calculation of scaled angle values for these configurations. However, this preference might not be the best suited scale for the configurations considered in this work, as most of the cases resulted in separation bubbles that occupy more than 80\% of the flat plate length and in some cases even more than 90\%, thus not allowing a developed boundary layer. We deem a physical length scale, such as the length from the tip of the plate to the separation point, would be more appropriate for these configurations. This change of length scale results in scaled angle quantities of $4<\alpha<6$ for the cases with secondary recirculation regions, and these values are closer to the limits described in previous studies~\cite{gai_khraibut_2019,hao_occurancer_2021}. Another option for a physics based length scale is the boundary layer thickness just prior to separation point. When this boundary layer thickness is used scaled angle quantities drop further to $3.3 < \alpha < 4.6$ and again are in better agreement with the values reported in literature.

 Similar to the analysis for the appearance of the secondary recirculation regions, there is no direct connection between the scaled angle quantity and the unsteadiness of the flow for the flow regime considered in this work at 40-60 km altitude or Reynolds number on the order of $10^4$, as can be seen in Fig.~\ref{fig:ScaledComparison} where the unsteady cases are marked by rectangular symbols. The unsteadiness characteristics of the double cone cases were discussed elsewhere~\cite{tumukluPoF,tumukluPoF2,KarpuzcuDCM6}, however the main unsteadiness mechanisms for these cases were the shock interactions with shocks or shear layers. For the compression ramp cases considered in this work, it is observed that whenever a secondary recirculation region is present, the flow is unsteady, i.e for the configurations M6, Tw175 and CR45 and the flow conditions given in Table~\ref{tab:CasesFreeScaled}. However, secondary recirculation regions cannot be linked directly to flow unsteadiness yet. The unsteadiness of the cases will be demonstrated later and given here for the completeness of the discussion of overall effect of the scaled angle.   The above discussion shows that the scaled angle is not comprehensive enough to define boundaries between steady and unsteady flowfields as well as the different mechanisms that are causing unsteadiness in such complex flowfields.   Instead we next turn to data driven methods to provide a better understanding of the spatial propagation of disturbances in massively unstable separation bubbles.

\section{Application of Data Driven Methods to the  Unsteady Flows Simulated with DSMC} \label{sec:datadrivenmethods}

Two data driven methods were utilized in this work, namely dynamic mode decomposition (DMD)~\citep{SCHMID_2010} and spectral orthogonal decomposition (SPOD)~\citep{Schmidt_Colonius_2020}.  Both methods were utilized because DMD is the more general approach of the two especially for determining the mode rank, however, it requires more computational DSMC data to resolve the spatial distribution of flows with multiple disturbances.  For both of these methods we use the method of snapshots, i.e. snapshots of flowfield at discrete time steps. The time interval between two snapshots defines the sampling frequency and the maximum frequency that can be resolved. The number of total snapshots or the total number of time steps included in the analysis defines the lowest frequency that can be resolved with the data driven analysis. In this work we use a time interval $\Delta t=5\mu s$ between snapshots and 2,000 snapshots in total with each of these snapshots having 1500 by 550 spatial points that results in
 a sampling frequency of $f_s=200kHz$, the highest frequency of $f_{max}=100$~kHz and the lowest frequency of $f_{min}=0.1kHz$ for both DMD and SPOD analyses.  This frequency range and resolution is determined based on power spectral density (PSD) analyses of DSMC numerical probe data that is taken from the locations shown in Fig.~\ref{fig:Schlieren}. The number of DSMC snapshots and the time interval between them were chosen to resolve frequencies in the 0.1–100 kHz range, based on our PSD analysis of numerical probe data  that we will show later in Section V. This is also a part of the suggested methodology by Schmidt and Colonius~\cite{Schmidt_Colonius_2020}.
 Details of the formulations and steps for the calculation of unsteady modes for both methods are discussed below. Please note that in the following sections, we follow the formulations laid out by Towne et al.\cite{Towne_Schmidt_Colonius_2018}, further details can be found in that reference.

\subsection{Dynamic Mode Decomposition}

Dynamic Mode Decomposition (DMD) is one of the data driven methods that is utilized in this work~\cite{SCHMID_2010}. DMD decomposes spatio-temporal patterns into modes with specific frequencies and growth rates. Each mode represents a coherent flow structure, and DMD approximates the system evolution using a linear operator derived from the data. The idea behind the DMD is that there exists a linear operator $\mathbf{A}$ that approximately advances the system state forward in time:
\begin{equation}
    \mathbf{q}_{k+1} \approx \mathbf{A}\mathbf{q}_k
\end{equation}
where k indicates the time step and q is the snapshot of the DSMC macroparameter flowfields at kth time step, in this work every 5~$\mu$s . For each $q$ snapshot the flowfields that are of 1500x550 size matrices, flattened into 1-D vectors having 825,000 elements. Then each vector, denoted as $\mathbf{q}_1,\mathbf{q}_2,...\mathbf{q}_M$  containing the flowfield macroparameter data for kth time step is inserted into the matrices $X$ and $Y$ as columns, given as;
\begin{equation}
    \mathbf{X} = \begin{bmatrix}
    | & | & \cdots & | \\
    \mathbf{q}_1 & \mathbf{q}_2 & \cdots & \mathbf{q}_{M-1} \\
    | & | & \cdots & |
    \end{bmatrix}, \quad
    \mathbf{Y} = \begin{bmatrix}
    | & | & \cdots & | \\
    \mathbf{q}_2 & \mathbf{q}_3 & \cdots & \mathbf{q}_M \\
    | & | & \cdots & |
    \end{bmatrix}
\end{equation}
In each of these matrices, vectors q are the flowfield snapshots with N being the total number of spatial points and M is the total number of flowfield snapshots. This allows us to express the approximation as a linear mapping matrix $\mathbf{A}$ as
$\mathbf{Y} \approx \mathbf{A}\mathbf{X}$.
While the system matrix $\mathbf{A}$ would be prohibitively large ($N \times M$), we can extract the key dynamics through a low-rank approximation. This begins with the Singular Value Decomposition (SVD) of $\mathbf{X}$:
\begin{equation}
\label{eqn:DMDSingularV}
    \mathbf{X} = \mathbf{U}\boldsymbol{\Sigma}\mathbf{V}^*
\end{equation}
where $\mathbf{U} \in \mathbb{R}^{N \times r}$ and $\mathbf{V} \in \mathbb{R}^{M \times r}$ contain the left and right singular vectors respectively, and $\boldsymbol{\Sigma} \in \mathbb{R}^{r \times r}$ contains the singular values and the superscript $^*$ denotes the complex conjugate. The rank $r$ is chosen to balance computational efficiency with accuracy. As will be shown in Section \ref{sec:FlowUnsteadiness}, we found in our analyses that
 $r = 20$ captures the dominant dynamics. After obtaining the SVD of matrix X, reduced operator $\tilde{\mathbf{A}}$ is constructed that best estimates dynamics of the system:
$\tilde{\mathbf{A}} = \mathbf{U}^*\mathbf{Y}\mathbf{V}\boldsymbol{\Sigma}^{-1}$
where the reduced operator $\tilde{\mathbf{A}} \in \mathbb{R}^{r \times r}$ is a lower-dimensional representation of the full system dynamics. The eigen-decomposition of this reduced operator
\begin{equation}
\label{eqn:DMDEval}
\tilde{\mathbf{A}}\mathbf{W} = \mathbf{W}\boldsymbol{\Lambda}
\end{equation}
yields the DMD eigenvalues $\boldsymbol{\Lambda} \in \mathbb{C}^{r \times r}$ where $\Lambda $ is a diagonal matrix and eigenvectors $\mathbf{W} \in \mathbb{C}^{r \times r}$. The diagonal matrix $\boldsymbol{\Lambda}$ containing the complex eigenvalues for each of the modes is:
    \begin{equation}
    \boldsymbol{\Lambda} = \begin{bmatrix}
    \lambda_1 & 0 & \cdots & 0 \\
    0 & \lambda_2 & \cdots & 0 \\
    \vdots & \vdots & \ddots & \vdots \\
    0 & 0 & \cdots & \lambda_r
    \end{bmatrix}
    \end{equation}

Each complex eigenvalue $\lambda_j$ can be expressed in polar form as $\lambda_j = |\lambda_j|e^{i\theta_j}$
where $|\lambda_j|$ represents the magnitude and $\theta_j$ represents the phase angle. To align the discrete eigenvalues obtained from the method of snapshots with the continuous-time behavior of the physical system, we convert the former into continuous-time growth rates and frequencies for each mode $j$ using the equations below,
\begin{align}
\label{eqn:DMDfrequency}
\Omega_j &= \sigma_j + i\omega_j && \text{(continuous-time eigenvalue)}\\
\sigma_j &= \frac{\ln|\lambda_j|}{\Delta t} && \text{(growth or decay rate)} \\
\omega_j &= \frac{\theta_j}{\Delta t} && \text{(angular frequency)}
\end{align}
where the matrix $\boldsymbol{\Omega}$ is given as;
    \begin{equation}
    \boldsymbol{\Omega} = \frac{\ln(\boldsymbol{\Lambda})}{\Delta t} = \begin{bmatrix}
    \sigma_1 + i\omega_1 & 0 & \cdots & 0 \\
    0 & \sigma_2 + i\omega_2 & \cdots & 0 \\
    \vdots & \vdots & \ddots & \vdots \\
    0 & 0 & \cdots & \sigma_r + i\omega_r
    \end{bmatrix}
    \end{equation}
and $\Delta t = 5\mu $s.

The DMD modes are then reconstructed as:
\begin{equation}
\label{eqn:DMDmodes}
\boldsymbol{\Phi} = \mathbf{Y}\mathbf{V}\boldsymbol{\Sigma}^{-1}\mathbf{W}
\quad,\quad
\boldsymbol{\Phi} = \begin{bmatrix}
| & | & \cdots & | \\
\boldsymbol{\phi}_1 & \boldsymbol{\phi}_2 & \cdots & \boldsymbol{\phi}_r \\
| & | & \cdots & |
\end{bmatrix}
\end{equation}
where $\boldsymbol{\Phi} \in \mathbb{C}^{N \times r}$. Each column vector $\boldsymbol{\phi}_j \in \mathbb{C}^N$ represents a spatial DMD mode that
has the same dimension as the original state vector. Each DMD mode contains the spatial structure of a coherent dynamic pattern and is associated with a specific eigenvalue $\lambda_j$ that describes its temporal behavior in terms of frequency and growth or decay rate.

The dynamic mode decomposition workflow in this study was implemented through a process of combining data, pre-processing, dimensional reduction, and modal analysis. As mentioned above, the DSMC macroparameter flowfield data consisted of at least 2,000 snapshots of $1,500 \times 550$ spatial grids taken $\Delta t=5\mu s$ apart.  However, it was observed that the DSMC macroparameter values that are the sampled bulk properties such as density, velocity and temperature were too noisy for meaningful DMD analysis.  To address this, 2$\times$ spatial coarsening was applied through block averaging:
\begin{equation*}
\tilde{x}_i = \frac{1}{2}\sum_{k=2i}^{2i+1}x_k, \quad \tilde{y}_j = \frac{1}{2}\sum_{k=2j}^{2j+1}y_j
\end{equation*}
that reduced the grid to  $750 \times 275$.  This coarsening lowered the noise level while maintaining a spatial resolution sufficiently fine to capture coherent wave structures.  Also,  edge padding ensured integer divisibility, maintaining geometric consistency between the original and reduced grids. Finally, a 300-snapshot moving average was computed using cumulative summation for efficient windowed averaging when necessary:
\begin{equation*}
\overline{\mathbf{X}}_k = \frac{1}{300}\left(\mathbf{C}_{k+150} - \mathbf{C}_{k-150}\right)
\end{equation*}
where $\mathbf{C}$ is the cumulative sum along time. While ultimately retained, this baseline could be subtracted to isolate fluctuation components in macroparameters $\mathbf{X}' = \mathbf{X} - \overline{\mathbf{X}}$.

\subsection{Spectral Proper Orthogonal Decomposition}

Spectral proper orthogonal decomposition (SPOD) is a data driven method to extract coherent structures from an unsteady flowfield, if the flow is stationary, i.e. oscillates  around a mean, and not in a transient phase~\citep{Schmidt_Towne_Rigas_Colonius_Brès_2018,Towne_Schmidt_Colonius_2018,Schmidt_Colonius_2020}. In this method discrete time snapshots of the flowfield  are used to calculate the spatio-temporal covariance of DSMC macroparameters at discrete frequency components. The SPOD modes are the eigenvalues of the cross-spectral density tensor at each frequency and harmonic in time.

To calculate SPOD from instantaneous DSMC flowfields, first a data matrix is constructed from each realization (or snapshot) of the flowfield~\cite{Towne_Schmidt_Colonius_2018}.
\begin{equation}
    \mathbf{Q} = \begin{bmatrix}
    | & | & \cdots & | \\
    \mathbf{q}_1 & \mathbf{q}_2 & \cdots & \mathbf{q}_{M} \\
    | & | & \cdots & |
    \end{bmatrix} \quad
    \in \mathbb{R}^{N \times M}
\end{equation}
where $\mathbf{q}_1,\mathbf{q}_2,...\mathbf{q}_M$ are the flowfield data for each snapshot in the vector form similar to the DMD analysis, N is the total number of spatial points in space and M is the total number of snapshots. One can directly estimate the cross-spectral density tensor by taking the discrete Fourier transform (DFT) of this data matrix, however this method is known to have very slow convergence. Instead, Welch's method~\cite{Welch1967} is used to group the data into blocks. Welch's method divides the original set of flowfield snapshots into smaller segments (blocks), typically overlapping, to improve the statistical convergence of spectral estimation. Given a total of $M$ snapshots, the data matrix $\mathbf{Q}$ is segmented into $N_b$ overlapping blocks, each consisting of $N_f$ snapshots.

Each block is formed as follows:
\begin{equation}
\label{eqn:WelchBlock}
\mathbf{Q}^{(n)} = \begin{bmatrix}
| & | & \cdots & | \\
\mathbf{q}^{(n)}_1 & \mathbf{q}^{(n)}_2 & \cdots & \mathbf{q}^{(n)}_{N_f} \\
| & | & \cdots & |
\end{bmatrix}
\quad \in \mathbb{R}^{N \times N_f}, \quad n = 1,2,\ldots,N_b
\end{equation}
where each block $\mathbf{q}^{(n)}$ contains snapshots starting from a shifted position in the original snapshot series such as the first block includes snapshots $1$ through $N_f$, the second block begins at snapshot $1 + N_f - N_o$ and ends at snapshot $2N_f - N_o$, where $N_o$ is the number of overlapping snapshots between blocks and this procedure is repeated until the last block is constructed. The overlap $N_o$ is chosen to balance spectral resolution with statistical reliability; commonly, a 50\% overlap is used.

The snapshot indices for each block $n$ are mathematically expressed as:
\begin{equation}
\label{eqn:BlockSnapshot}
\mathbf{q}^{(n)}_j = \mathbf{q}_{(n-1)(N_f - N_o) + j}, \quad j = 1,2,\ldots,N_f
\end{equation}

This procedure transforms the original snapshot matrix into multiple smaller segments suitable for subsequent windowing, discrete Fourier transform, and accurate estimation of the cross-spectral density tensor in SPOD analysis. Then the DFT is applied to each of these blocks, which can be treated as the ensemble of the flow realizations as;

\begin{equation}
\begin{aligned}
&\hat{\mathbf{Q}}^{(n)}=\left[\hat{\mathbf{q}}_1^{(n)}, \hat{\mathbf{q}}_2^{(n)}, \ldots, \hat{\mathbf{q}}_{N_f}^{(n)}\right],\\
&\text { with }\\
&\hat{\mathbf{q}}_k^{(n)}=\frac{1}{\sqrt{N_f}} \sum_{j=1}^{N_f} w_j \mathbf{q}_j^{(n)} \mathrm{e}^{-\mathrm{i} 2 \pi(k-1)\left[(j-1) / N_f\right]}
\end{aligned}
\end{equation}
where subscript k denotes the discrete frequency component. The scalar weights $w_j$ correspond to the values of a windowing function at specific nodes. The Fourier transformed data matrix can be written as:

\begin{equation}
\hat{\mathbf{Q}}_{f_k}=\sqrt{\kappa}\left[\hat{\mathbf{q}}_k^{(1)}, \hat{\mathbf{q}}_k^{(2)}, \ldots, \hat{\mathbf{q}}_k^{\left(N_b\right)}\right] \in \mathbb{R}^{N \times N_b},
\end{equation}
where $N_b$ is the number of blocks and it will defined further in Eqn.~\ref{eq:Nblock} at the end of this section.
Then the cross spectral density tensor can be estimated for each frequency with;
\begin{equation}
\mathbf{S}_{f_k}=\hat{\mathbf{Q}}_{f_k} \hat{\mathbf{Q}}_{f_k}^* .
\end{equation}
With that, SPOD problem reduces to an eigenvalue problem for each discrete frequency with the following form;
\begin{equation}
\label{eqn:SPODEvalEvec}
\mathbf{S}_{f_k} \mathbf{W} \boldsymbol{\Psi}_{f_k}=\boldsymbol{\Psi}_{f_k} \boldsymbol{\Lambda}_{f_k}
\end{equation}
and SPOD modes are given by the eigenvectors of this eigenvalue problem, namely the columns of $\boldsymbol{\Psi}_{f_k}$. While interpreting the SPOD spectra, low-rank behavior is an important concept. Low rank behavior happens when the modes have distinguishable difference in energy for certain frequency bands, similar to PSD where some frequencies having higher energy, thus being dominant, and generally represent dominant physical mechanisms present in the flow for these frequencies. The energy in the SPOD spectra for each frequency is determined from the absolute value of the eigenvalue at that frequency, calculated from the eigenvalues determined from Eqn. \ref{eqn:SPODEvalEvec}.

It should be noted that unlike the DMD analysis, no pre-processing was necessary for the SPOD analysis employed in this work, meaning that SPOD is more robust than DMD for the same stochastic fluctuations consisting of numerical noise and thermal fluctuations present in the flowfield, as expected. We are following the methodology laid out and the software provided by Schmidt and Colonius~\citep{Schmidt_Colonius_2020} for the SPOD analysis in this work. We use a time interval $\Delta t=5\mu s$ between snapshots and 2000 snapshots in total. This results in a sampling frequency of $f_s=200kHz$, the highest frequency of $f_{max}=100kHz$ and the lowest frequency of $f_{min}=0.1kHz$. With 50\% overlap and the number of FFT frequencies $N_{FFT}=128$, results in a total number of 30 blocks. The relation between these parameters are given as;
\begin{equation}
\label{eq:Nblock}
N_{\mathrm{b}}=\left\lfloor\frac{N_t-N_{\mathrm{o}}}{N_{\mathrm{FFT}}-N_{\mathrm{o}}}\right\rfloor
\end{equation}
where $N_t$ is the total number of flowfield snapshots.  In this work, the values of $N_{b}$, $N_o$, and $N_t$ used were 30, 64, and 2,000 for the SPOD analyses.

\section{Data Driven Analysis of Unsteady High Scaled-Angle Flowfields}\label{sec:FlowUnsteadiness}

\begin{figure}[h!]
    \centering
       \subfigure[]{\label{fig:rmsNdenM6}\includegraphics[trim=100 10 100 200,clip,width=0.47\linewidth]{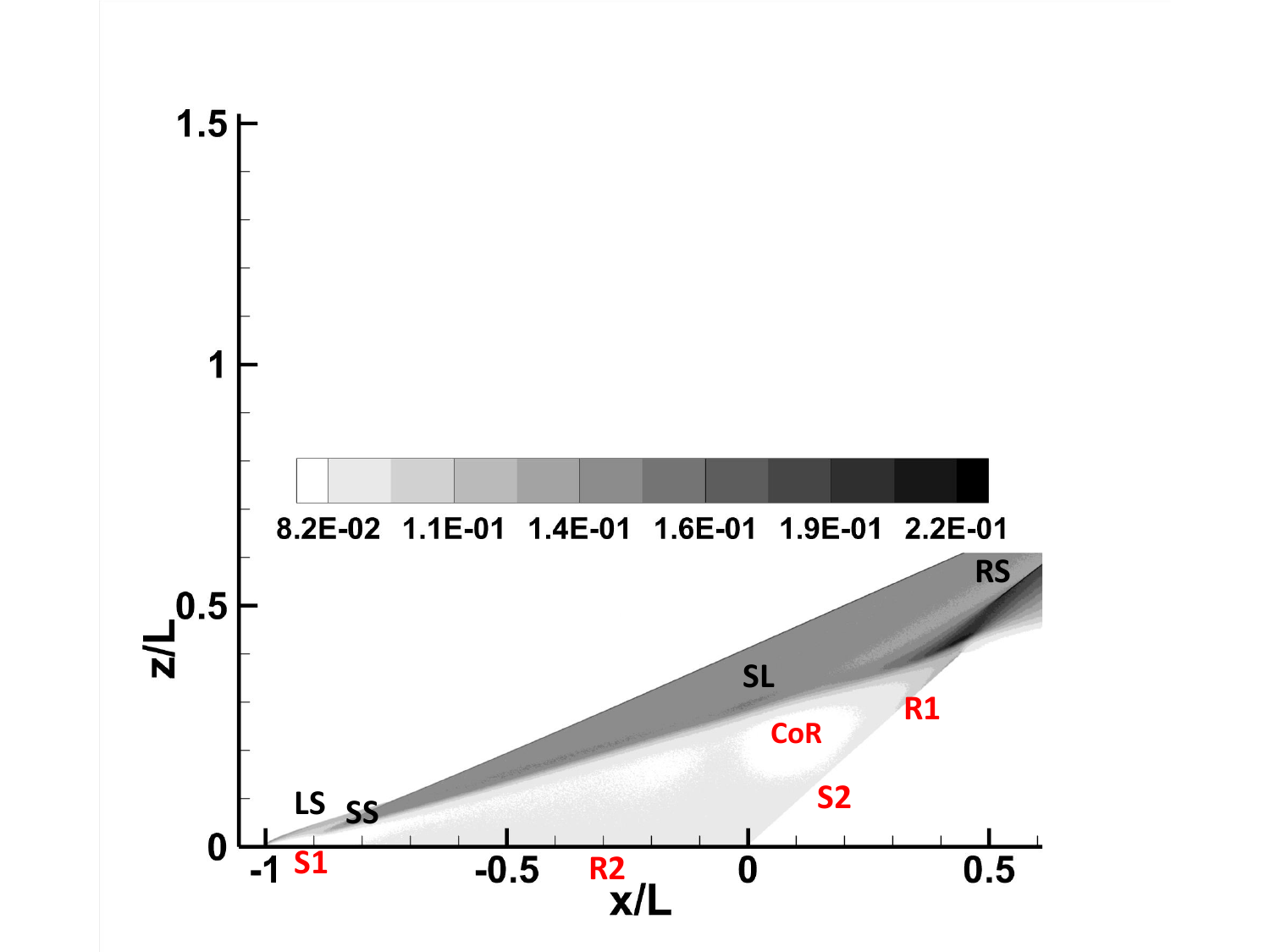}}
        \subfigure[]{\label{fig:rmsNdenTw175}\includegraphics[trim=100 10 100 200,clip,width=0.47\linewidth]{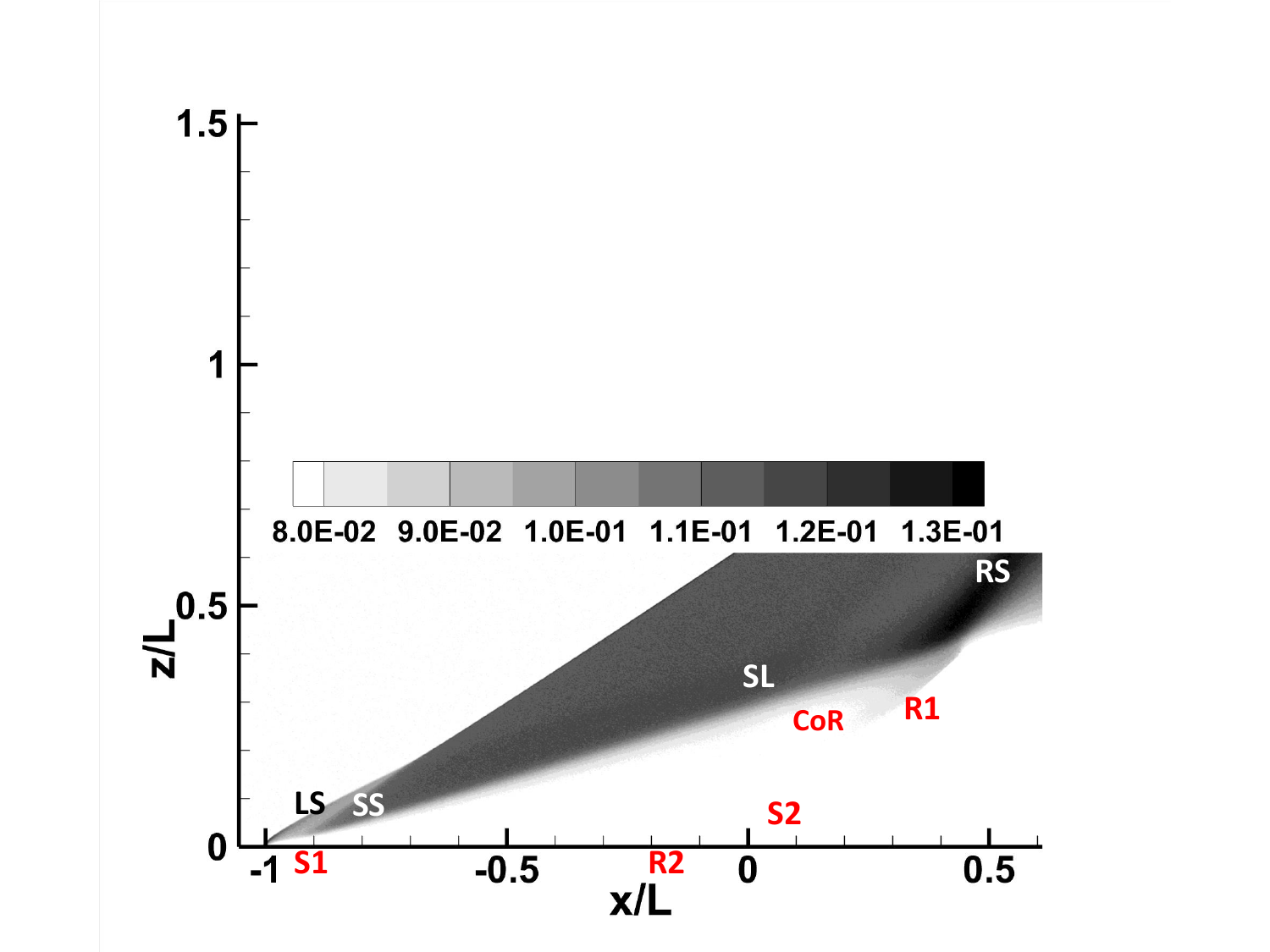}}
        \subfigure[]{\label{fig:rmsUvelM6}\includegraphics[trim=100 10 100 200,clip,width=0.47\linewidth]{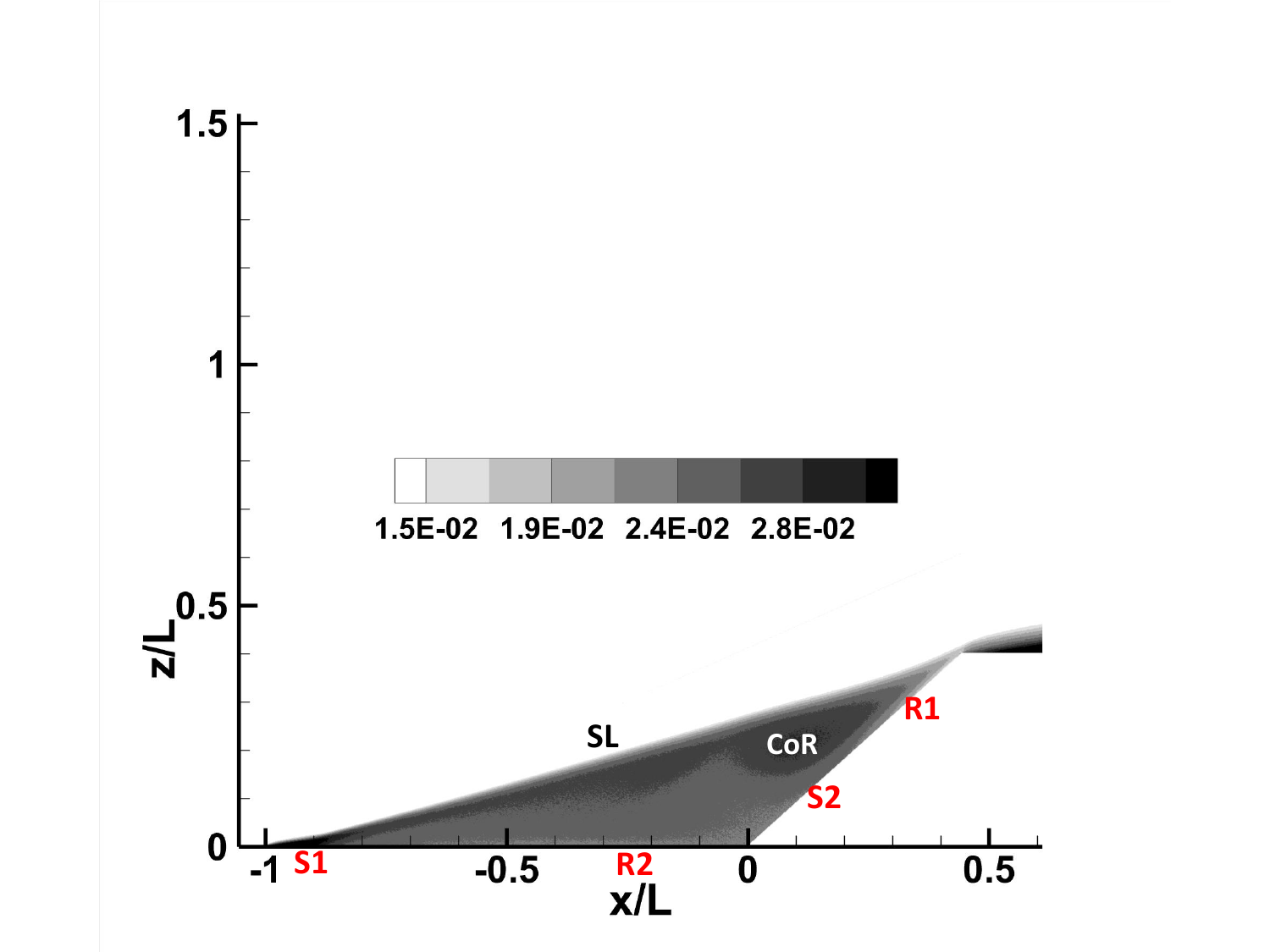}}
        \subfigure[]{\label{fig:rmsUvelTw175}\includegraphics[trim=100 10 100 200,clip,width=0.47\linewidth]{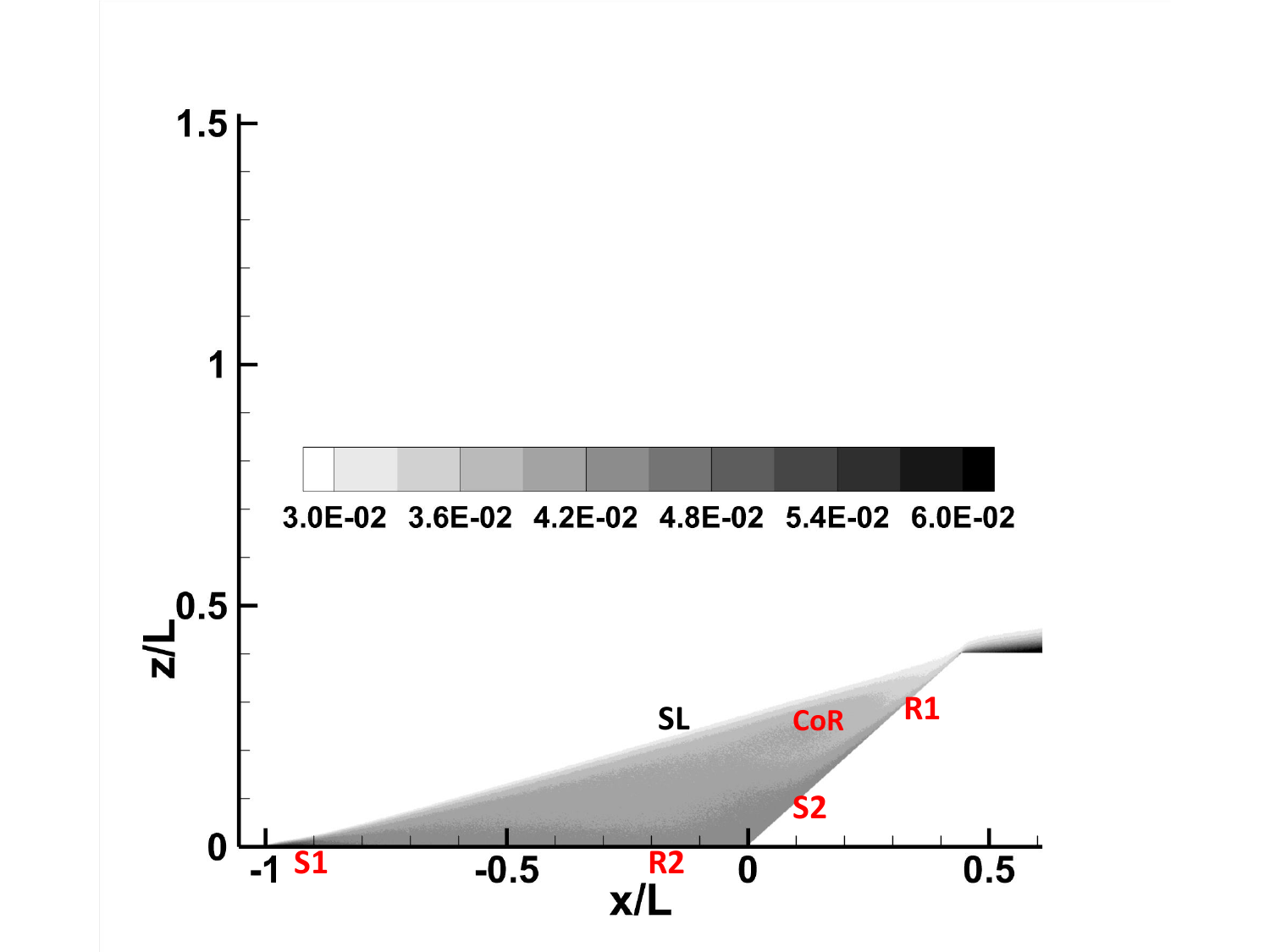}}
        \caption{Standard deviation (root-mean-square) of the macro-parameters. Top row; number density, bottom row; streamwise velocity. (a), (c) for M6 configuration, (b), (d) for Tw175 configuration. CoR indicates the center of recirculation.
        }
    \label{fig:rms}
\end{figure}

Using the window averaging SPOD and DMD approaches discussed above, we
 next investigate the unsteadiness of the destabilized recirculation region for the M6 and Tw175 configurations as they show the strongest unsteadiness characteristics. CR45 configuration also showed unsteadiness, however the data driven analysis results were quite similar to those of Tw175. Therefore data driven analysis results for the CR45 configuration are not reported in this work for the sake of brevity. In this section, data is extracted from the snapshots of instantaneous flowfields. The DSMC macroparameter flowfields used for the data driven methods in this section are instantaneous snapshots of the flow taken $\Delta t=5\mu s$ apart, {\em i.e.,} the flowfields are
 only sampled in space and used at  one DSMC time step of  $t=50ns$. This is equivalent to a high speed Schlieren imaging system having a camera with a frame interval of $\Delta t=5\mu s$ and an exposure time of $t=50ns$.  To start with, we show the standard deviation of the flowfields for number density and streamwise velocity in Fig.~\ref{fig:rms} calculated by the formula;

\begin{equation}
\sigma = \sqrt{\frac{1}{N} \sum_{i=1}^{N} (q_i - q_{ave})^2}
\end{equation}
where $q_i$ is an instantaneous flowfield macroparameter, $q_{ave}$ is its arithmetic mean of the flowfield taken over the flow field snapshots and N is the total number of snapshots. The standard deviation gives information regarding where the unsteadiness is the strongest on average. Note that standard deviation and variation is used interchangeably in this section. In Figs.~\ref{fig:rmsNdenM6} and ~\ref{fig:rmsNdenTw175}, the variation of number density peaks at the reattachment location for both M6 and Tw175 configurations. For the M6 configuration, the Edney interaction of the transmitted leading edge shock and the shear layer that creates compression waves also caused the variation to be more structured. Streamwise velocity variation peaks at the developing boundary layer after the expansion corner for both cases, as can be seen in Figs.~\ref{fig:rmsUvelM6} and ~\ref{fig:rmsUvelTw175}. More importantly, the variation streamwise velocity follow the inner structure of the recirculation region and also peak near the center of recirculation regions. The standard deviation contours for number density and streamwise velocity for both configurations in summary indicate the  reattachment shock and recirculation vortices are the important flow structures that cause variation in the flowfield.

\subsection{Configuration M6: The Effect of Increased Free Stream Velocity}
\begin{figure}[htbp]
\center
\subfigure[PSD for the DSMC pressure data]{\label{fig:M6PSD}\includegraphics[trim=30 30 30 30,clip,width=0.45\linewidth]{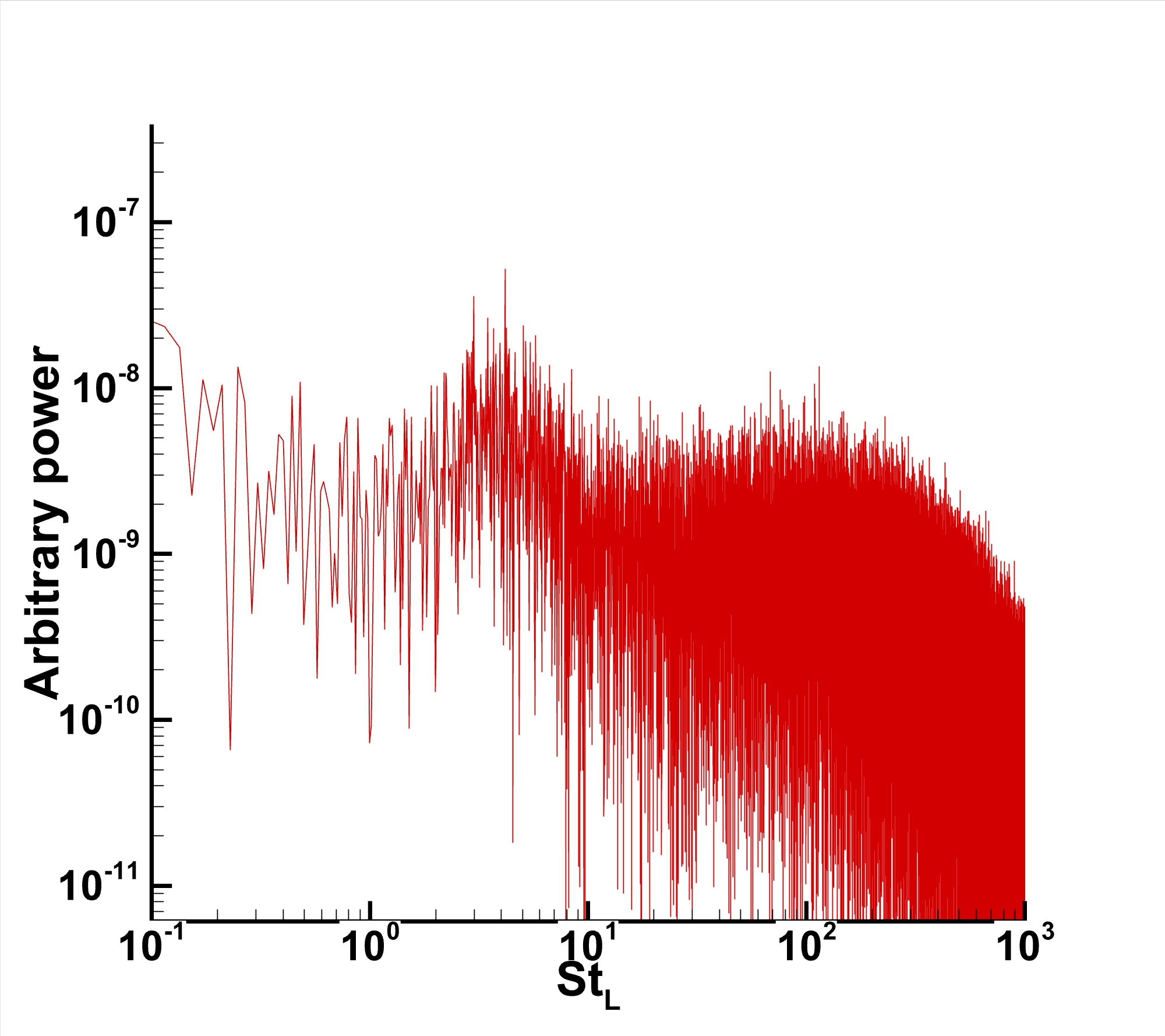}}
\subfigure[SPOD spectra for number density]{\label{fig:SPODM6NdenSpectra}\includegraphics[trim=10 1 30 30,clip,width=0.45\linewidth]{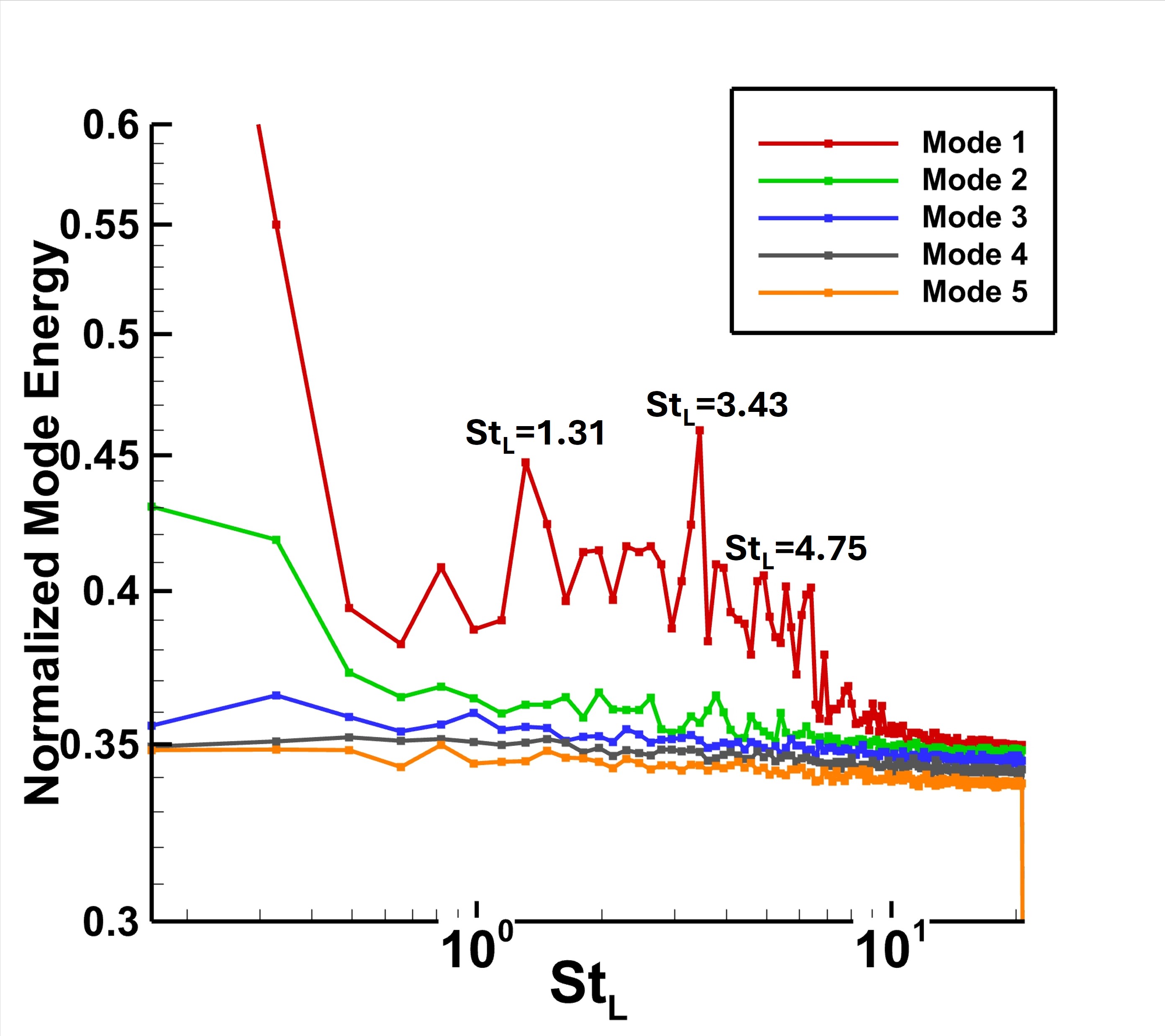}}
\subfigure[SPOD spectra for streamwise velocity]{\label{fig:SPODM6UvelSpectra}\includegraphics[trim=1 10 30 30,clip,width=0.45\linewidth]{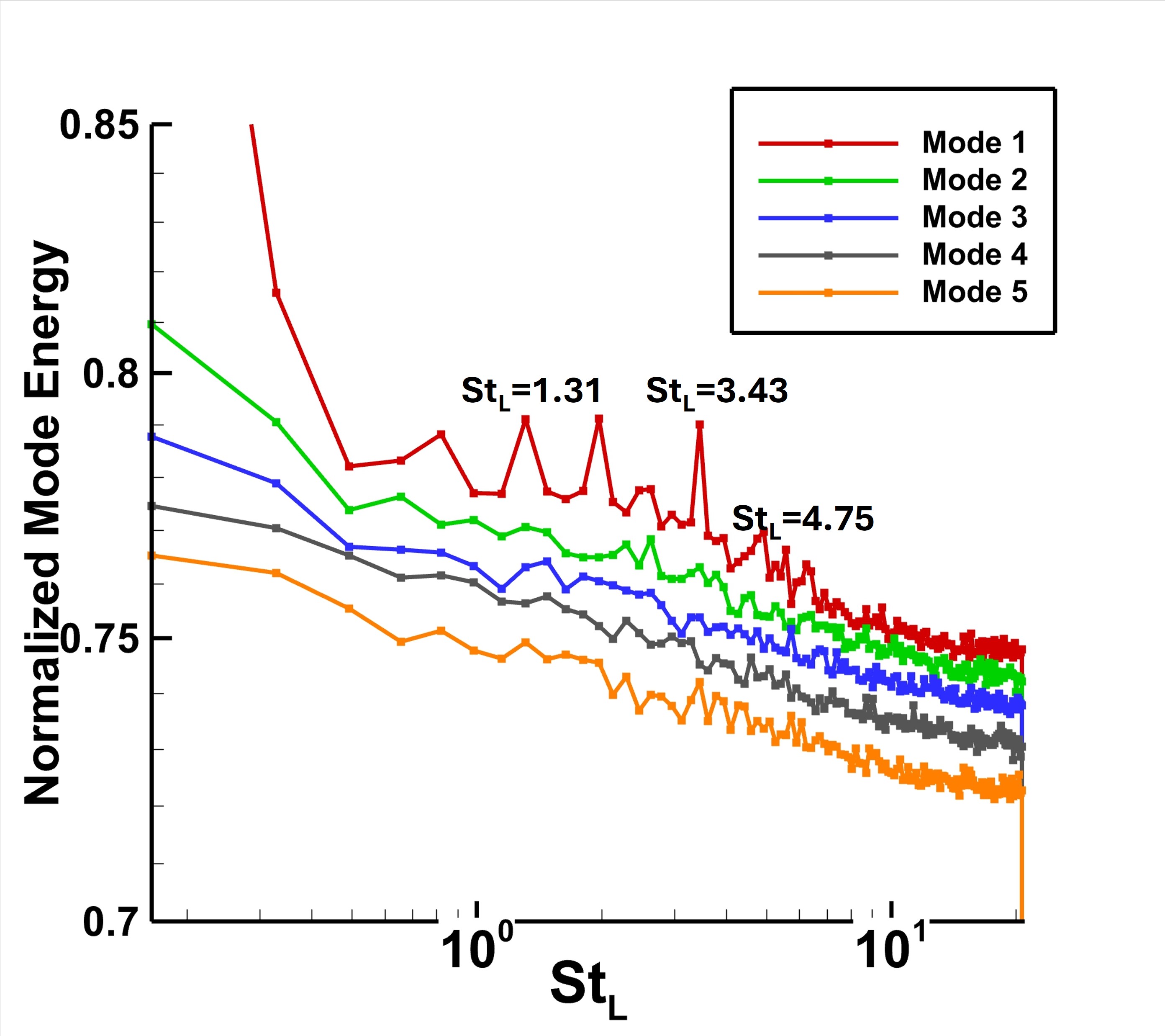}}
\subfigure[SPOD spectra of mode 1 for number density by varying N$_{FFT}$]{\label{fig:SPODNfft}\includegraphics[trim=10 10 30 30,clip,width=0.45\linewidth]{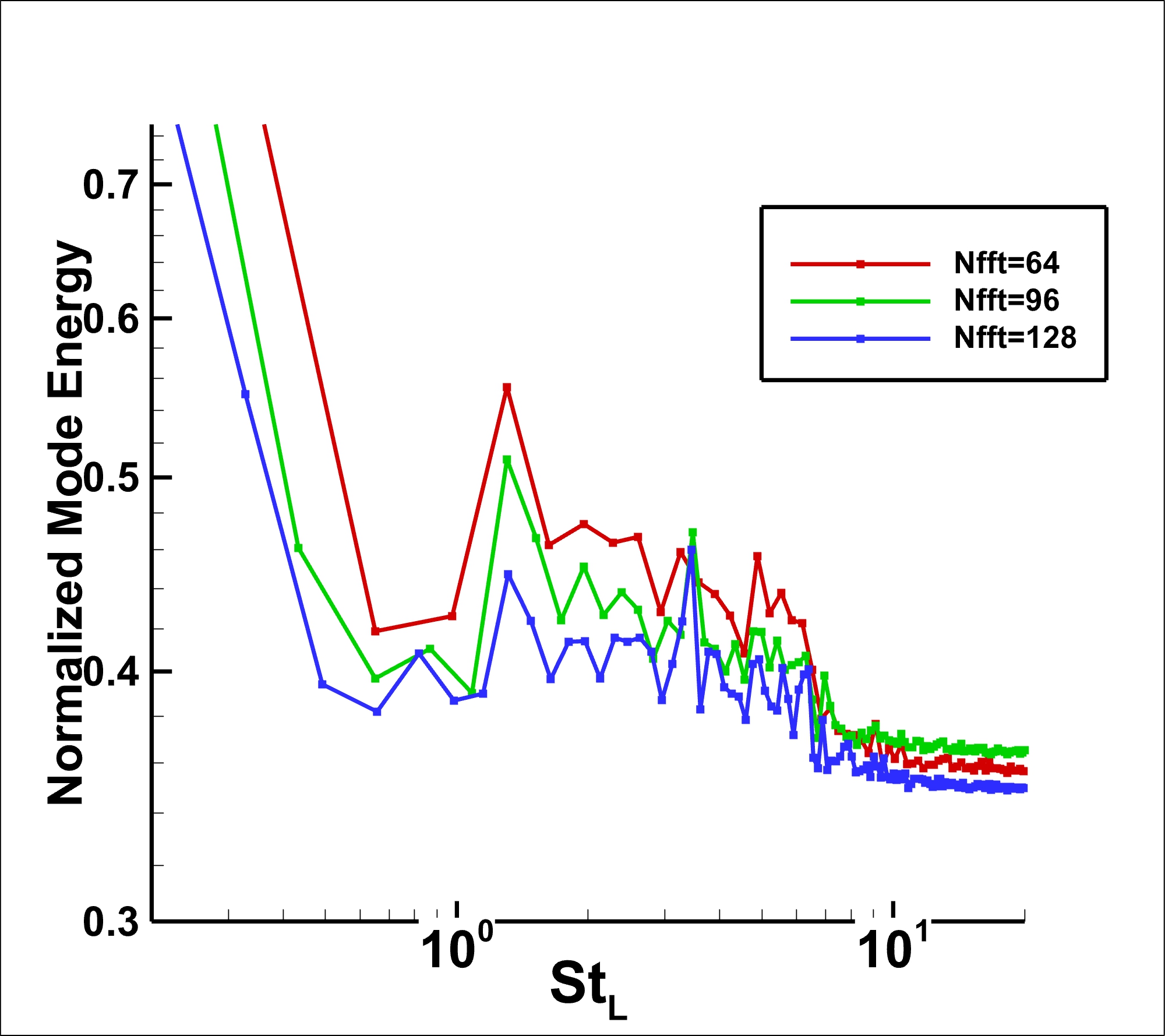}}
\caption{PSD of pressure fluctuations, SPOD spectra modes for number density and streamwise velocity, and the comparison of SPOD spectra for different number of FFT frequencies (N$_{FFT}$) for M6 configuration. The PSD is calculated at the probe location shown in Fig.~\ref{fig:M6}.}
\label{fig:M6SPODSpectra}
\end{figure}

The detailed investigation for the unsteadiness characteristics of these flowfields is first done for the M6 configuration that assumes a free stream velocity doubled from the baseline configuration.
First, a numerical probe is located near the foot of the reattachment shock to obtain DSMC time accurate data at each time step. Then using the mean subtracted pressure values at this location, power spectral density analysis is utilized over a time period of 25 ms corresponding to 500,000 time steps. We will use Strouhal number $St_L$ to characterize the frequencies defined as;
\begin{equation}
St_L=fL/U_\infty
\end{equation}
where f is the frequency of the unsteady mode, $U_\infty$ is the free stream velocity and L is the flat plate length. The power spectral density (PSD) is firstly utilized to get the range of unsteady frequencies in the flowfield. PSD is based on fast Fourier transform (FFT) and shows how the energy is distributed among different frequency bands and frequencies having higher energy (or power) are the important ones. PSD analysis of the pressure data at this location reveals a low frequency broadband region centered around $St_L=4.0$ as shown in Fig.~\ref{fig:M6PSD}. This is predicted to be caused by the reattachment shock and expansion waves being in close proximity near the corner location, and in agreement with the results in Fig.~\ref{fig:rmsNdenM6}, in which the standard deviation peaks near this location. Another important outcome of the PSD analysis shows that the power of fluctuations due to DSMC numerical noise and thermal fluctuations at frequencies higher than St$_L$>100 are much less than due to the physical unsteadiness fluctuations of the flow centered around St$_L$=4. This further demonstrating that there are enough statistics in the instantaneous DSMC flowfields to allow capturing of such unsteadiness characteristics.

 Using 2,000 snapshots of instantaneous flowfields, SPOD and DMD analyses were done for the M6 configuration to further analyze the predicted unsteadiness from the PSD analysis. One main difference between the SPOD and DMD is that in SPOD discrete frequencies are inputs and known a-priori, i,e, the spectra and modes are calculated for one known frequency at a time.  In contrast,  frequencies are obtained from  the output of  the DMD analysis. Thus, for the results  shown in this section comparisons between the SPOD and DMD modes cannot be done for the identical frequencies but rather those in the vicinity of each other.

 The SPOD analysis for number density and streamwise velocity also revealed low-rank-zones between $St_L=1-10$, as can be seen in the SPOD spectra plots in Figs.~\ref{fig:SPODM6NdenSpectra} and ~\ref{fig:SPODM6UvelSpectra} where mode 1 has distinctively higher mode energy compared to mode 2, which is consistent with the PSD analysis of the DSMC numerical probe data. The SPOD mode energy is the magnitude of the eigenvalues given by the Eqn.~\ref{eqn:SPODEvalEvec}, denoted by $\boldsymbol{\Lambda}_{f_k}$, normalized by the maximum value in Figs.~\ref{fig:SPODM6NdenSpectra} and ~\ref{fig:SPODM6UvelSpectra}. The shapes of the SPOD spectra shown in  Figs.~\ref{fig:SPODM6NdenSpectra} and ~\ref{fig:SPODM6UvelSpectra} are similar to each other since  both parameters are expected to have similar frequency bands with higher power, in agreement with the PSD plot shown in Fig.~\ref{fig:M6PSD}. To check for the numerical convergence of SPOD analysis, two parameters can be varied; the overlap (N$_o$) and number of FFT frequencies (N$_{FFT}$). Since 50\% overlap is well tested and its effects on the SPOD spectra are known~\citep{Schmidt_Colonius_2020}, we are not showing results for the variation of N$_o$. However, to make sure that the observed frequencies are not depending on the number of FFT frequencies chosen for the SPOD analysis, N$_{FFT}$ is varied and the results are shown in Fig.~\ref{fig:SPODNfft}. It can be clearly seen that variation of FFT frequencies does not affect the unsteadiness frequencies captured in this work at St$_L=1.31, 3.43, 4.75$.

\begin{figure}[htbp]
\center
\subfigure[DMD spectra of number density]{\label{fig:DMDM6NdenSpectra}\includegraphics[trim=70 10 110 100,clip,width=0.45\linewidth]{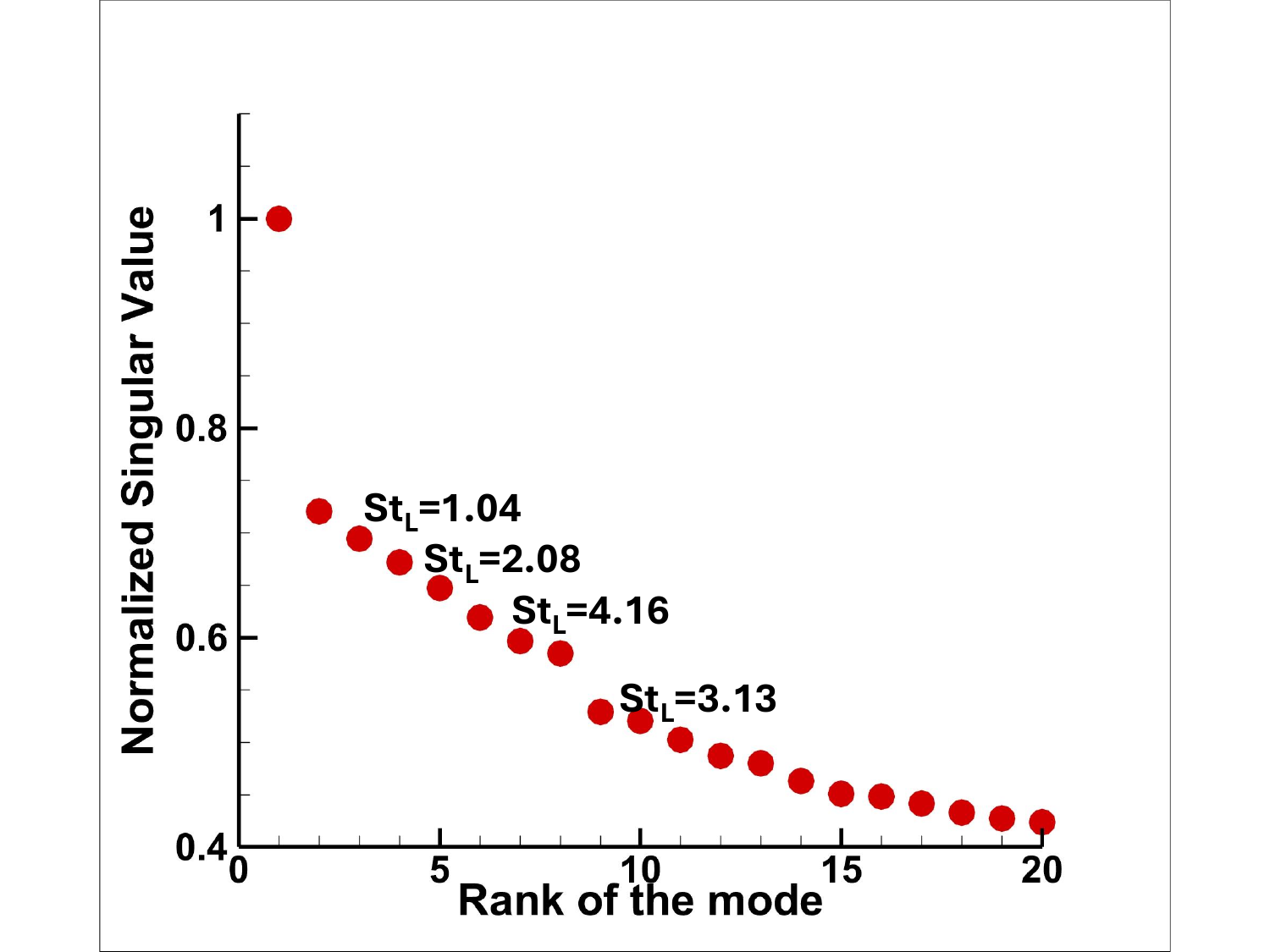}}
\subfigure[DMD spectra of streamwise velocity]{\label{fig:DMDM6UvelSpectra}\includegraphics[trim=80 10 110 100,clip,width=0.45\linewidth]{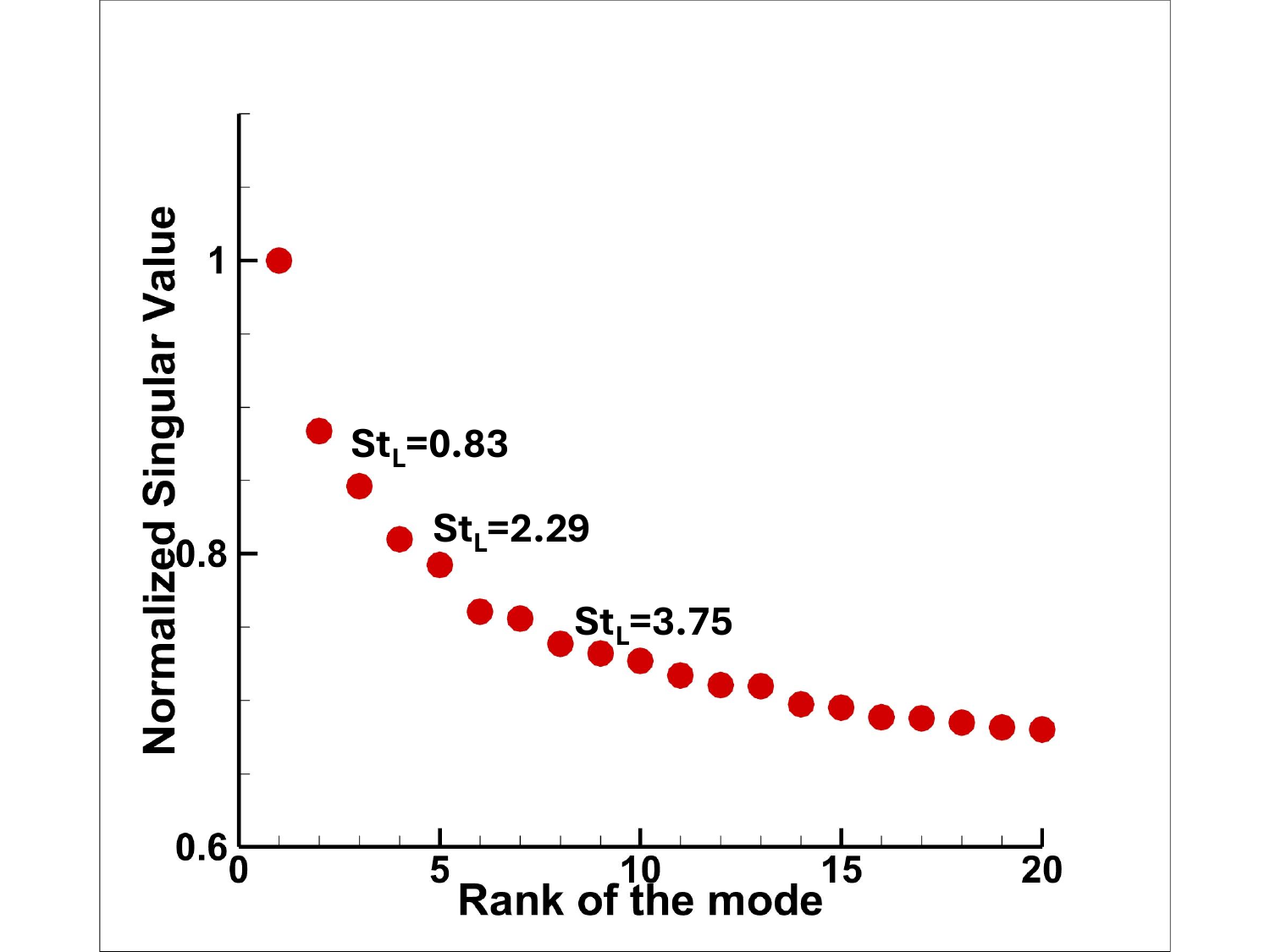}}
\subfigure[Singular Decay]{\label{fig:SingDecayM6}\includegraphics[trim=10 10 30 30,clip,width=0.45\linewidth]{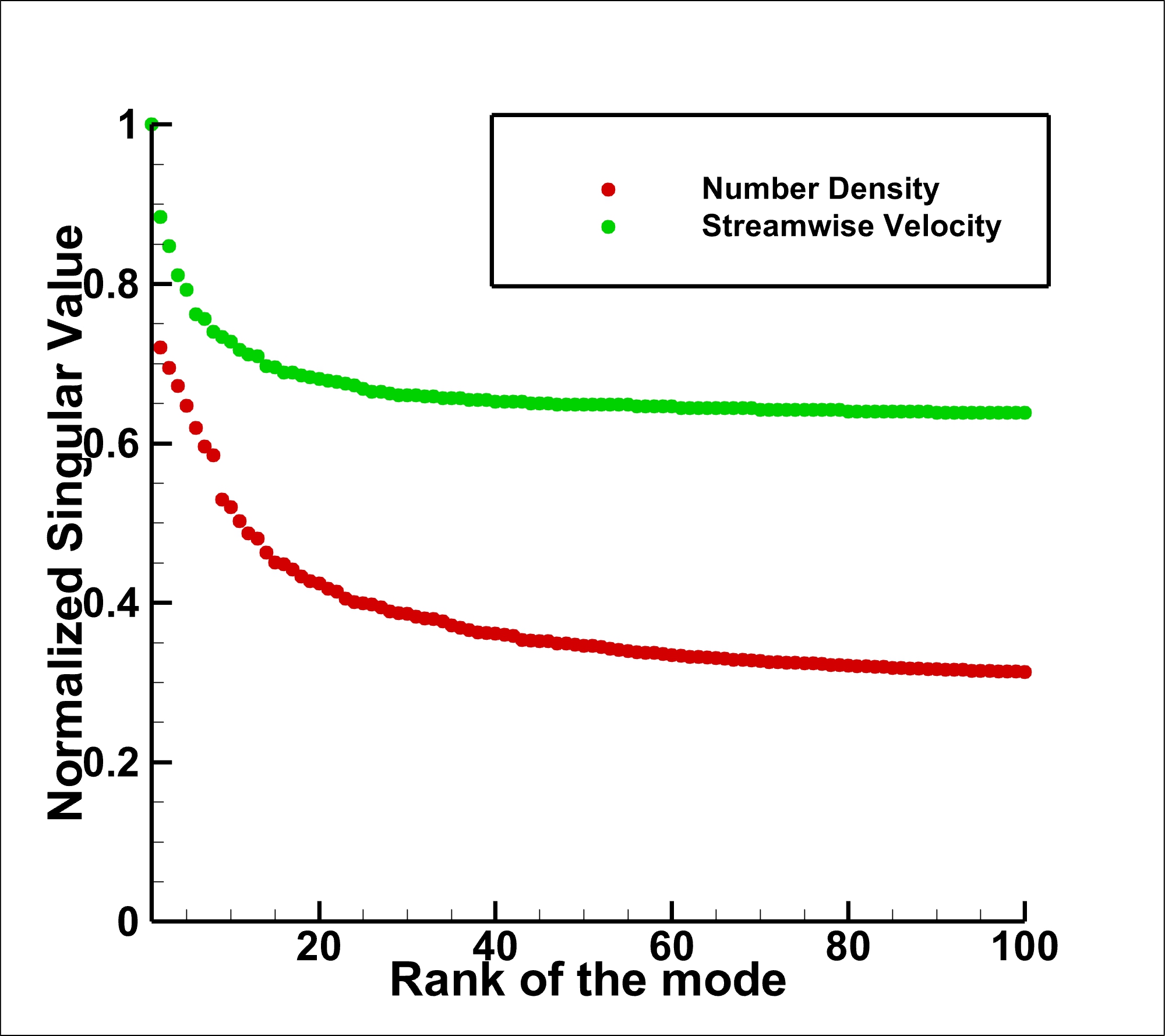}}
\caption{Leading DMD modes for number density and streamwise velocity and the singular value decay for the M6 configuration. In the sub-figures the horizontal axis shows the mode rank number and the vertical axis denotes the power or importance of the mode with its corresponding Strouhal number.  The mode power is self-normalized with the maximum of singular value.}
\label{fig:M6DMDSpectra}
\end{figure}

Also shown in Figs.~\ref{fig:DMDM6NdenSpectra} and ~\ref{fig:DMDM6UvelSpectra} the DMD modes sorted by the power of the mode, for which the first three modes are identified for $St_L=1.04$ to $St_L=4.16$, similar to the SPOD analysis. The singular values are calculated using Eqn.~\ref{eqn:DMDSingularV} and the frequency for each of these singular values are given by Eqn.~\ref{eqn:DMDfrequency}. For DMD, the truncation rank r=20 was selected based on the singular value decay spectrum as shown, where the majority of the perturbation energy is captured by the first ~13 modes (including 5 conjugate pairs corresponding to the same frequencies). However, due to the inherent stochastic noise in DSMC flowfields, which amplifies numerical and thermal fluctuations, using only r=13 or lower allows noise modes to contaminate the physical instability modes. To mitigate this, we extended to r=20, which captures and isolates additional noise while preserving the desired physical modes; ranks higher than 20 primarily add more noise-dominated modes without improving the representation of underlying dynamics. In addition, singular value decay plot in Fig.~\ref{fig:SingDecayM6} shows that our choice of r=20 is enough to capture the necessary mode energy as the singular values after r=20 starts to have a constant energy and very low distinction from each other for both number density and streamwise velocity fluctuations.

\begin{figure}[h!]
    \centering
        \subfigure[St$_L$=1.31]{\label{fig:SPODM6NdenMode1}\includegraphics[trim=80 10 110 250,clip,width=0.47\linewidth]{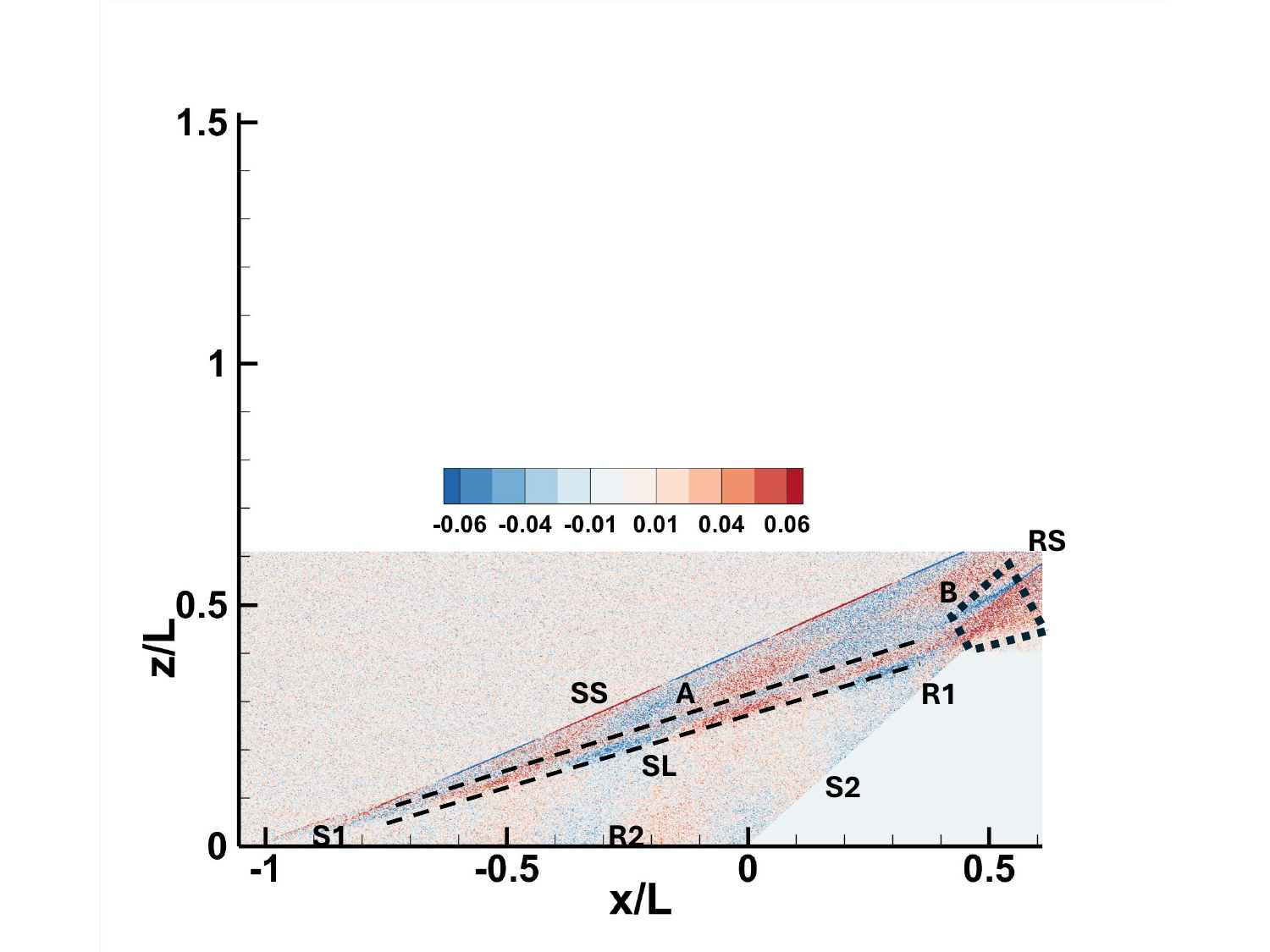}}
        \subfigure[St$_L$=1.04]{\label{fig:DMDM6NdenMode1}\includegraphics[trim=80 10 110 250,clip,width=0.47\linewidth]{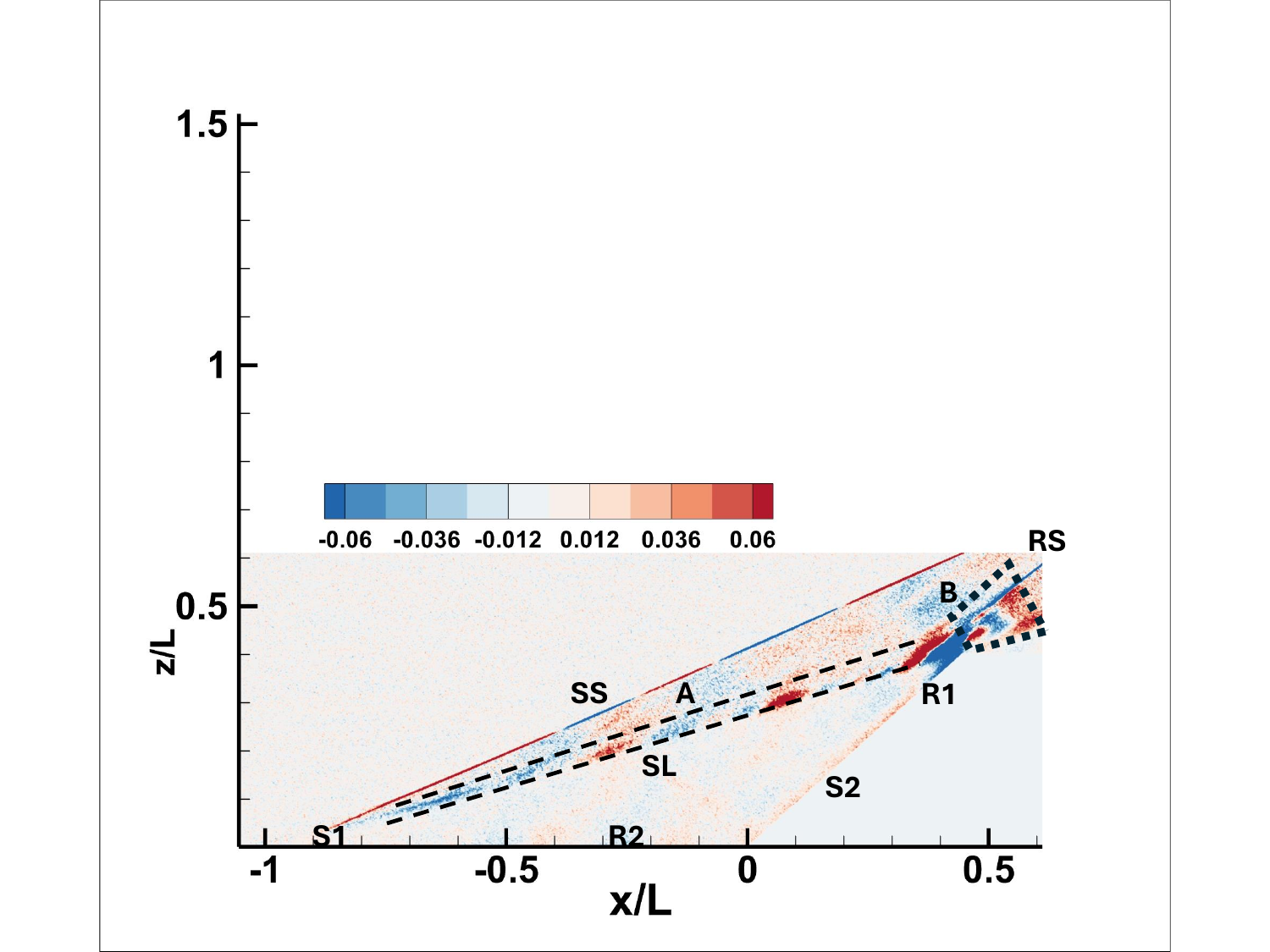}}
        \subfigure[St$_L$=3.43]{\label{fig:SPODM6NdenMode2}\includegraphics[trim=80 10 110 250,clip,width=0.47\linewidth]{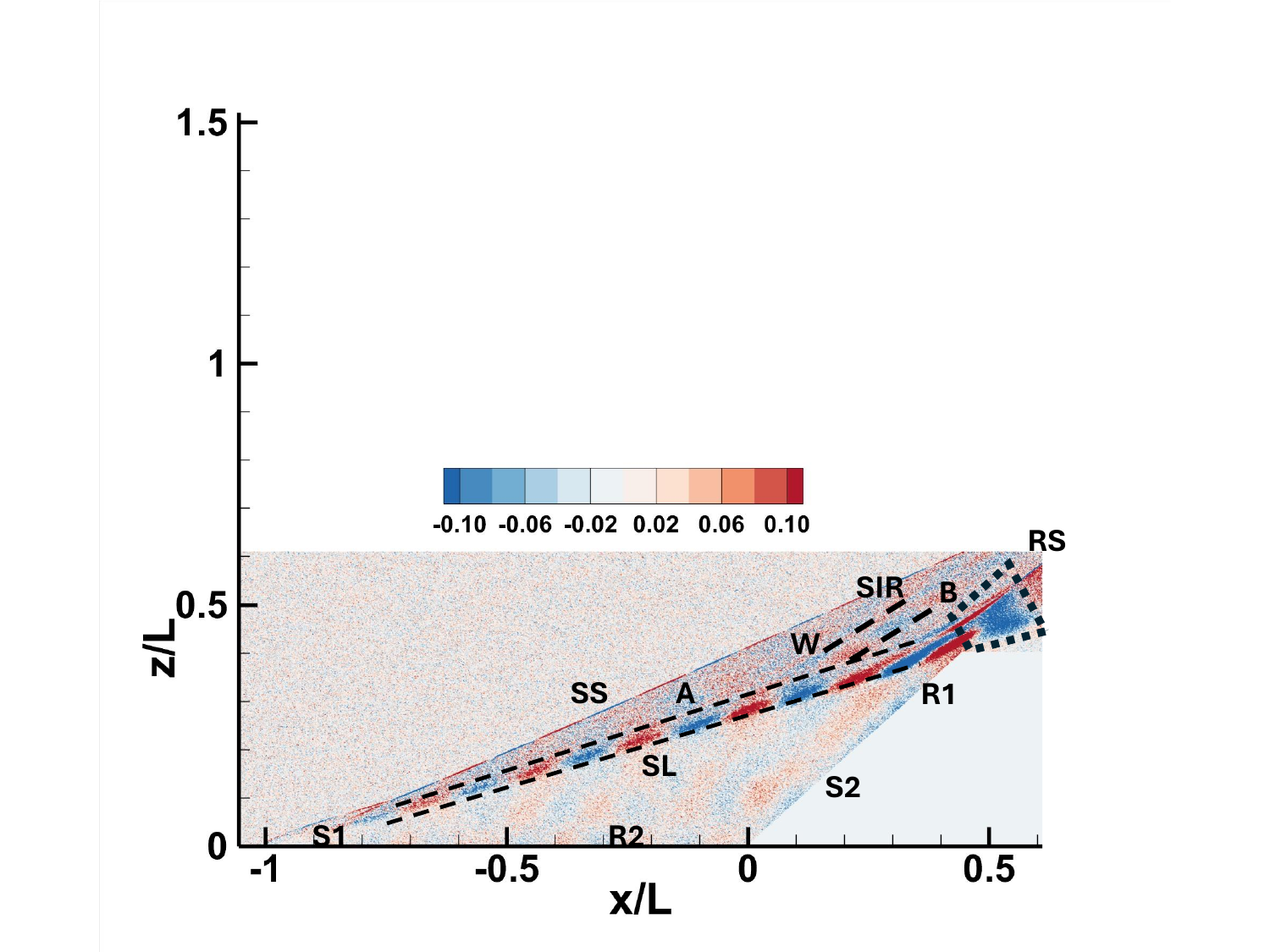}}
        \subfigure[St$_L$=2.08]{\label{fig:DMDM6NdenMode2}\includegraphics[trim=80 10 110 250,clip,width=0.47\linewidth]{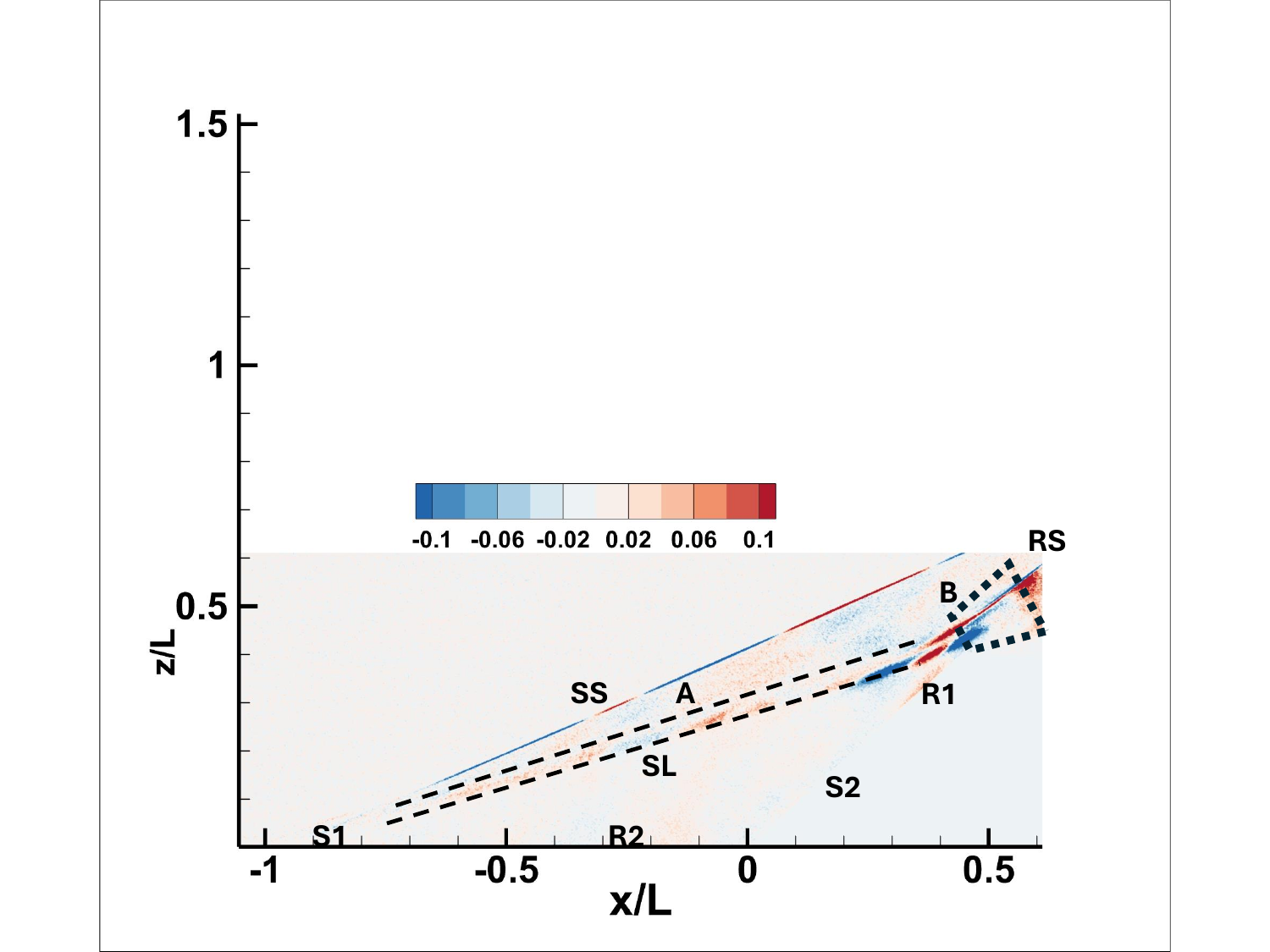}}
        \subfigure[St$_L$=4.75]{\label{fig:SPODM6NdenMode3}\includegraphics[trim=80 10 110 250,clip,width=0.47\linewidth]{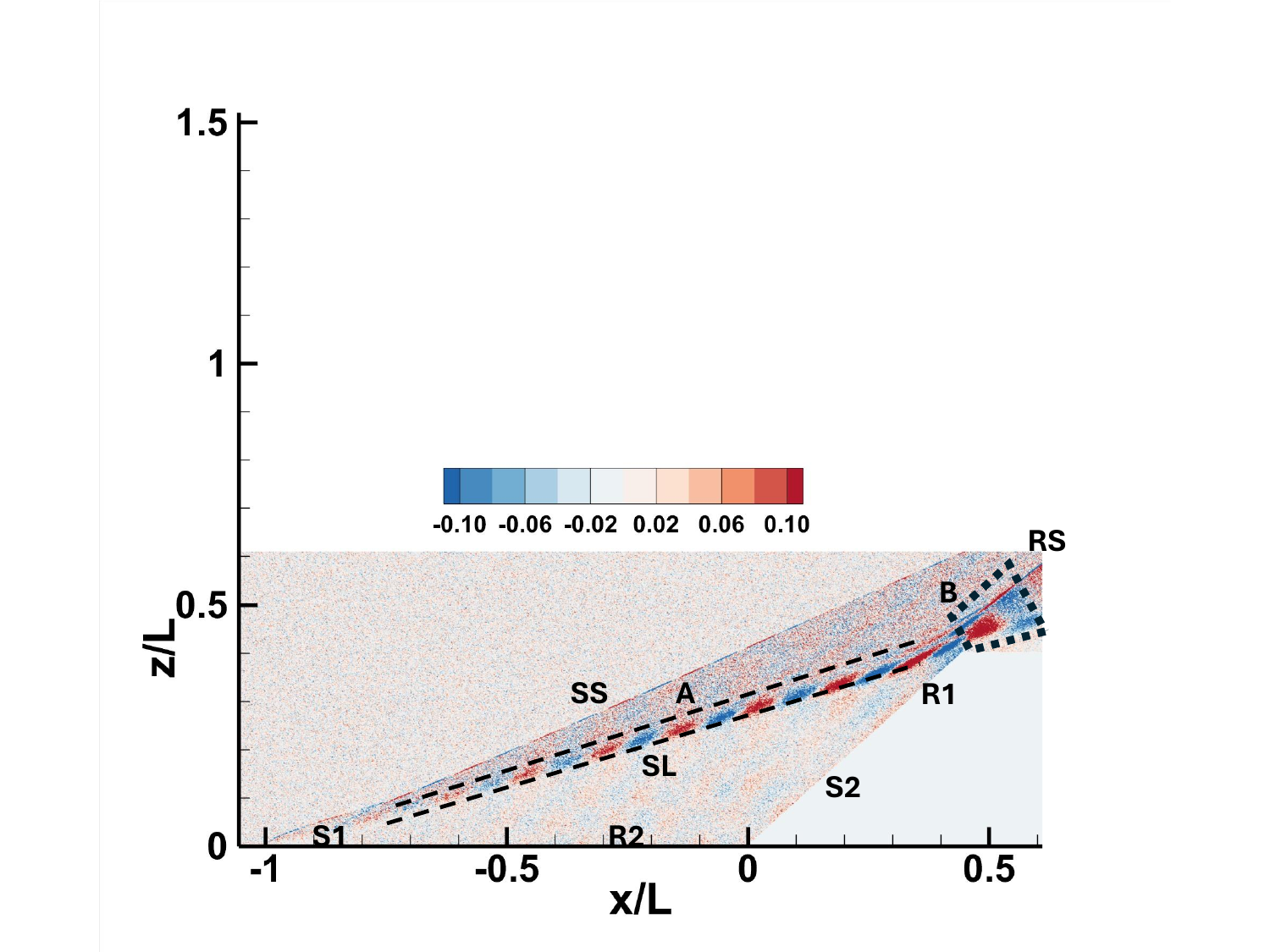}}
        \subfigure[St$_L$=4.16]{\label{fig:DMDM6NdenMode3}\includegraphics[trim=80 10 110 250,clip,width=0.47\linewidth]{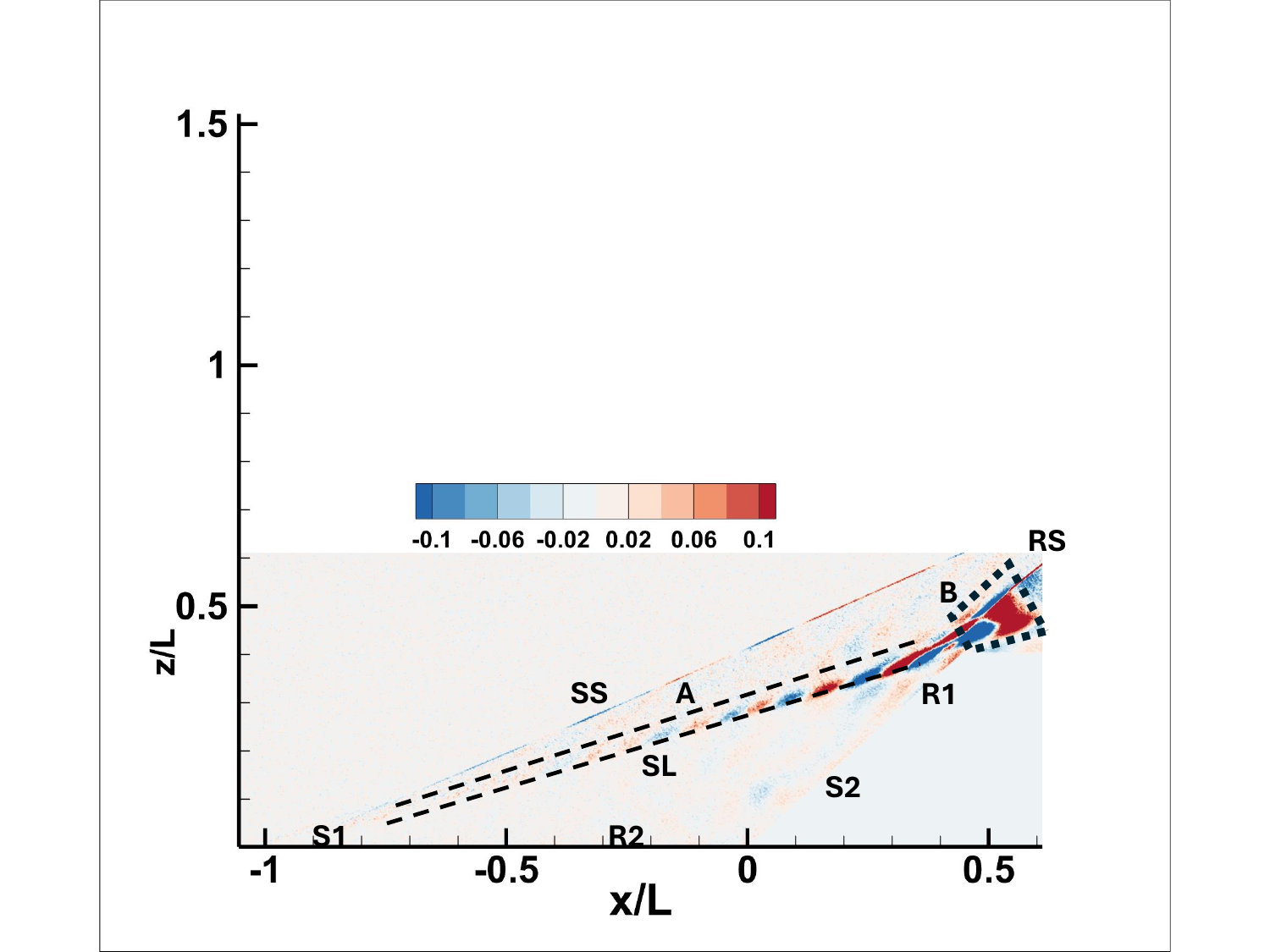}}
    \caption{Comparison of SPOD (left) and DMD (right) modes for number density fluctuations for M6 configuration.  The subfigure legends correspond to
    the real part of $\boldsymbol{\Psi}_{f_k}$ of Eq.~\ref{eqn:SPODEvalEvec} for SPOD and
     the real part of $\mathbf{W} $ given by Eq.~\ref{eqn:DMDmodes} for DMD modes, each self-normalized.
    S1: primary separation, R1: primary reattachment, S2: secondary separation, R2: secondary reattachment, SS: separation shock, SL: shear layer, RS: reattachment shock, SIR: shock interaction region, in this figure and for similar subsequent figures. A, B and W indicate the regions with the most prominent unsteady structures.}
    \label{fig:SPODDMDModesNdenM6}
\end{figure}

Turning to the coherent structures seen in the real parts of the eigenvectors from the SPOD and DMD analyses,
Fig.~\ref{fig:SPODDMDModesNdenM6} shows their spatial distribution based on number density fluctuations.  The modes corresponding to $St_L=1.31, 3.43, 4.75$ for SPOD in Figs.~\ref{fig:SPODM6NdenMode1}, ~\ref{fig:SPODM6NdenMode2} and ~\ref{fig:SPODM6NdenMode3} and $St_L=1.04, 2.08, 4.16$ for DMD analysis in Figs.~\ref{fig:DMDM6NdenMode1}, ~\ref{fig:DMDM6NdenMode2} and ~\ref{fig:DMDM6NdenMode3} are presented first.
A number of  common structures can be observed in all the sub-figures of Fig.~\ref{fig:SPODDMDModesNdenM6} especially where the modes peak within the shear layer and near the reattachment shock foot,  but with slight differences for each frequency component. First of all, in all frequencies and with both methods coherent structures exist in the shear layer and move upstream and peak at the reattachment/expansion location immediately downstream of  the expansion corner point for the number density fluctuations, (labeled A), but, they are  more powerful at  higher frequencies. These propagating disturbances within the shear layer resemble Kelvin-Helmholtz type instabilities~\cite{Schmidt_Towne_Rigas_Colonius_Brès_2018}. Another common mechanism observed at all frequencies and with both methods is the structure due to the reattachment shock oscillations (labeled B). Comparing the modes captured by SPOD and DMD for these three frequencies, it can be seen that SPOD captures significantly more structures than the DMD analysis.  For example structures are observed within the entire shear layer for the SPOD modes whereas they are only observed closer to the expansion corner in the DMD modes. Another point worth noting for this M6 configuration is that the shock interaction caused by the transmitted leading edge shock and shear layer (labeled SIR) also contributes to the unsteadiness for the frequency $St_L=3.43$ as can be seen from the SPOD mode shown in Fig.~\ref{fig:SPODM6NdenMode2}, (labeled W).

\begin{figure}[h!]
    \centering
        \subfigure[St$_L$=1.31]{\label{fig:SPODM6UvelMode1}\includegraphics[trim=80 10 110 250,clip,width=0.47\linewidth]{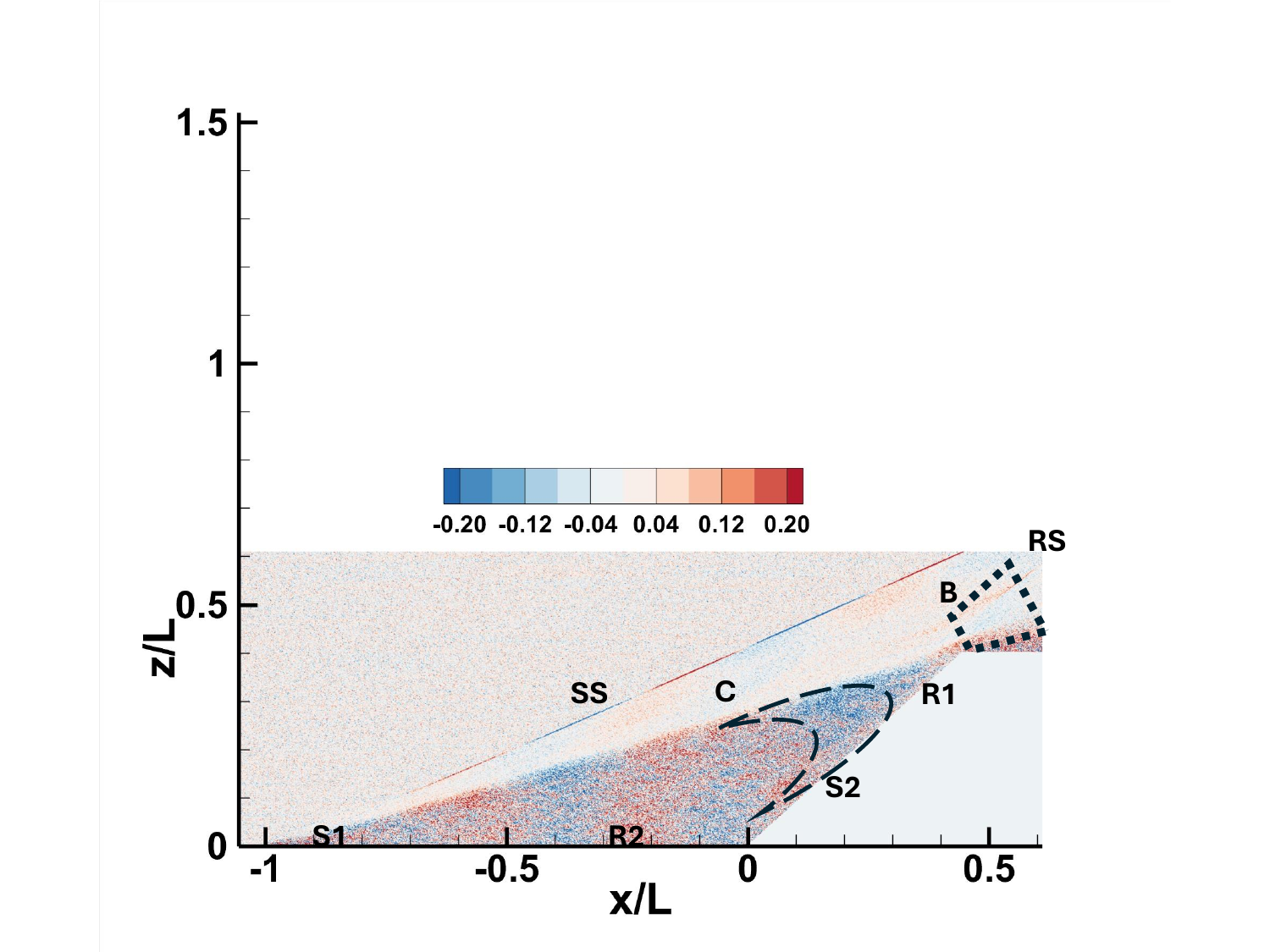}}
        \subfigure[St$_L$=0.83]{\label{fig:DMDM6UvelMode1}\includegraphics[trim=80 10 110 250,clip,width=0.47\linewidth]{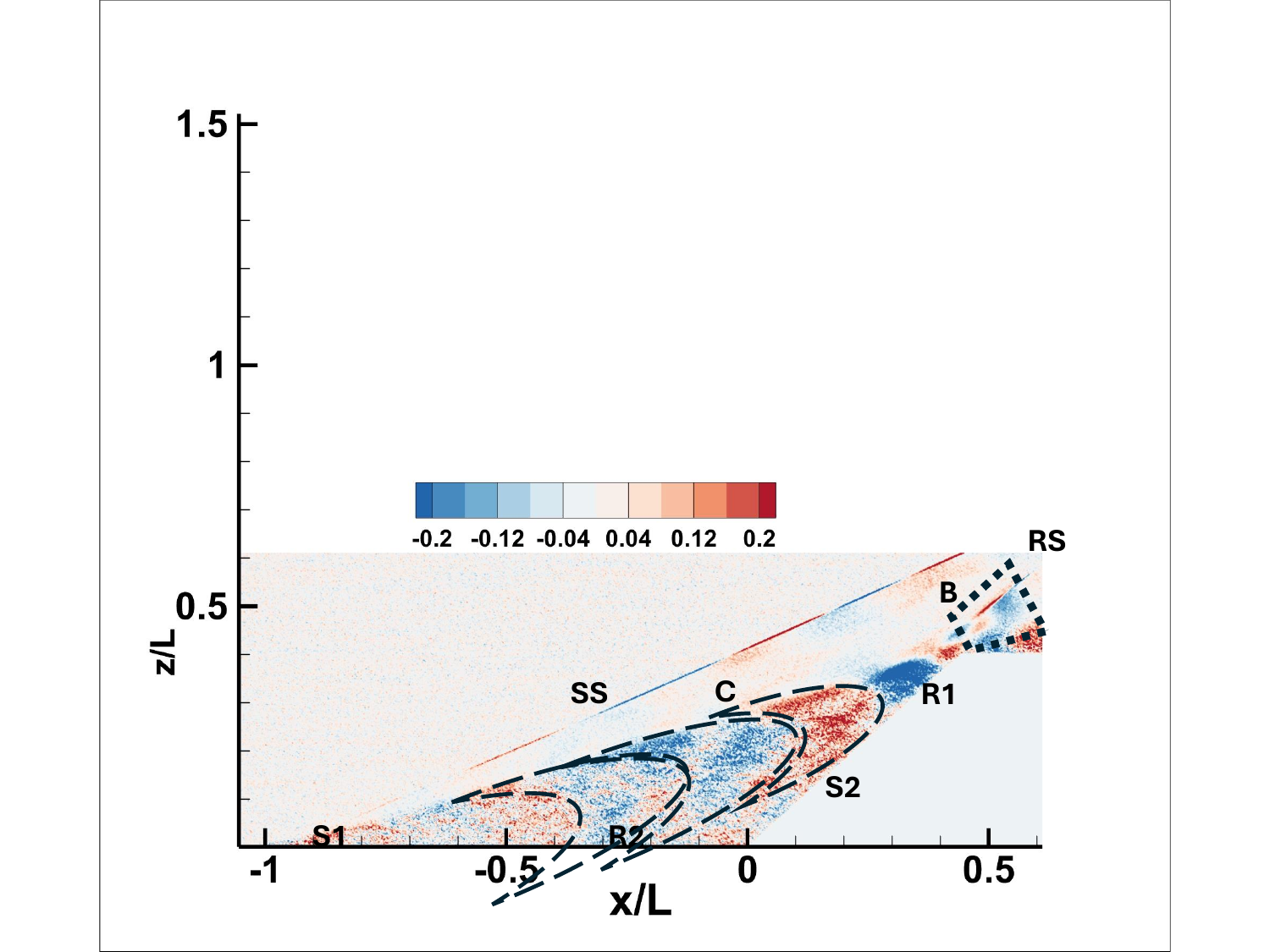}}
        \subfigure[St$_L$=3.43]{\label{fig:SPODM6UvelMode2}\includegraphics[trim=80 10 110 250,clip,width=0.47\linewidth]{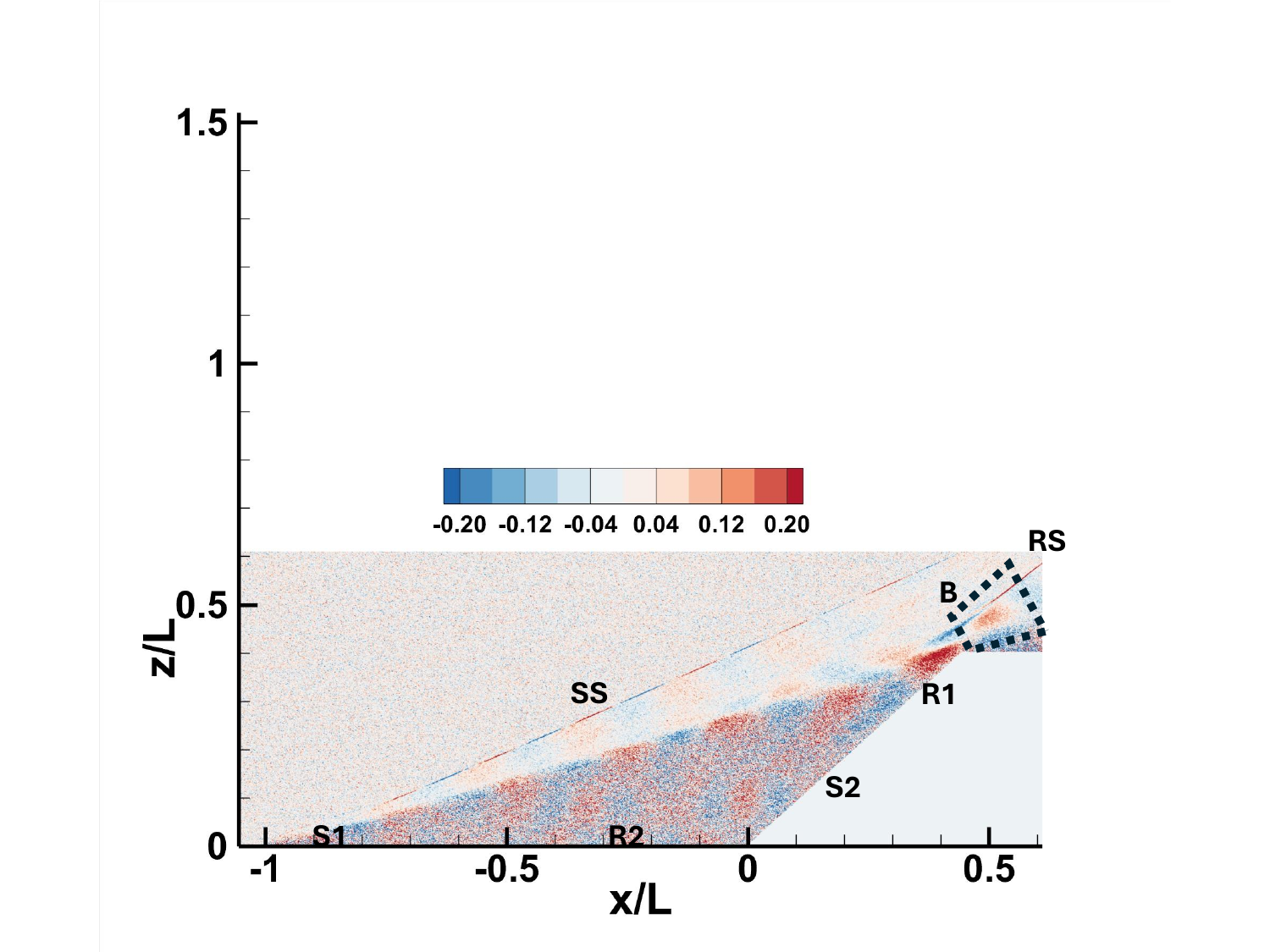}}
        \subfigure[St$_L$=3.75]{\label{fig:DMDM6UvelMode3}\includegraphics[trim=80 10 110 250,clip,width=0.47\linewidth]{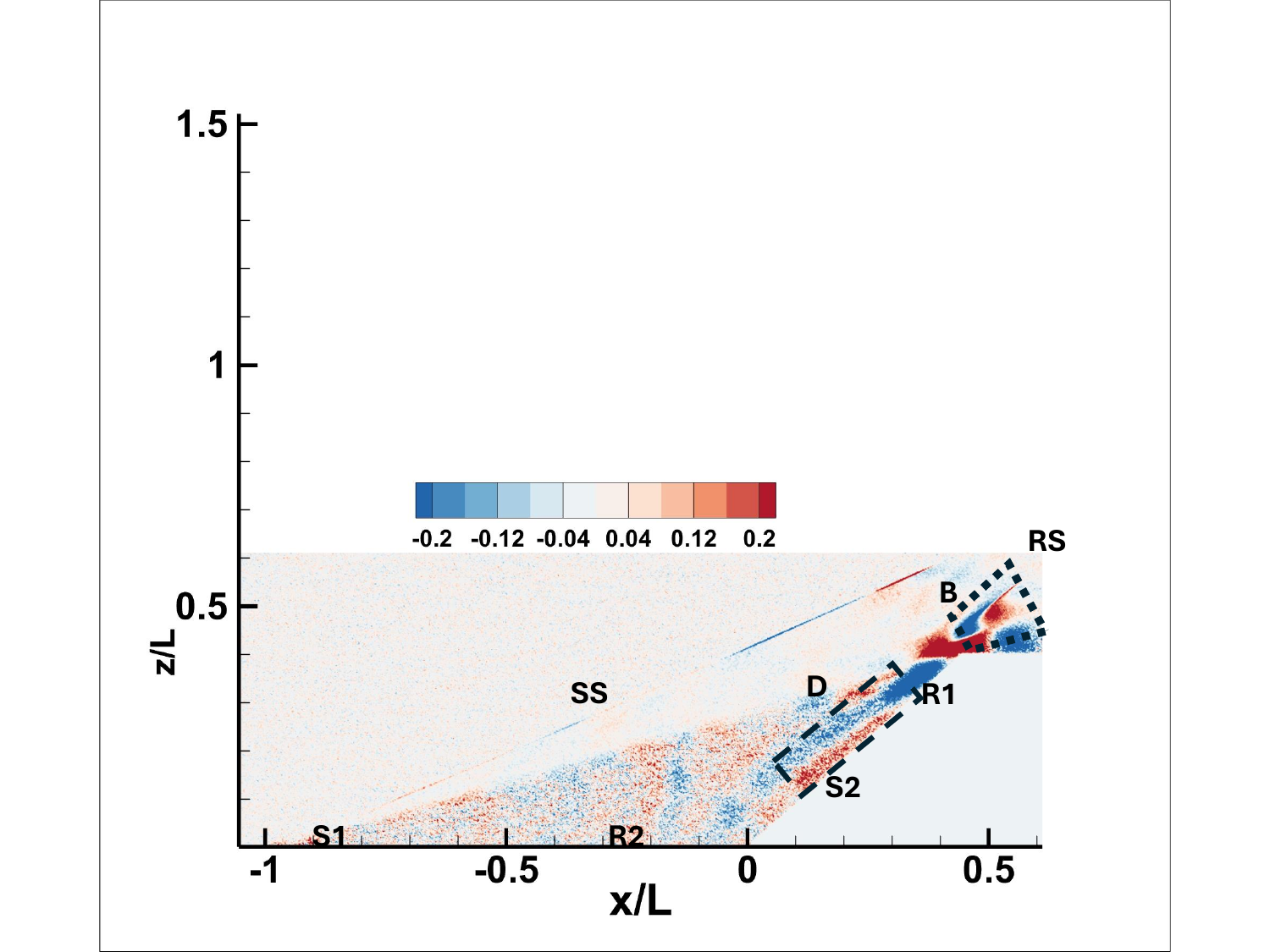}}
    \caption{Comparison of SPOD (left) and DMD (right) modes for streamwise velocity fluctuations for M6 configuration. Black dashed lines added to emphasize presence of c-shaped mode.}
    \label{fig:SPODDMDModesUvelM6}
\end{figure}

Similarly, the modes peak at the expansion corner locations for the streamwise velocity component, as shown in Fig.~\ref{fig:SPODDMDModesUvelM6} for all frequencies  for both SPOD and DMD modes. The frequency values are the same as before for SPOD modes in Figs.~\ref{fig:SPODM6UvelMode1} and ~\ref{fig:SPODM6UvelMode2} but the results are shown for $St_L=0.83, 3.75$ for DMD analysis in in Figs.~\ref{fig:DMDM6UvelMode1} and ~\ref{fig:DMDM6UvelMode3}. In addition to the contribution of the reattachment shock oscillations, there is a global structure existing within the separation bubble region, that is especially clear in Figs.~\ref{fig:SPODM6UvelMode1} and ~\ref{fig:DMDM6UvelMode1}, (labeled C). This mode is analogous to the c-shaped mode seen in the linear stability analysis of the baseline configuration~\cite{karpuzcu2024linear}. Moreover, the oblique waves have distinct phase shift going through the shear layer. For the higher frequencies the structures become more elongated and for the case of $St_L=3.43$ and $St_L=3.75$ in Figs.~\ref{fig:SPODM6UvelMode2} and ~\ref{fig:DMDM6UvelMode3} there is a distinct elongated structure along the primary reattachment region of the flow (labeled D). In the lower frequencies, both the leading edge shock and separation shock has more dominant coherent structures for both number density and streamwise velocity component in both data-driven methods. DMD analyses resulted in fewer meaningful modes for the streamwise velocity fluctuations compared to SPOD modes, demonstrating once again it requires more data than SPOD to capture similar coherent structures.

In summary, key mechanisms for the unsteadiness in this M6 configuration are identified as the shock oscillations of the reattachment shock, Kelvin-Helmholtz type instability within the shear layers and c-shaped mode within the separation bubble. It is important to note that each of these types of unsteadiness/instability mechanisms are related to different flow physics and the appearance of a range of unsteady frequencies rather than a single frequency band is justified as there are competing physical mechanisms that drives the unsteadiness in this complex flowfield.

\subsection{Configuration Tw175: The Effect of Reduced Wall Temperature}

\begin{figure}[htbp]
\center
\subfigure[PSD of DSMC pressure data]{\label{fig:Tw175PSD}\includegraphics[trim=30 30 30 30,clip,width=0.45\linewidth]{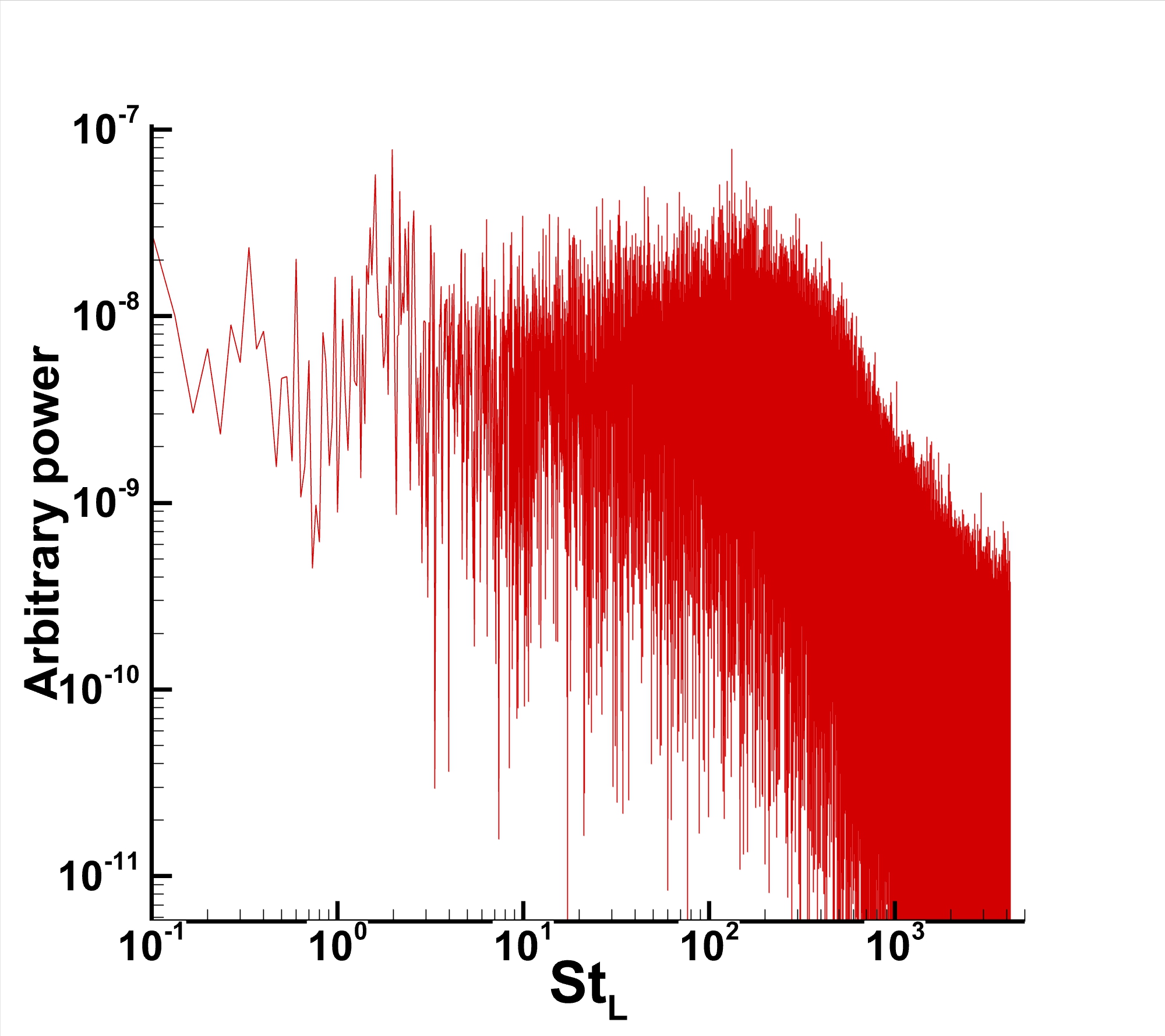}}
\subfigure[SPOD spectra of number density]{\label{fig:SPODSpectraNdenTw175}\includegraphics[trim=60 10 80 30,clip,width=0.45\linewidth]{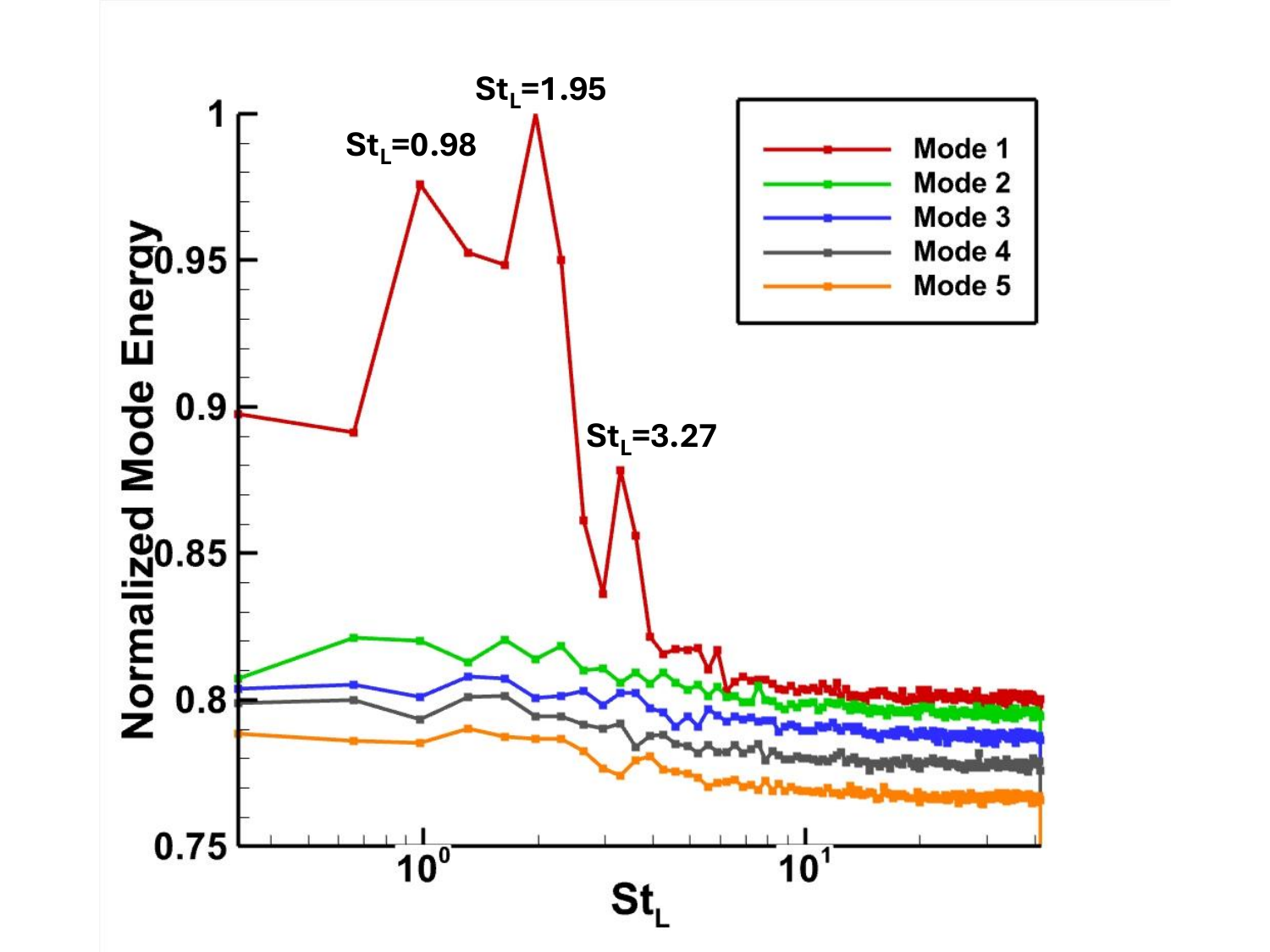}}
\subfigure[SPOD spectra of streamwise velocity]{\label{fig:SPODSpectraUvelTw175}\includegraphics[trim=60 10 80 30,clip,width=0.45\linewidth]{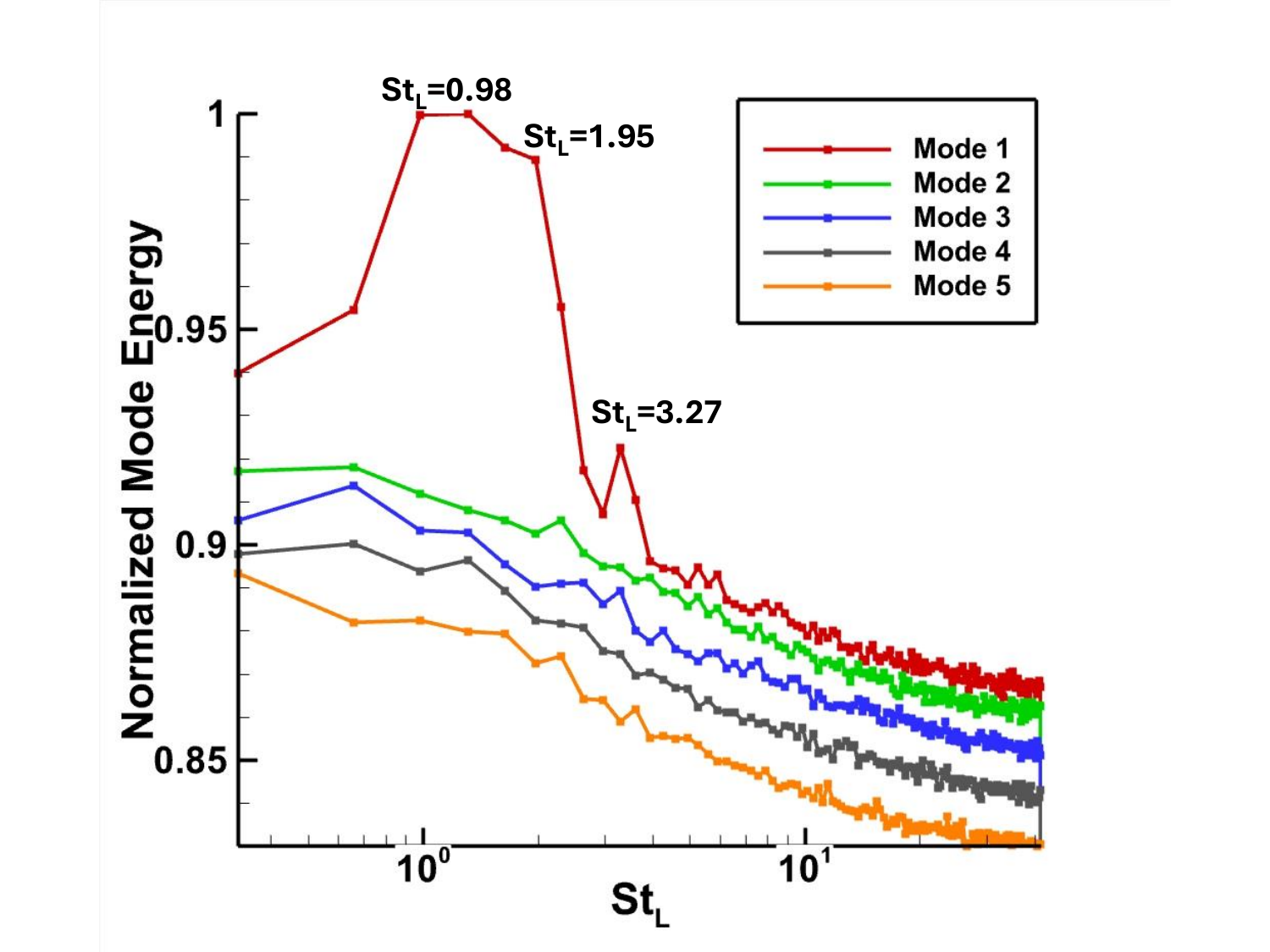}}
\subfigure[SPOD spectra of mode 1 for number density by varying N$_{FFT}$]{\label{fig:SPODTw175Nfft}\includegraphics[trim=10 10 30 30,clip,width=0.45\linewidth]{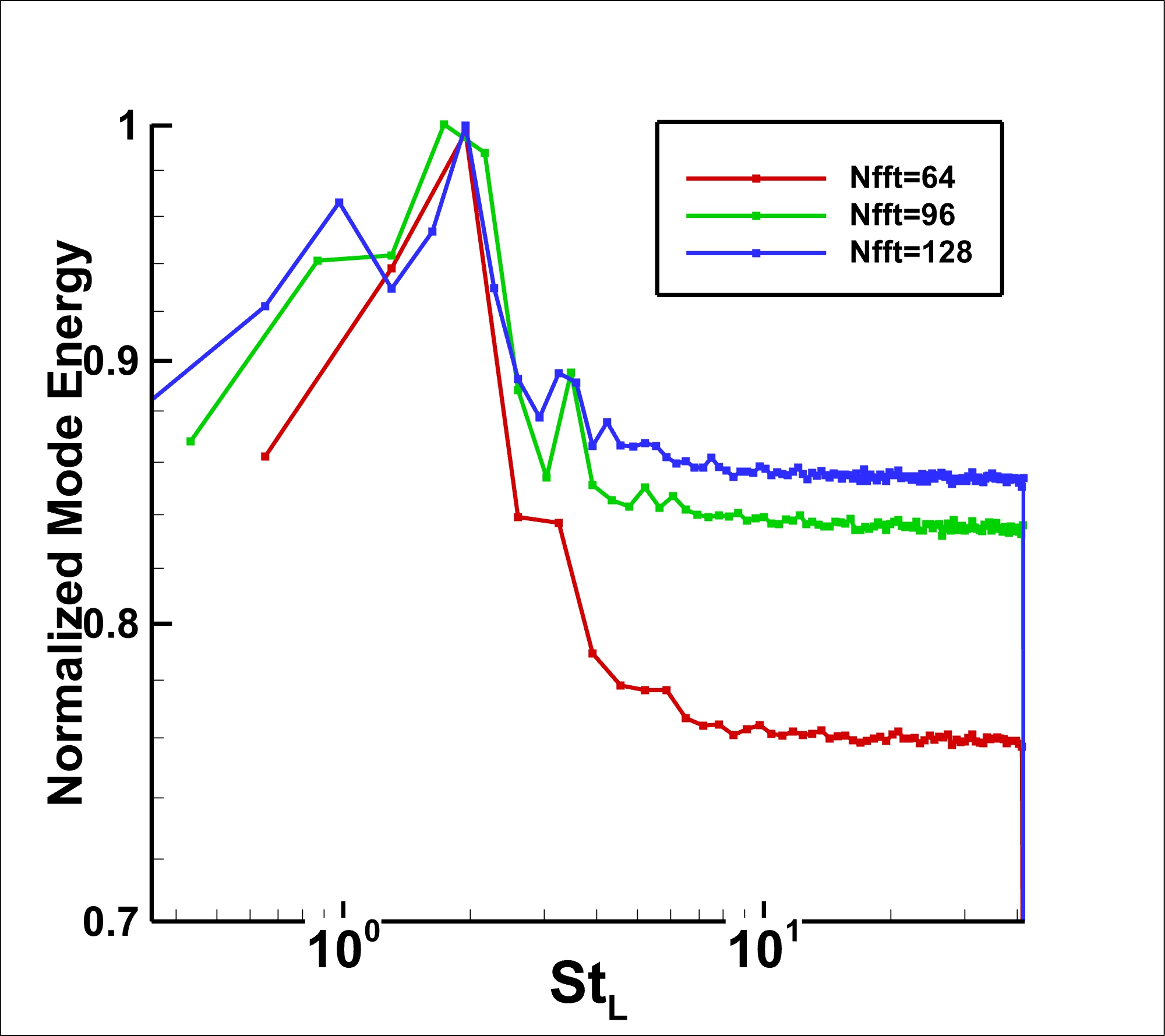}}
\caption{Pressure fluctuations obtained from PSD, SPOD spectra of number density and streamwise  velocity and the comparison of SPOD spectra for different number of FFT frequencies (N$_{FFT}$) for the Tw175 configuration.}
\label{fig:Tw175SPODSpectra}
\end{figure}

The second configuration that is considered is the case where the  wall temperature is reduced, the Tw175 configuration.
Since this configuration is at M=3, it does not have the Edney type interactions caused by the transmitted leading edge shock that was observed in the M6 configuration. Similar to the M6 configuration, first the PSD of the pressure signal obtained from a DSMC numerical probe placed near the expansion corner was investigated and it was observed that there is a low frequency unsteadiness present around   the $St_L=1-3$ band in this configuration too, as can be  seen in Fig.~\ref{fig:Tw175PSD}.

\begin{figure}[htbp]
\center
\subfigure[DMD spectra of number density]{\label{fig:DMDSpectraNdenTw175}\includegraphics[trim=80 10 110 100,clip,width=0.45\linewidth]{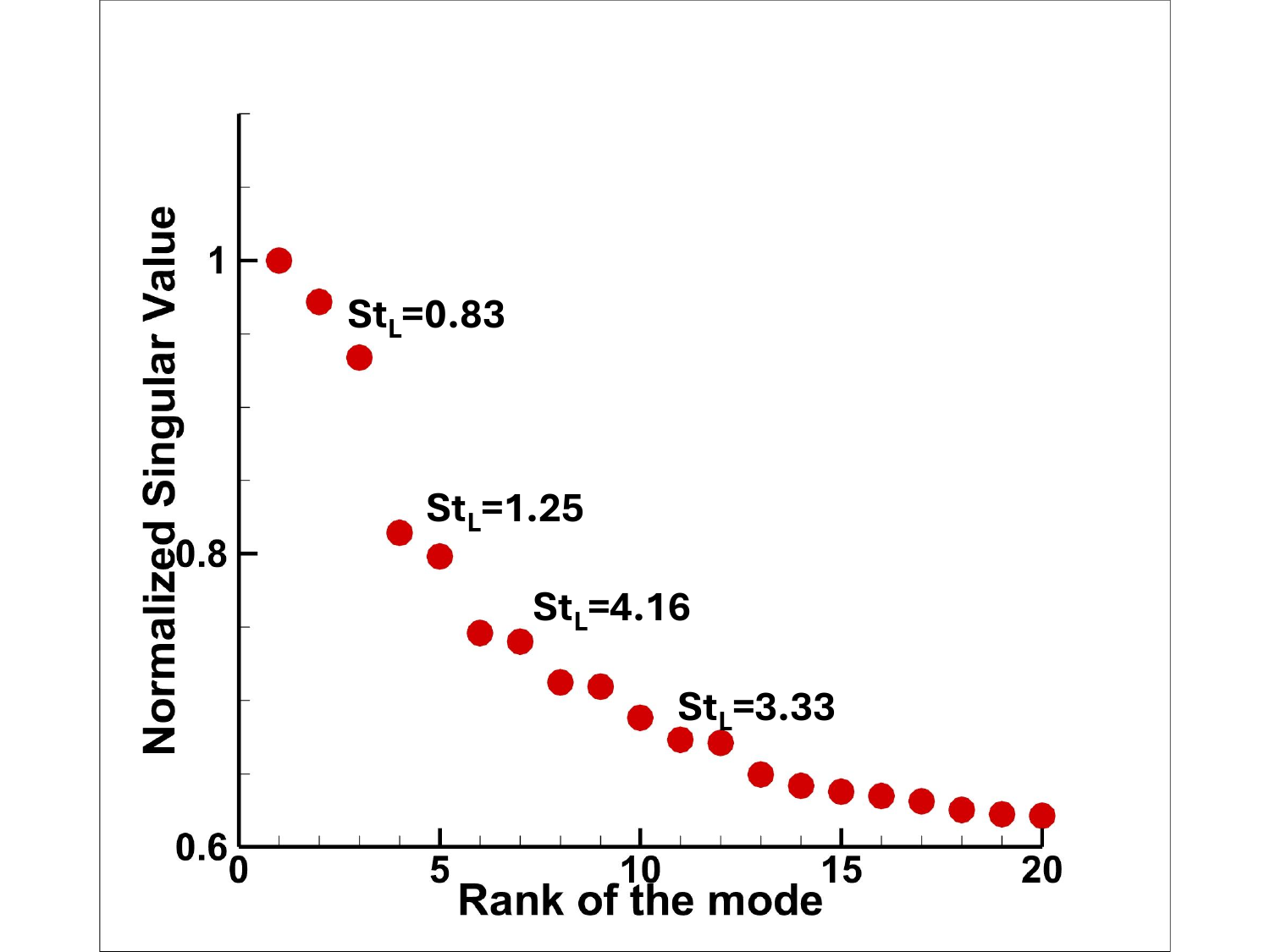}}
\subfigure[DMD spectra of streamwise velocity]{\label{fig:DMDSpectraUvelTw175}\includegraphics[trim=80 10 110 100,clip,width=0.45\linewidth]{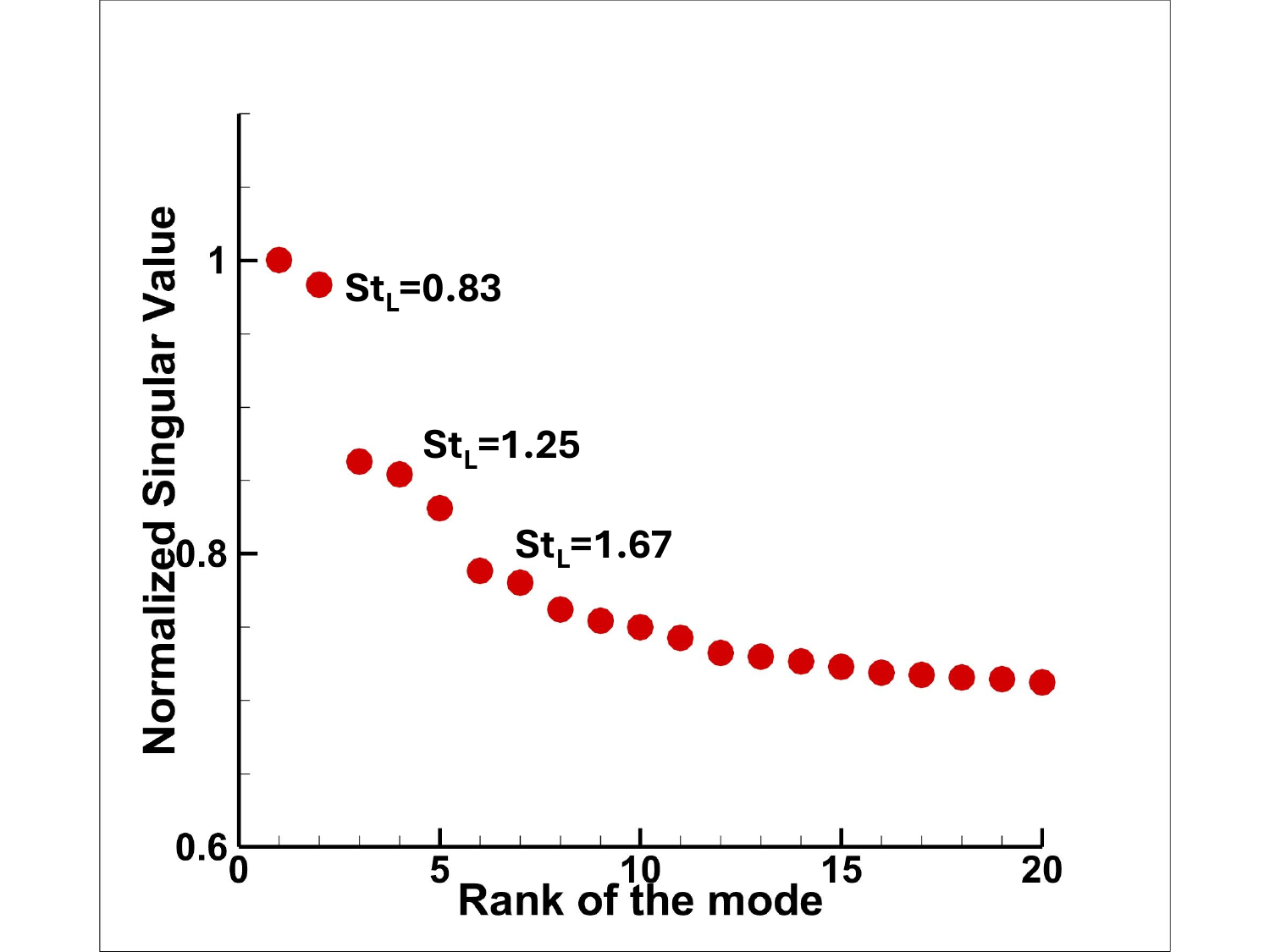}}
\subfigure[Singular Decay]{\label{fig:SingDecayTw175}\includegraphics[trim=10 10 30 30,clip,width=0.45\linewidth]{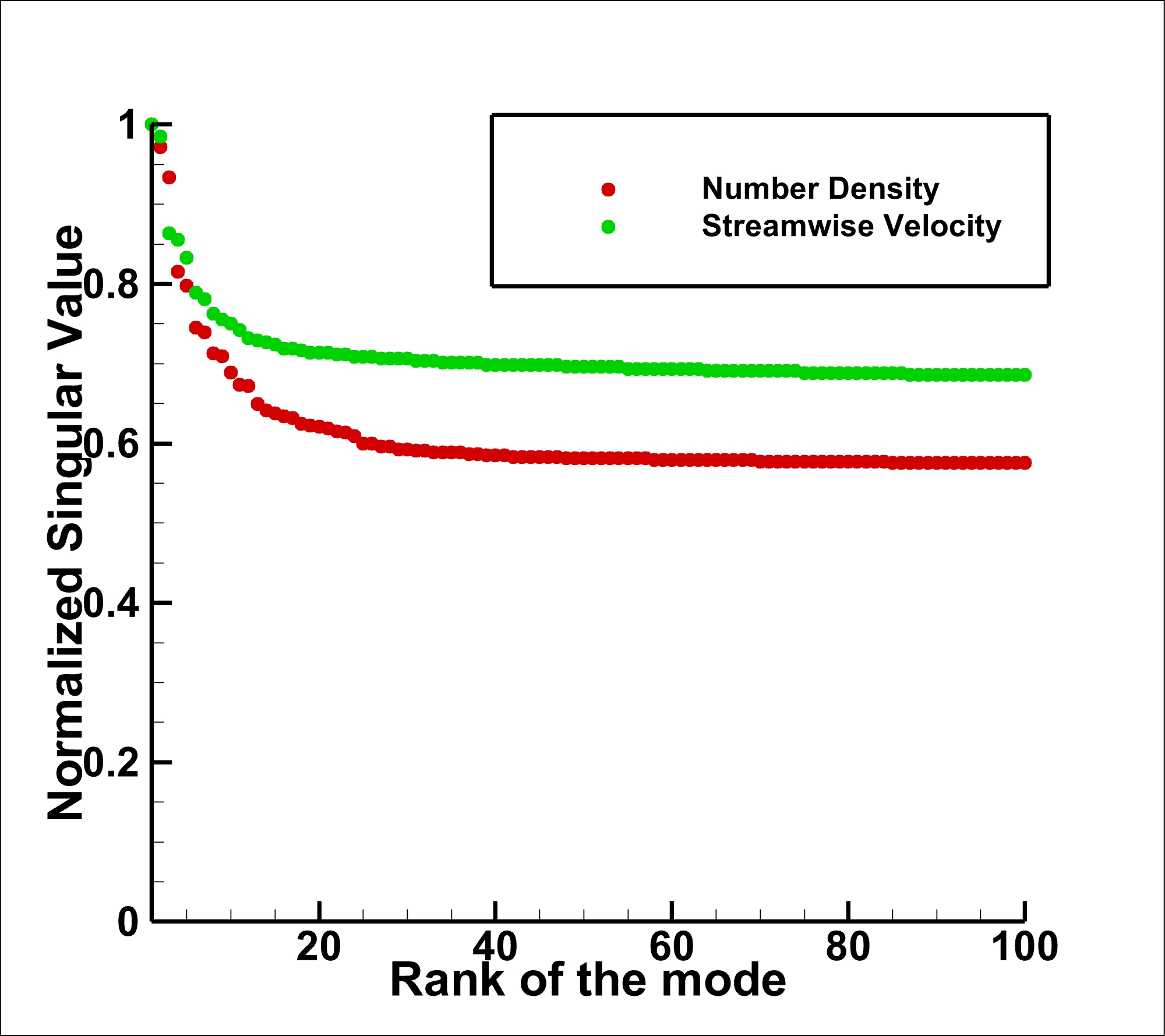}}
\caption{Leading DMD modes for number density and streamwise velocity and the singular value decay for Tw175 configuration and corresponding frequencies are inserted near symbols to the modes as well.}
\label{fig:Tw175DMDSpectra}
\end{figure}

To further quantify this unsteadiness, both SPOD and DMD analyses were performed using 2,000  instantaneous DSMC flowfields. From the frequency spectra shown in Figs.~\ref{fig:SPODSpectraNdenTw175} and ~\ref{fig:SPODSpectraUvelTw175} low rank zones can be identified between $St_L=1-3$ in the regions between modes 1 and 2 for both the number density and streamwise velocity values in the SPOD analysis.   Their counterparts are also  observed to be leading modes in the DMD analysis marked in Figs.~\ref{fig:DMDSpectraNdenTw175} and ~\ref{fig:DMDSpectraUvelTw175}. This further confirms that the unsteadiness seen in a single DSMC probe indeed exists across the flowfield. The shapes and peaks of the SPOD spectra in Figs.~\ref{fig:SPODSpectraNdenTw175} and \ref{fig:SPODSpectraUvelTw175} are similar to the PSD spectra of Fig.~\ref{fig:Tw175PSD}., i.e. there is a broadband frequency region observed in PSD near St$_L=1$ and there are peaks in SPOD spectra at St$_L$= 0.98 and 1.95. To check for the numerical convergence of SPOD analysis for this configuration as well, N$_{FFT}$ is varied and the results are shown in Fig.~\ref{fig:SPODTw175Nfft}. It can be clearly seen that variation of FFT frequencies does not affect the unsteadiness frequencies captured in this work as all of them captures St$_L=0.98, 1.95, 3.27$.  In addition, singular value decay plot in Fig.~\ref{fig:SingDecayTw175} shows that our choice of r=20 of the DMD analysis is enough to capture the necessary mode energy as the singular values after r=20 starts to have a constant energy and very low distinction from each other for both number density and streamwise velocity fluctuations.

\begin{figure}[h!]
    \centering
        \subfigure[St$_L$=0.98]{\label{fig:SPODTW175NdenMode1}\includegraphics[trim=80 10 110 250,clip,width=0.47\linewidth]{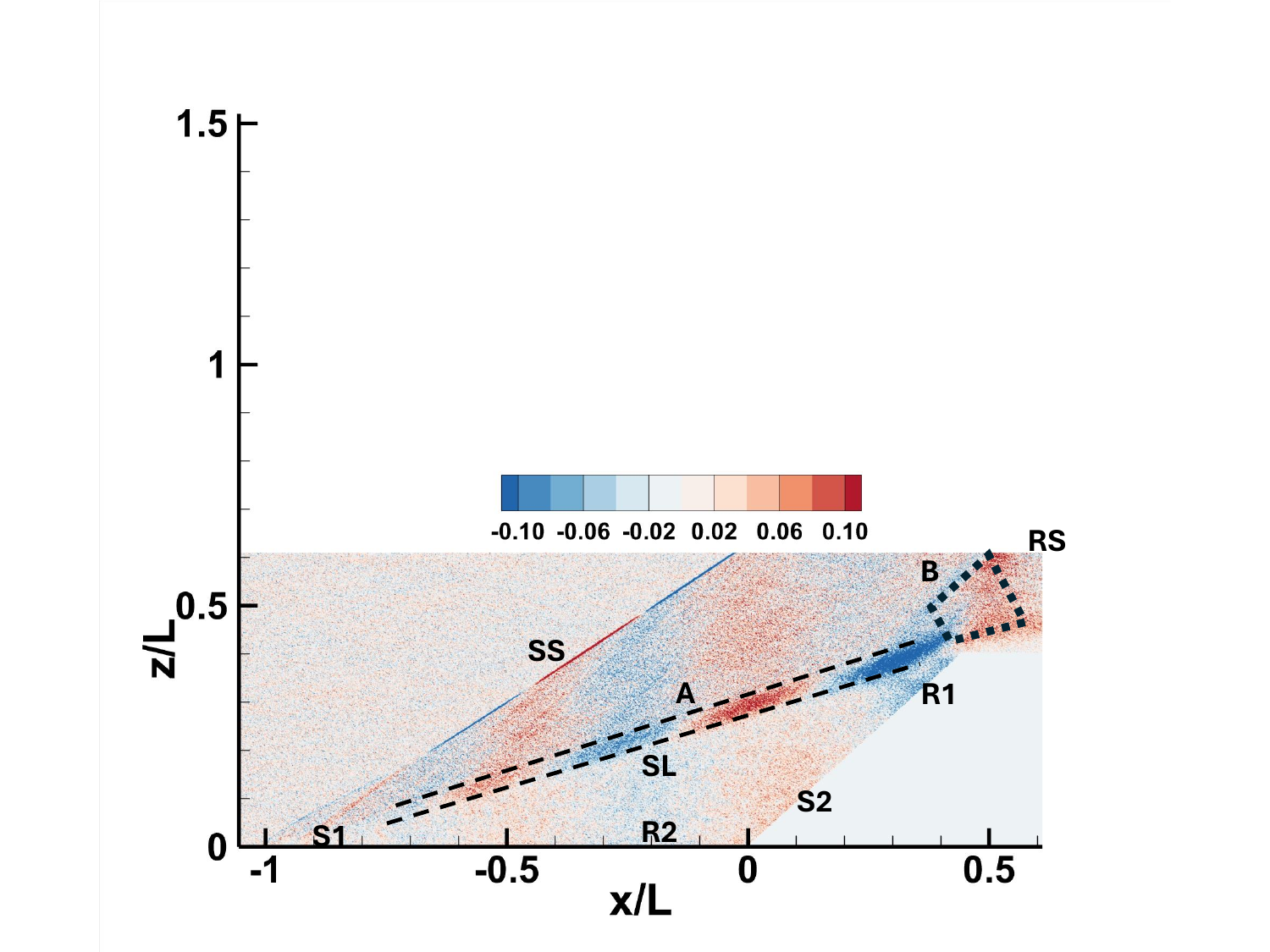}}
        \subfigure[St$_L$=1.25]{\label{fig:DMDTW175NdenMode1}\includegraphics[trim=80 10 110 250,clip,width=0.47\linewidth]{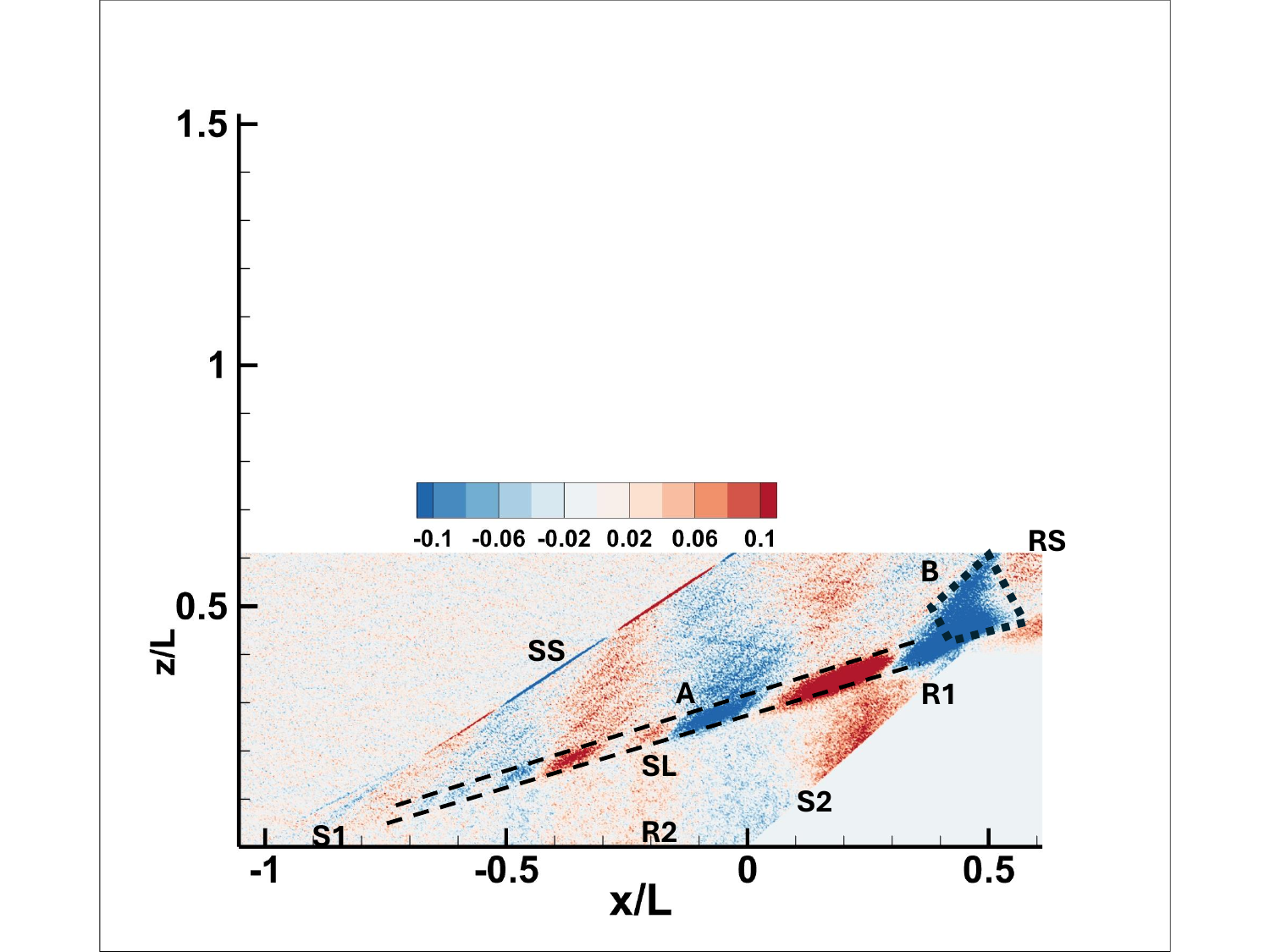}}
        \subfigure[St$_L$=1.95]{\label{fig:SPODTW175NdenMode2}\includegraphics[trim=80 10 110 250,clip,width=0.47\linewidth]{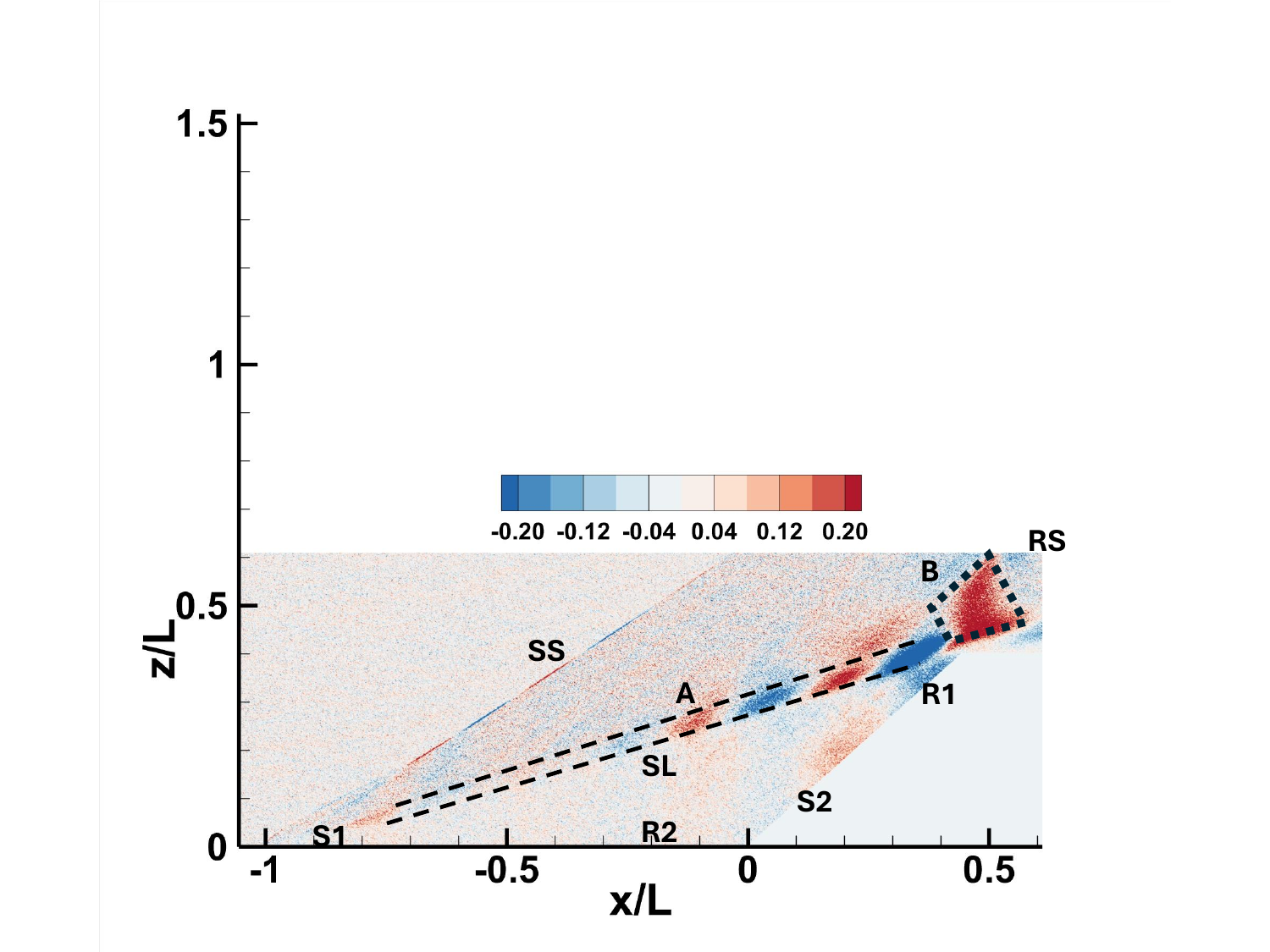}}
        \subfigure[St$_L$=2.08]{\label{fig:DMDTW175NdenMode2}\includegraphics[trim=80 10 110 250,clip,width=0.47\linewidth]{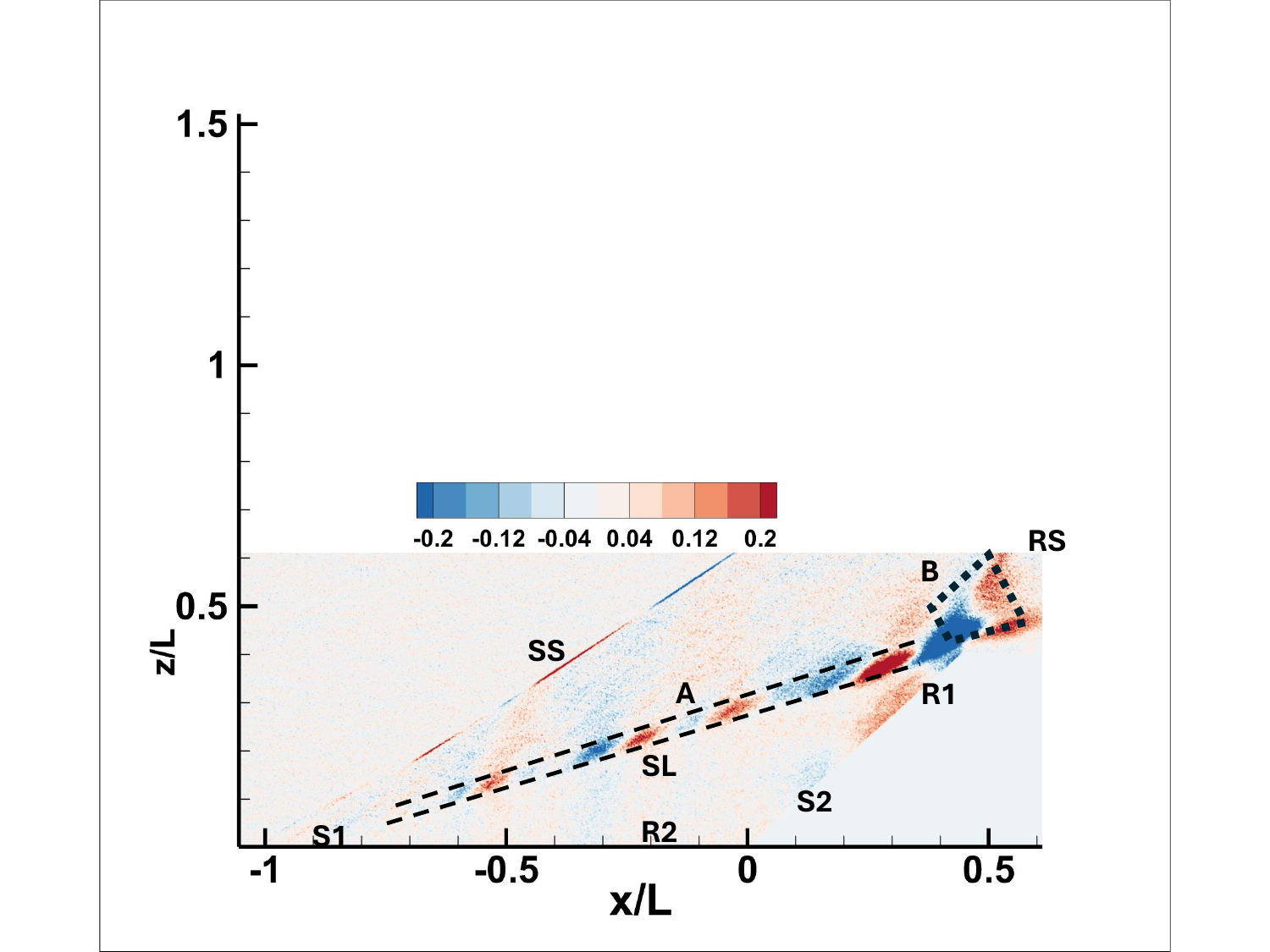}}
        \subfigure[St$_L$=3.27]{\label{fig:SPODTW175NdenMode3}\includegraphics[trim=80 10 110 250,clip,width=0.47\linewidth]{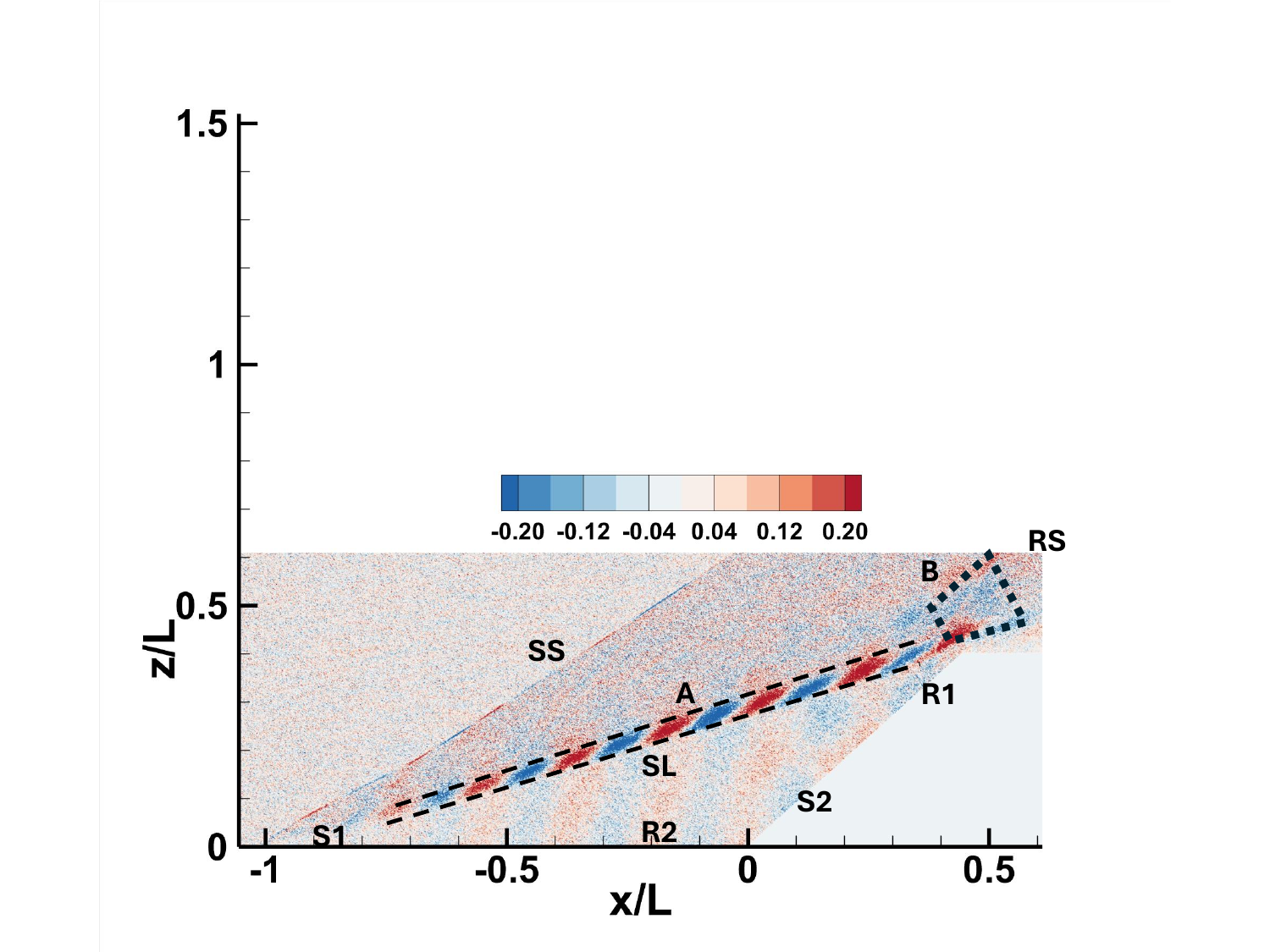}}
        \subfigure[St$_L$=3.33]{\label{fig:DMDTW175NdenMode3}\includegraphics[trim=80 10 110 250,clip,width=0.47\linewidth]{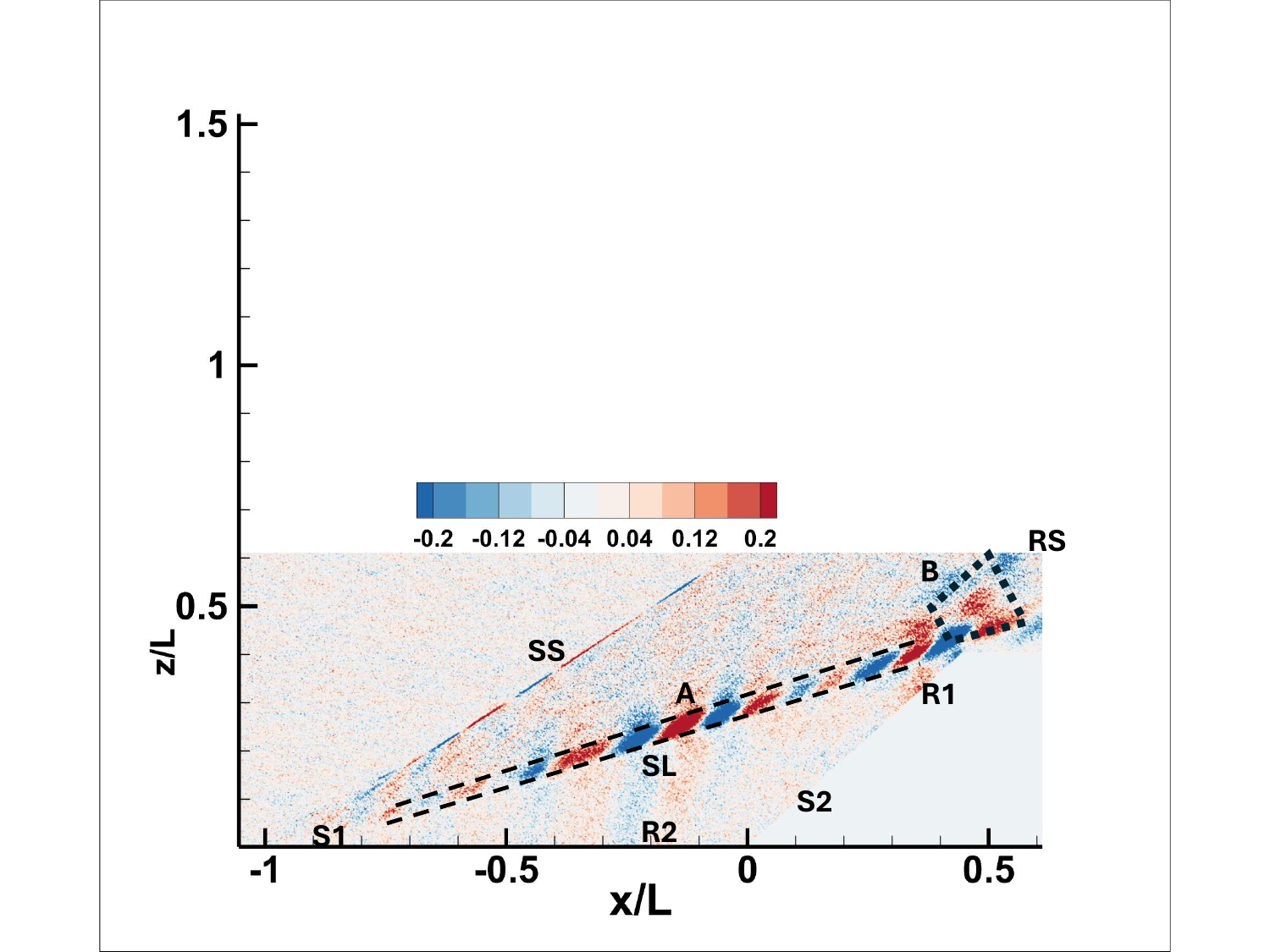}}
    \caption{Comparison of SPOD (left) and DMD (right) modes for number density fluctuations for Tw175 configuration. }
    \label{fig:SPODDMDModesNdenTw175}
\end{figure}

Spatio-temporally coherent structures for the number density fluctuations are shown in Fig.~\ref{fig:SPODDMDModesNdenTw175} for both SPOD and DMD modes. We show the modes corresponding to $St_L=0.98, 1.95, 3.27$ for SPOD in Figs.~\ref{fig:SPODTW175NdenMode1}, ~\ref{fig:SPODTW175NdenMode2} and ~\ref{fig:SPODTW175NdenMode3} and $St_L=1.25, 2.08, 3.33$ of DMD analysis in  Figs.~\ref{fig:DMDTW175NdenMode1}, ~\ref{fig:DMDTW175NdenMode2} and ~\ref{fig:DMDTW175NdenMode3} respectively. Some common features at all frequencies, such as peak disturbances near the reattachment shock and the expansion region at the corner location (labeled B),
can be identified for this configuration as well.
Similar to the M6 case, there is also contribution from the shear layer, (labeled A) in Fig.~\ref{fig:SPODDMDModesNdenTw175}. For the low frequency modes, coherent structures elongate towards the primary separation location, however, at  the higher frequencies they peak more downstream. Also, as mentioned earlier, there is a distinct contribution from the secondary separation location for this case for all the leading modes seen near the S2 location in Figs.~\ref{fig:SPODTW175NdenMode2} and ~\ref{fig:DMDTW175NdenMode2}. Comparison of SPOD and DMD modes for this case shows that DMD seems to perform better compared to the M6 configuration as it can capture almost as many structures as SPOD for all three frequencies of  number density disturbances.

\begin{figure}[h!]
    \centering
    \subfigure[St$_L$=1.95]{\label{fig:SPODTW175UvelMode2}\includegraphics[trim=80 10 110 250,clip,width=0.47\linewidth]{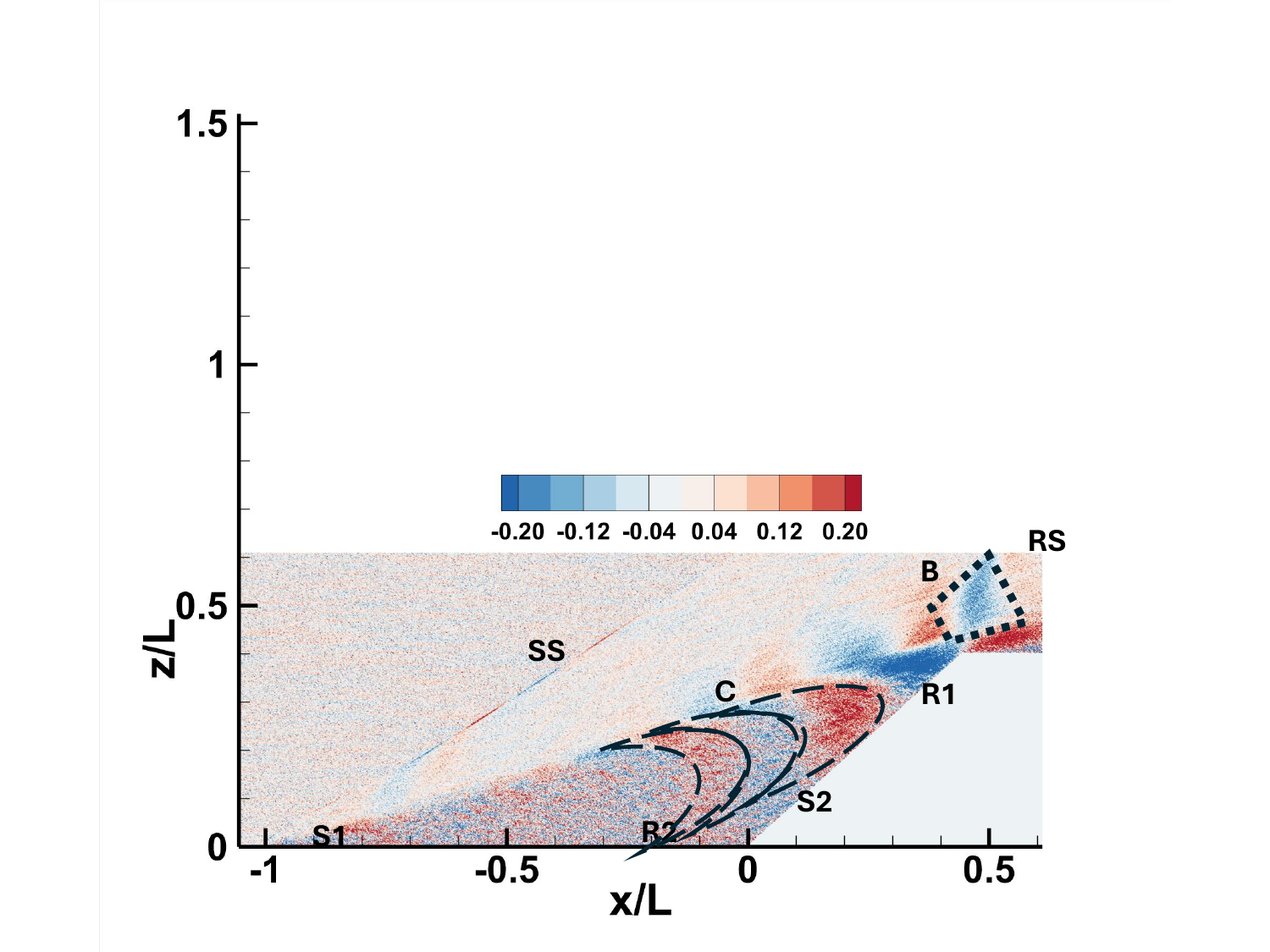}}    \subfigure[St$_L$=2.08]{\label{fig:DMDTW175UvelMode2}\includegraphics[trim=80 10 110 250,clip,width=0.47\linewidth]{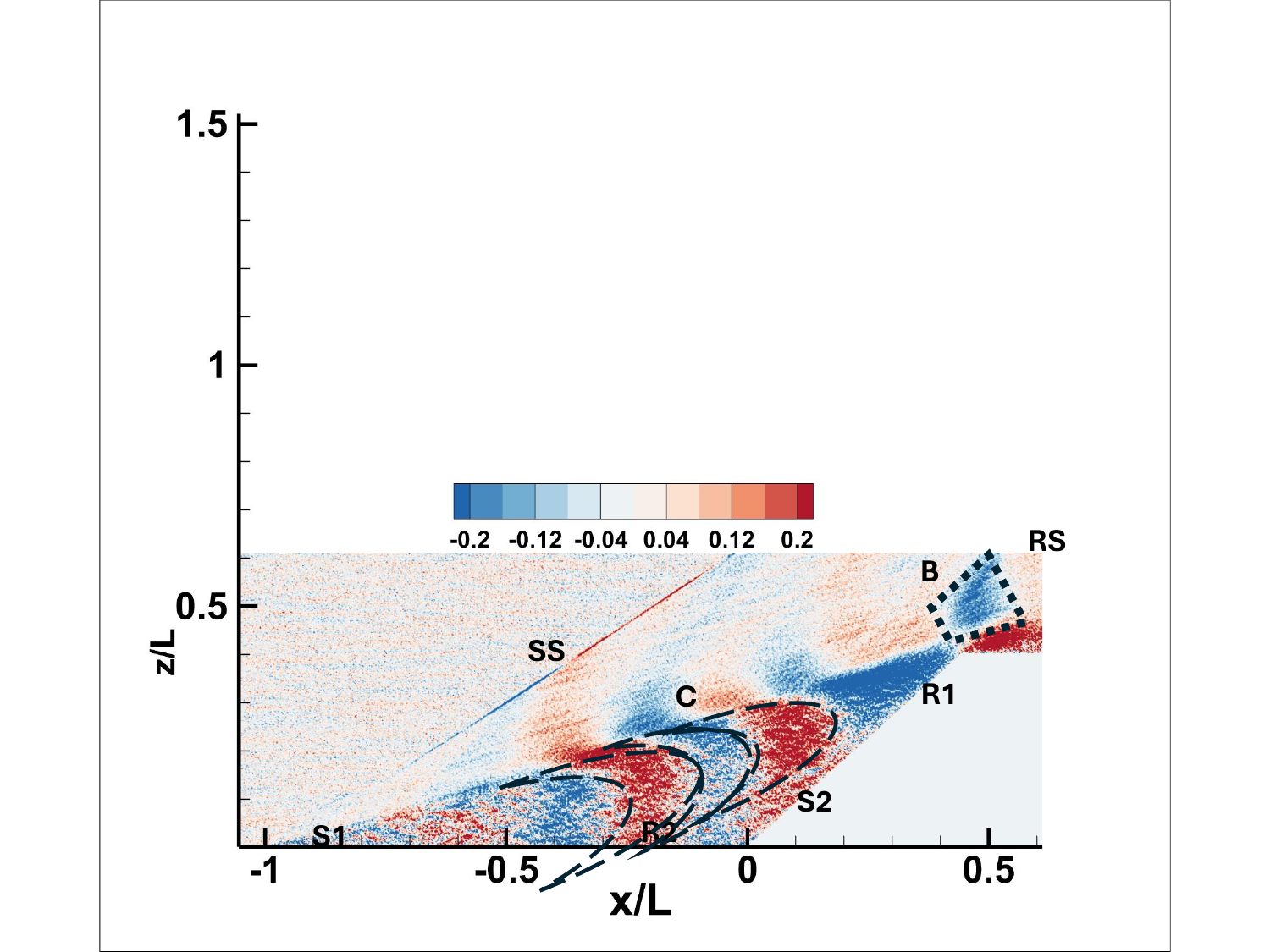}}
    \caption{Comparison of SPOD (left) and DMD (right) modes for streamwise velocity fluctuations for the Tw175 configuration. Black dashed lines added to emphasize presence of c-shaped mode.}
    \label{fig:SPODDMDModesUvelTw175}
\end{figure}

However, for the streamwise velocity fluctuations, DMD struggles to capture distinct frequencies and the captured ones seem to be lumped around $St_L=2.08$, therefore, we only show comparison for one mode.  In Fig.~\ref{fig:SPODDMDModesUvelTw175}, there is a global mode within the separation bubble, with peaks near the expansion corner location. This global mode, a mode that has a single coherent structure with an oscillating pattern derived from the DSMC flowfield,
 again has a c-shaped structure, and is even more distinct than the ones observed in the M6 case in Figs.~\ref{fig:SPODM6UvelMode1} and ~\ref{fig:DMDM6UvelMode1}. It is even more clear in this configuration than the M6 configuration that the c-shaped modes are the dominant mechanisms within the separation bubble, as is well known in the linear stability analysis framework.\cite{karpuzcu2024linear}.
 Also, even though the oblique waves  between the shear layer and the leading edge shock are present in both M6 and Tw175 cases, their appearance is stronger in the M6 case, indicative of more dominant inviscid forces.

In summary, the key mechanisms for the unsteadiness in the Tw175 flow are similar to the M6 configuration and have been identified as the shock oscillations of the reattachment shock, Kelvin-Helmholtz type instabilities within the shear layers and the c-shaped mode within the separation bubble. In this case however, the separation bubble instabilities have similar power of fluctuations as the shear layer instability, as is well demonstrated in the streamwise velocity modes. This outcome is expected as well since for the lower Mach numbers, viscous and inviscid mechanisms are known to have a similar influence on the stability of such flowfields.

\section{Conclusions}\label{sec:conc}

In this study, high speed flows over compression expansion corners for geometries that create  large separation regions were studied using the high fidelity particle-kinetic DSMC and data driven methods.  By varying several free stream parameters, the effect of scaled angle on flow properties was investigated and it was found that the single large recirculation region of the baseline configuration was destabilized with increase in the free stream speed and ramp angle and decrease in the wall temperature. This destabilization was characterized by the appearance of the secondary vortices.

Using  the configurations considered in this work as well as several other previous flow configurations, correlation of the scaled angle with separation length, recirculation strength, secondary recirculation regions and unsteadiness was investigated. It was found that  there is a general trend with increasing scaled angle where both separation length and recirculation strength increases. If the separation length is limited by the geometric constrains of the ramp, configurations with similar scaled angle values result in vastly different flowfields.  This general trend suggested that for such large separation regions, the best length scale to use in the calculation of the scaled angle should be one of physical importance, such as, boundary layer thickness. For similar scaled angle values, destabilization of the flow was achieved once the separation length is maximized, i.e. when it encompasses the whole ramp geometry, and further destabilization of the flow resulted in several recirculation regions in the separation bubble.
It was also shown that several distinct unsteadiness mechanisms can dominate for similar large scaled angle values because use of the scale angle alone elucidate the separate contributions due to free stream Reynolds number, Mach number, and wall temperature. To obtain a  better understanding of the spatial propagation of disturbances in massively unstable separation bubbles data driven methods are required.

To provide  information regarding the spatio-temporal coherence of unsteady flow structures SPOD and DMD were  performed with time accurate DSMC flowfield simulations for the first time.   Two of the four scaled angle cases were selected to study the frequency spectra of density and velocity disturbances for an increased Mach number and a higher wall temperature.  SPOD and DMD showed  similar frequency bands with unsteady modes and were found to also agree with  the numerical probe data from the DSMC simulations.   Three main mechanisms were detected from the modes extracted using SPOD and DMD; {\em i.e.,} Kelvin-Helmholtz, shock oscillations and c-shaped disturbances, each originating from different flow features. Of these two methods, SPOD proved to be the more robust method because it did not require any pre-conditioning of the DSMC flowfields whereas DMD required more sampling of the DSMC results.  The data driven methods employed in this work showed different success levels for the kinetic particle method base flows; however their utilization proved to be useful to differentiate the relative importance of the three mechanisms for different initial conditions demonstrated in this work.

\vspace{-2.5em}
\section*{Acknowledgments}
 The research conducted in this paper is supported by the Office of Naval Research under Grant No. N00014-23-1-2839 entitled "Kinetic Treatment of Sources and Mechanisms that Drive Unsteady, Shock-dominated Flow Instability" with Dr. Eric Marineau as the Program Officer. This research was also supported by computational resources provided by the Texas Advanced Computing Center (TACC) at the University of Texas, which allowed access to the Frontera supercomputer with project number CTS23002. TACC's Frontera is funded by the National Science Foundation (NSF). The authors are grateful to Prof. Vassilis Theofilis for fruitful discussions and guidance on selection of configurations.

\section*{Author Declarations}
\subsection*{Conflict of Interest}
The authors have no conflicts to disclose.

\subsection*{Author Contributions}
\textbf{Irmak T. Karpuzcu:} Conceptualization (equal); Data curation (equal); Formal analysis (equal); Investigation (equal); Methodology (equal); Software (equal); Validation (equal); Visualization (equal); Writing -- original draft (equal). \textbf{Mert Senkardesler:} Formal analysis (equal); Investigation (equal); Methodology (equal); Software (equal); Writing -- original draft (equal). \textbf{Deborah A. Levin:} Funding acquisition (equal); Project administration (equal); Resources (equal); Software (equal); Supervision (equal); Writing -- original draft (equal); Writing -- review \& editing (equal).

\section*{Data Availability}
The data that support the findings of this study are available from the corresponding author upon reasonable request.

\bibliography{DataDrivenCRRefs}
\end{document}